\documentclass[12pt]{article}

\usepackage{latexsym,amssymb,epsfig}
\input{diagrams}

\DeclareFontFamily{OT1}{rsfs10}{}
\DeclareFontShape{OT1}{rsfs10}{m}{n}{ <-> rsfs10 }{}
\DeclareMathAlphabet{\mathscript}{OT1}{rsfs10}{m}{n}

\newcommand{\eref}[1]{(\ref{#1})}

\newcommand{\tref}[1]{Table~\ref{#1}}
\newcommand{\fref}[1]{Figure~\ref{#1}}
\newcommand{\cref}[1]{Chapter~\ref{#1}}
\newcommand{\bcenter}{\begin{center}}
\newcommand{\ecenter}{\end{center}}
\newcommand{\beq}{\begin{equation}}
\newcommand{\eeq}{\end{equation}}
\newcommand{\bea}{\begin{eqnarray}}
\newcommand{\eea}{\end{eqnarray}}
\newcommand{\bean}{\begin{eqnarray*}}
\newcommand{\eean}{\end{eqnarray*}}
\newcommand{\ba}{\begin{array}}
\newcommand{\ea}{\end{array}}
\newcommand{\ben}{\begin{enumerate}}
\newcommand{\een}{\end{enumerate}}
\newcommand{\bi}{\begin{itemize}}
\newcommand{\ei}{\end{itemize}}
\newcommand{\bd}{\begin{description}}
\newcommand{\ed}{\end{description}}
\def\fnote#1#2{\begingroup\def\thefootnote{#1}\footnote{#2}
     \addtocounter{footnote}{-1}\endgroup}

\def\IC{\mathbb{C}}
\def\IE{\mathbb{E}}

\def\IF{\mathbb{F}}
\def\IZ{\mathbb{Z}}

\def\IP{\mathbb{P}}

\def\Hom{{\rm Hom}}
\def\dim{{\rm dim}}
\def\cN{{\mathcal N}}
\def\cM{{\mathcal M}}
\def\cE{{\mathcal E}}

\def\cO{{\mathcal O}}
\def\cC{{\mathcal C}}
\def\cF{{\mathcal F}}
\def\cP{{\mathcal P}}
\def\cL{{\mathcal L}}
\def\nn{\nonumber}
\def\td{\mbox{td}}
\def\ch{\mbox{ch}}
\def\rk{\mbox{rk}}

\def\dirac{\slash{\! \! \! \! D}}
\newcommand{\gen}[2]{{\rm span}_{#1}\left\{#2\right\}}
\def\point{\mbox{pt}}
\def\av{\wedge^2 V}
\def\vv{V \otimes V^*}

\def\to{\rightarrow}
\def\sC{\bar{c}_U}

\def\sym{{\rm Sym}}
\newcommand{\mat}[1]{\left( \matrix{#1} \right)}
\newcommand{\smat}[1]{{\scriptsize \mat{#1}}}
\newcommand{\tmat}[1]{{\tiny \mat{#1}}}
\def\II{{\rlap{1} \hskip 1.6pt \hbox{1}}}
\newarrow{DDTo}{}{dash}{}{dash}>
\def\ol{\overline}

\begin{document}

\begin{titlepage}

\vspace{-2cm}

\title{
   \hfill{\normalsize  UPR-1080-T} \\[1em]
   {\LARGE The Particle Spectrum of Heterotic Compactifications}
\author{Ron Donagi$^1$, Yang-Hui He$^2$, Burt A.~Ovrut$^2$, 
	and Ren\'{e} Reinbacher$^3$
	\fnote{~}{donagi@math.upenn.edu;
	yanghe,	ovrut@physics.upenn.edu;
	rreinb@physics.rutgers.edu}\\[0.5cm]
   {\normalsize $^1$ 
	Department of Mathematics, University of Pennsylvania} \\
	{\normalsize Philadelphia, PA 19104--6395, USA} \\
   {\normalsize $^2$
	Department of Physics, University of Pennsylvania} \\
   {\normalsize Philadelphia, PA 19104--6396, USA} \\
   {\normalsize $^3$
	Department of Physics and Astronomy, Rutgers University}\\
   {\normalsize Piscataway, NJ 08855-0849, USA}}
}
\date{}

\maketitle

\begin{abstract}
Techniques are presented for computing the cohomology of stable,
holomorphic vector bundles over elliptically fibered Calabi-Yau
threefolds. These cohomology groups explicitly determine the spectrum
of the low energy, four-dimensional theory. Generic points in vector
bundle moduli space manifest an identical spectrum. However, it is
shown that on subsets of moduli space of co-dimension one or higher,
the spectrum can abruptly jump to many different values. Both analytic
and numerical data illustrating this phenomenon are presented. This
result opens the possibility of tunneling or phase transitions between
different particle spectra in the same heterotic compactification. In
the course of this discussion, a classification of
$SU(5)$ GUT theories within a specific context is presented.
\end{abstract}

\thispagestyle{empty}

\end{titlepage}

\section{Introduction}
The work of Ho\v{r}ava and Witten \cite{HW} opened the door to
constructing phenomenologically realistic $N=1$ supersymmetric vacua
of strongly coupled heterotic string theory. A key ingredient in such
constructions is the method presented in \cite{fmw1,don,fmw2} 
and \cite{dlow,holo} for finding stable, holomorphic vector bundles on
elliptically fibered Calabi-Yau
threefolds. Using this technique, a large number of GUT theories with
gauge groups $SU(5)$ and $SO(10)$ were produced in
\cite{dlow,holo,Bjorn1}. These
ideas were generalized in \cite{dopw-i,opr-i,dp}, 
where it was shown how to construct
stable, holomorphic vector bundles on torus-fibered Calabi-Yau
threefolds. Using these generalized
techniques, standard-like models were produced using Wilson lines to
spontaneously break both $SU(5)$ \cite{dopw-i} 
and $SO(10)$ \cite{opr-i} GUT
groups. Within this context, a number of new phenomena were discussed
such as small instanton phase transitions \cite{transition},
non-perturbative superpotentials \cite{super,super2}, 
five-brane moduli space
\cite{dow}, brane worlds \cite{bworld} 
and a new theory of the Big Bang \cite{bb}.

One aspect of such theories that was left unsolved was how to compute
the complete particle spectrum of the low-energy, four-dimensional
theory. The existence of precisely three families of quarks and
leptons was guaranteed in these theories by choosing the third Chern
class of the holomorphic vector bundle appropriately. However, the
number of other particles, such as the Higgs or exotic particles, was
not specified. To compute their spectrum, it is necessary to construct
the complete cohomology of the vector bundle $V$ on the Calabi-Yau
threefold $X$. This was carried out \cite{GSW} in the case of the
``standard embedding,'' that is, when $V=TX$, where $TX$ is the
tangent bundle. In this case, the spectrum is directly related to
Betti numbers of $TX$, which are
known. However, this is manifestly not the case for the bundles
discussed above. For these more general vector bundles, the relevant
cohomology groups are unrelated to the Dolbeault cohomology and,
hence, are much more difficult to compute.

We discussed a general approach to this problem in \cite{letter}.
It is the purpose of this paper to give explicit techniques for
calculating the cohomology of stable, holomorphic vector bundles on
elliptically fibered Calabi-Yau threefolds. That
is, we will show how to compute the complete spectrum for the low
energy GUT theories that arise in this context. In the process of
doing this, we have found what we believe to be an interesting
phenomenon in particle physics. This is the following. Holomorphic
vector bundles have complex moduli. In the present context, 
these have been discussed and
enumerated in \cite{super,dow,Evgeny}. For generic values of these
moduli, we find a specific particle spectrum. However, on loci of
co-dimension one or higher in the vector bundle moduli space we find
that the spectrum ``jumps,'' changing abruptly by integer values. In
this paper, we will conclusively demonstrate that this phenomenon
exists. We will discuss the mathematical underpinnings of this result
and give a concrete example using both analytic and numerical
techniques.

Specifically, in this paper we will do the following. In section 2, we
briefly review some salient facts about elliptically fibered
Calabi-Yau threefolds. These spaces are fibered over base surfaces
$B$, whose properties are discussed in Section 3. Section 4 is devoted
to a short discussion of the method of constructing stable,
holomorphic vector bundles from spectral data via the Fourier-Mukai
transformation. In Section 5, we present the three physical
constraints required of any phenomenologically relevant heterotic
string vacuum. Using the mathematical constraints arising from these
conditions, we present a classification of $SU(5)$ GUT
theories that can arise in our context. This is given in Section 6. In
Section 7, we present the general techniques for computing the low
energy spectrum from the cohomology of the holomorphic vector
bundle. First, we discuss the relationship between the spectrum and
cohomology, as well as the constraints on the spectrum arising from
the index theorem. We then present methods for computing the
cohomology based on Leray spectral sequences and the Riemann-Roch
theorem. For convenience, this is carried out within the context of
vector bundles with an $SU(5)$ structure group. Section 8 is devoted
to using these techniques to compute the complete cohomology of a
specific $SU(5)$ GUT model satisfying the three physical
constraints. The spectrum of this theory is then presented.
We note and discuss the
phenomenon that part of the cohomology and, hence, some of the
spectrum is dependent upon the vector bundle moduli for which they are
evaluated. Finally, in Section 9 we indicate why one expects the
spectrum to be moduli dependent for general holomorphic bundles, as
opposed to the standard embedding where this phenomenon does not occur.
Appendices A and B present various aspects of topological
data required in the text. Appendix C gives a general method for
computing a large set of cohomology groups that are required in our
discussion. Several matrices that are central to the calculation of
the spectrum are defined and explicitly
constructed in Appendix D, including their exact dependence on the
vector bundle moduli. The spectra of these representations depend on
the rank of one of these matrices. The rank is computed both
analytically and numerically in Appendix E. Explicit data is
presented, showing that the rank of this matrix is dependent upon
where in moduli space it is evaluated.

Although much of this paper is presented within 
the context of $SU(5)$ GUT
theories, the techniques introduced are completely general. They can
be used to compute the spectrum of any heterotic vacuum.

\section{Elliptically Fibered Calabi-Yau Threefolds}\label{s:CY}
We will consider elliptically fibered threefolds $X$.
Each such manifold has a base surface $B$ and a mapping
$\pi: X\to B$ such that $\pi^{-1}(b)$ is a smooth torus, 
$E_b$, for each generic point $b \in B$. Additionally, there are
special points in the base over each of which the fiber is
singular. These fibers are typically of type $I_1$, in the Kodaira
classification, but may be more singular.
What makes this torus fibration elliptic is the existence of
a zero section; that is, there 
exists an analytic map $\sigma : B\to X$ that assigns to 
every element $b$ of $B$ an element $\sigma(b) \in E_b$. The
point $\sigma(b)$ acts as the zero element for an Abelian group
which turns $E_b$ into an elliptic curve and $X$ into an elliptic
fibration. We will denote the fiber class by $F$.

In terms of explicit coordinates, one can express $X$ as a
Weierstrass model
\beq
\label{weier}
y^2 z = x^3 + g_2 x z^2 + g_3 z^3
\eeq
which describes $X$ as a divisor 
in a $\IP^2$-bundle $P$ over $B$.
The coefficients $g_2$ and $g_3$ are
sections of line bundles on the base.
The bundle $P$ is the
projectivization $\IP(\cL^2 \oplus \cL^3 \oplus \cO_B)$, where $\cL$
is a line bundle on $B$ which is the conormal bundle to the section
$\sigma$. Subsequently, we have
\beq\label{weier2}
x \sim \cO_P(1) \otimes \cL^2, \quad
y \sim \cO_P(1) \otimes \cL^3, \quad
z \sim \cO_P(1)
\eeq
and
\beq
g_2 \sim \cL^4, \quad g_3 \sim \cL^6 \ ,
\eeq
where we have used $\sim$ to denote ``global section of.'

An important property of elliptic fibrations is that
$X$ has a $\IZ_2$ symmetry $\tau = (-1)_X$, which, on the
Weierstrass coordinates defined in \eref{weier} acts as
\beq\label{tau}
\tau : y \rightarrow -y
\eeq
while leaving $x$ and $z$ invariant. Clearly this action leaves the
Weierstrass equation \eref{weier} unchanged. In other words $\tau$ is
a natural involution on $X$. It 
acts trivially on the base $B$ and
maps each element $b \in E_b$ to its inverse $-b$.

$N=1$ supersymmetry in four-dimensions demands that $X$ be
a Kahler manifold with vanishing first Chern class of its tangent
bundle $TX$; that is,
\beq\label{c1TX}
c_1(TX) = 0 \ .
\eeq
Such manifolds always
admit a Kahler metric of $SU(3)$ holonomy and are called Calabi-Yau
manifolds. Henceforth, we will choose $X$ to be a Calabi-Yau threefold.
In general, the Chern classes of $X$
can be conveniently expressed in terms of
those of the base $B$ \cite{don,Bjorn}. In addition to \eref{c1TX},
one finds that
\bea
\label{chernX}
&&c_2(TX)= 12 \sigma \cdot \pi^*(c_1(TB)) + 
	\pi^*(c_2(TB) + 11 c_1^2(TB)) , \nn \\
&&c_3(TX)= -60 (c_1(TB)^2 \cdot B) \mbox{pt} \ ,
\eea
where $c_1(TB)$ and $c_2(TB)$ are the first and second Chern classes
of $B$ respectively and pt is the class of a point.
When $X$ is a Calabi-Yau threefold, severe restrictions are placed
on the base surface $B$. It turns out that $B$ can only be
Enriques, del Pezzo, Hirzebruch and blowups of Hirzebruch
surfaces \cite{MV}.
We will present some relevant properties of these surfaces
shortly.

Throughout this paper we will make frequent 
use of the intersection relation
\beq
\label{intersig}
\sigma \cdot \sigma = - \pi^*(c_1(TB)) \cdot \sigma \ ,
\eeq
which follows from the adjunction formula.

\section{Properties of the Base Surface}
We now present the requisite properties, such as Chern classes and 
homology groups, of the surfaces $B$. Before doing so, however,
it is helpful to define some fundamental notions.

Consider a complex surface $B$ and its second homology group $H_2(B,
\IZ)$. Let $C \subset B$ be a holomorphic curve in $B$ and $[C] \in
H_2(B,\IZ)$ the class of curves equivalent to $C$. Then $[C]$ is
called an ``effective'' class. Clearly, not every class, such as
$-[C]$, is effective. If $[C]$ and $[D]$ are two effective classes,
then so is $m[C]+n[D]$ where $m,n \in \IZ_{\ge 0}$. Therefore, the
subset of effective classes forms a cone in $H_2(B,\IZ)$, called the
Mori cone. The Mori cone is spanned by a countable number of
generators, $[C_i]$, where $C_i \subset B$ are irreducible
curves. That is, any effective class $[C]$ can be expressed as
\beq
[C] = \sum_{i} r_i [C_i], \quad r_i \in \IZ_{\ge 0} \ .
\eeq
The reader is referred to
\cite{hart,griffith,mori}, for example, for details. 
The Mori cone is not necessarily finitely generated over $\IZ_{\ge
0}$, although for all surfaces discussed below, 
with the exception of $d\IP_9$, the
associated Mori cones are indeed finitely generated. We will shortly
present their bases explicitly. For $d\IP_9$, the Mori cone has an
infinite number of generators. Nevertheless, there is a convenient
description of them.


Let $C \subset B$ be a holomorphic curve in a complex surface
$B$. Since $C$ is a divisor of $B$, there exists a line bundle
$\cO_B(C)$ which has a section $s_C$, unique up to scalar
multiplication, whose zero locus is $C$. Now, consider another curve
$C' \ne C$ with the property that $\cO_B(C') \simeq \cO_B(C)$. Then, there
exists a section $s_{C'}$ of $\cO_B(C)$ whose zero locus is
$C'$. Note that $s_C/s_{C'}$ is a meromorphic function $f$ on $B$.
Two such
divisors $C$ and $C'$ are said to be linearly equivalent. The set of
all divisors linearly equivalent to $C$, denoted by $|C|$, is called the
linear system associated with $C$. A crucial property of linear
systems is the following. A base point of a linear system $|C|$ of
curves on $B$ is the intersection of all its members. If there is no
such common point, then $|C|$ is called base point free.
Furthermore, note that all numerical properties of a divisor $C$, such
as its self-intersection number, are completely determined by its
linear system.

We have restricted the discussion in this section to divisor classes
$[C]$, divisors $C$, and to bundles $\cO_B(C)$ associated with
$B$. However, all of our remarks apply to classes, divisors and line
bundles of any complex manifold, such as the threefold $X$.
  
%
\subsection{Hirzebruch Surfaces}
The Hirzebruch surfaces $\IF_r$
are $\IP^1$ fibrations over $\IP^1$. There is
an infinite family of such surfaces indexed by $r \in \IZ_{\ge 0}$. 
The second homology group is
\beq
H_2(\IF_r, \IZ) = \gen{\IZ}{S, \ \cE} \ ,
\eeq
where the generators $S$ and $\cE$ are effective classes with
the intersection numbers
\beq
\label{interFr}
S \cdot S = -r, \quad \cE \cdot \cE = 0, \quad S \cdot \cE = 1 \ .
\eeq
All effective classes are of the form
\beq\label{eff-Fr}
a S + b \cE, \qquad a,b \in \IZ_{\ge 0} \ .
\eeq
The aggregate of these is called the Mori cone of $\IF_r$.
The Chern classes are given by
\bea
\label{chernFr}
&&c_1(T\IF_r) = -c_1(K_{\IF_r}) = 2 S + (r+2) \cE \nn \\
&&c_2(T\IF_r) = 4 \ ,
\eea
where $K_{\IF_r}$ is the canonical bundle.
Finally, on $\IF_r$, the linear system $|a S + b \cE|$ is base-point
free if
\beq\label{Frbpf}
b \ge a \ r \ .
\eeq
\subsection{del Pezzo Surfaces}
\label{sec:dP}
There are, in all, nine del Pezzo surfaces, which we denote as $d
\mathbb{P}_r$ for $r=1,\ldots,9$. Each $d\IP_r$ is the $\IP^2$ surface
blown up at $r$
generic points. The second homology group for $d\mathbb{P}_r$ is
\beq
H_2(d\mathbb{P}_r, \IZ) = \gen{\IZ}{\ell, \ E_{i=1,\ldots,r}} \ ,
\eeq
where $\ell$ is the hyperplane class in $\IP^2$ and $E_{i=1,\ldots,r}$
are the $r$ exceptional divisors. Each $E_i$ corresponds to the $\IP^1$
blowup of a point in $\IP^2$. 
These classes have the following intersections
\beq
\label{interdPr}
\ell \cdot \ell = 1, \quad \ell \cdot E_i = 0, \quad
E_i \cdot E_j = - \delta_{ij} \ .
\eeq
The Chern classes are given by
\bea
\label{cherndPr}
&&c_1(Td\mathbb{P}_r) = -c_1(K_{d\mathbb{P}_r}) 
	= 3 \ell - \sum_{i=1}^r E_i
\nn \\
&&c_2(Td\mathbb{P}_r) = r+3 \ ,
\eea
where $K_{d\mathbb{P}_r}$ is the canonical bundle.
We now study the Mori cone of $d\mathbb{P}_r$.
The effective classes in $H_2(d\mathbb{P}_r, \IZ)$, that is, those which
can be expressed as non-negative integral combinations of classes of
irreducible curves, are tabulated in \cite{demazure}.
Here, we re-cast them into a more convenient form and present
the generators for the Mori cones in \tref{tab:dPr}.
\begin{table}[h]
\hspace{-1cm}
\begin{tabular}{|c|c|c|c|}
\hline
r & Generators & Distinct Indices & Number \\ \hline
1 & $E_1, \  \ell - E_1$ & & 2 \\ \hline
2 & $E_i, \  \ell - E_1 - E_2$ & $i=1, \ 2$ & 3 \\ \hline
3 & $E_i, \ell - E_i - E_j$ & $i , j = 1, 2,3$ & 6 \\ \hline
4 & $E_i, \  \ell - E_i - E_j$ & $i , j = 1,\ldots, 4 $ & 10 \\ \hline
5 & $E_i, \  \ell - E_i - E_j, \  2\ell - E_i - E_j-E_k-E_l-E_m$ 
	& $i, j, k, l, m = 1,\ldots, 5$ & 16 \\ \hline
6 & $E_i, \  \ell - E_i - E_j, \  2\ell - E_i - E_j-E_k-E_l-E_m$ 
	& $i, j, k, l, m = 1,\ldots, 6$ & 27 \\ \hline
7 & $\ba{c}
	E_i, \  \ell - E_i - E_j, \  2\ell - E_i - E_j-E_k-E_l-E_m, \\
	3\ell  - 2 E_i - E_j - E_k - E_l -E_m - E_n - E_o
	\ea$  
	& $i, j, k, l, m, n, o = 1,\ldots, 7$ & 56 \\ \hline
8 & $\ba{c}
	E_i, \  \ell - E_i - E_j, \  2\ell - E_i - E_j-E_k-E_l-E_m, \\
	3 \ell - 2 E_i - E_j - E_k - E_l -E_m - E_n - E_o,\\
	4 \ell - 2(E_i+E_j+E_k) - \sum\limits_{i=1}^5 E_{m_i},\\
	5 \ell - 2\sum\limits_{i=1}^6 E_{m_i} - E_k - E_l, \ 
	6 \ell - 3 E_i - 2\sum\limits_{i=1}^7 E_{m_i}
	\ea$  
	& $\ba{c}i, j, k, l, m, n, o, m_i \\
		= 1,\ldots, 8
	\ea$ & 240 \\ \hline
9 & $\ba{c} f = 3 \ell - \sum\limits_{i=1}^9 E_i, \mbox{~and~}
\{ y_i \} \mbox{~such that~}
y_i^2 = -1, \ y_i \cdot f = 1
\ea$ & --- & $\infty$ \\ \hline
\end{tabular}
\caption{
{ The generators for the Mori cone of a generic
	$d\mathbb{P}_r$ for $r=1,\ldots,9$. All effective classes
	of curves in
	$H_2(d\mathbb{P}_r,\IZ)$ can be written as non-negative integral
	combinations of these generators.
	We emphasize that all indices are distinct.}
}
\label{tab:dPr}
\end{table}

We note that 
$d\IP_9$ is itself an elliptic
fibration over $\IP^1$ with fiber class
\beq\label{f-dP9}
f = c_1(Td\IP_9) \ . 
\eeq
As stated in \tref{tab:dPr}, this class is a generator of the Mori
cone of $d\IP_9$. It is useful to note from \eref{interdPr},
\eref{cherndPr} and \eref{f-dP9} that
\beq\label{f2dP9}
f^2 = 0 \ .
\eeq
The remaining generators, $y_i$, of $d\IP_9$ form an infinite, but
countable, set whose properties are listed in \tref{tab:dPr}.

We will also need the base-point free condition for linear systems
on del Pezzo surfaces. Here,
we are aided by Proposition 2.3 of \cite{BP} which states
the following.
Let $\eta$ be a divisor on a del Pezzo surface $d\IP_r$ for $2 \le r
\le 7$ such that $\eta \cdot E \ge 0$ for every curve $E$ for which
$E\cdot E =-1$ and $E \cdot c_1(Td\IP_r)=1$. Then the linear system
$|\eta|$ is base point free.
Note from \tref{tab:dPr} that
the bases for the Mori cones are precisely
all curves satisfying the two
conditions $E\cdot E =-1$ and $E \cdot c_1(Td\IP_r)=1$.
As a last remark, note that
the surface $d\IP_1$ is actually isomorphic to
$\IF_1$.
One can see this, for example, from the fact that $d\IP_1$ fibers over
its unique exceptional divisor $S$ where $S^2=-1$ and each fiber is a
$\IP^1$. Alternatively, these two surfaces are
toric varieties with identical toric diagrams.

\subsection{Enriques Surfaces}
The Enriques surface $\IE$ is obtained from a K3 surface modulo
involutions. Its canonical bundle is torsion, that is, 
\beq\label{torKE}
K_{\IE} \otimes K_{\IE} = \cO_{\IE} \ .
\eeq
This implies that
\beq
2 c_1(T\IE) = 0 \ .
\eeq
The second Chern class is given by
\beq
c_2(T\IE) = 12 \ .
\eeq
The second homology group is
\beq
H_2(\IE, \IZ) \simeq \IZ^{10} \oplus \IZ_2 \ .
\eeq
We will not present the explicit generators for the Mori cone of $\IE$
since these will not be used in this paper.

\section{Vector Bundles on Elliptically Fibered Calabi-Yau
Threefolds}\label{s:V}
We consider rank $n$ stable holomorphic vector bundles $V$ on
$X$. These bundles have a convenient description, known as the
spectral cover construction \cite{fmw1,don,fmw2,dlow,holo}. The
spectral data is given by two objects, an effective 
divisor $\cC_V$ of $X$, 
called the spectral cover, and a spectral
line bundle $\cN_V$ on $\cC_V$. 
The spectral cover, $\cC_V$, is a surface in
$X$ that is an $n$-fold cover, $p: \cC_V \rightarrow B$,
of the base $B$. Its general form is
\beq
\label{coverX}
\cC_V \in |n\sigma + \pi^*\eta| \ ,
\eeq
where $\sigma$ is the zero section associated with $\pi$, and $\eta$ is
some effective curve in $B$. 
The spectral line bundle $\cN_V$ is defined by its first Chern class
\begin{equation}
\label{c1N}
c_{1}(\cN_V)=n(\frac{1}{2}+\lambda)\sigma+(\frac{1}{2}-\lambda)
\pi^{*}\eta+(\frac{1}{2}+n\lambda)\pi^{*}c_{1}(TB) \ ,
\end{equation}
where $\lambda$ is a rational number such that
\beq
\label{lambda}
\ba{rcll}
\lambda &=& m, \qquad &n \mbox{~~even} \\
\lambda &=& m + \frac12, \qquad & n \mbox{~~odd}
\ea
\eeq
for some $m \in \IZ$. When $n$ is even, we must also impose that
\beq
\eta = c_1(TB) \bmod 2 \ .
\eeq
Note that, as defined in \eref{c1N}, $\cN_V$ is actually a line bundle
on $X$. It can, of course, be restricted to be a line bundle over
$\cC_V$. Throughout this paper, we will, in general, not distinguish
between $\cN_V$ and $\cN_V|_{\cC_V}$, denoting both by $\cN_V$.

\subsection{Fourier-Mukai Transformation}\label{s:FM}

Given the spectral data $(\cC_V,\cN_V)$,
one can construct the vector bundle $V$ explicitly, 
using the Fourier-Mukai transformation
\beq
(\cC_V, \cN_V) \stackrel{FM}{\longleftrightarrow} V \ .
\eeq
We briefly remind the reader of the structure of this transformation
\cite{fmw1,don,fmw2,dlow,holo}. Let us form the fiber product
$X \times_B X'$,
where $X' \simeq X$ is another copy of $X$. We let $\pi : X \rightarrow B$
and $\pi' : X' \rightarrow B$ be the projections onto the base $B$
with sections $\sigma$ and $\sigma'$
respectively. 
The fiber product is a four-dimensional space defined as
\beq
X \times_B X' = \{(x,x') \in X \times X' \ | \
\pi(x) = \pi'(x') \} \ .
\eeq
Therefore, over any generic point $b \in B$ we have
a fiber $E_b \times E'_b$, where $E_b$ and $E'_b$ are elliptic curves.
We define the Poincar\'e sheaf 
$\cP$ 
to be
\beq
\cP = \cO_{X \times_B X'}(
\Delta - \sigma \times_B X' - X \times_B \sigma'
) \otimes K_B \ ,
\eeq
where $\Delta$ is the (diagonal) divisor given by the 
set of points $(x,x)$ in $X \times_B X'$.
Recall \cite{dopw-i} that $\cP$ is a bundle except at points $(x,x')
\in X \times_B X'$ where both $x$ and $x'$ are singular points of
their respective fibers.

Now, let us take the spectral cover 
$\cC_V \subset X$ and form the fiber
product $\cC_V \times_B X'$. Then, we have the following diagram,
with projection maps $\pi_1$ and
$\pi_2$ appropriately defined.
\beq
\label{fprod}
\ba{ccccc}
\cN_V &&&& \\
\downarrow &&&& \\
\cC_V & \stackrel{\pi_2}{\longleftarrow} & \cC_V \times_B X' &
	\stackrel{\pi_1}{\longrightarrow} & X'
\ea
\eeq
The Fourier-Mukai transformation is the explicit map that re-constructs
the vector bundle $V$ from the spectral data
in accordance with \eref{fprod}
\beq
\label{FM}
V = \pi_{1*} \left( (\pi^*_2 \cN_V) \otimes \cP \right) \ .
\eeq
We emphasize that we are using, as it is standard in the literature,
the canonical isomorphism of $X \simeq X'$ so that saying $V$ is a
vector bundle on $X'$ is equivalent to saying that $V$ is a vector
bundle on $X$. In the same way, we could have defined the spectral
data $(\cC'_V, \cN'_V)$ in $X'$ and produced a vector bundle $V$ on
$X$.

Altogether, we have the following commutative diagram
\beq\label{commU}
\ba{ccccc}
& \pi_2^*\cN_V \otimes \cP & 
\stackrel{\pi_{1*}}{\longrightarrow} & V & \\
& \downarrow & & \downarrow  & \\
& \cC_V \times_B X' & \stackrel{\pi_{1}}{\longrightarrow} 
	& X' & \\
& \pi_2 \downarrow && \downarrow \pi' & \\
& \cC_V & \stackrel{\pi_{\cC}}{\longrightarrow} & B & \ .
\ea
\eeq

The Chern classes of $V$ are found to be \cite{fmw1,don,fmw2}
\bea
\label{chernV}
c_1(V)&=&0,\nn \\
c_2(V)&=&\sigma \cdot \pi^*(\eta) - \frac{1}{24} c_1(TB)^2 (n^3-n) F 
	+\frac{1}{2}(\lambda^2-\frac{1}{4})
	n \ \eta \cdot (\eta - n \ c_1(TB)) F , \nn \\ 
c_3(V) &=& 2\lambda \eta \cdot (\eta - n \ c_1(TB)) \point \ .
\eea

\section{The Physical Constraints}\label{s:cons}
The requirements of particle physics phenomenology put three strong
constraints on both the Calabi-Yau threefold $X$ and the holomorphic
vector bundle $V$. These arise from the necessity that the theory be
consistent quantum mechanically, that there be three families of
quarks and leptons and that the theory preserve $N=1$
supersymmetry. Let us examine each of these constraints.

\subsection{Anomaly Cancellation Condition}
The anomaly cancellation condition is given by
\beq
\label{anom1}
c_2(TX) - c_2(V) = W \ ,
\eeq
where $W$ is the five-brane class in the vacuum. Furthermore,
since these must be physical five-branes, the class $W$ must be
effective, that is, in the Mori cone of $H_2(X, \IZ)$. The five-brane
class can be written as
\beq
W = W_B + a_F F \ ,
\eeq
where, using \eref{chernX}, \eref{chernV} and \eref{anom1},
\bea
\label{anom}
W_B &=& \sigma \cdot \pi^*(12 c_1(TB) - \eta) \ , \nn \\
a_F &=& c_2(TB) + (11 + \frac{n^3-n}{24}) c_1(TB)^2 -
	\frac{1}{2}n(\lambda^2 - \frac14)\eta \cdot (\eta - n \
	c_1(TB)) \ .
\eea
It was shown \cite{dlow,holo} that $W$ is an effective class in $X$ if and
only if
\beq
\label{anom-eff}
W_B \mbox{~is effective in~} B, \quad a_F \ge 0 \ .
\eeq
Henceforth, we will require the expressions in \eref{anom} to satisfy
the constraints given in \eref{anom-eff}.
\subsection{Three Family Models}
The number of generations, $N_{gen}$, 
is related to the zero-modes of the Dirac
operator $\dirac$
in the presence of the vector bundle. Specifically, it is
given by
\beq
\label{Ngen1}
N_{gen} = \mbox{index}(V, \dirac) = \int_X \td(TX)\ch(V) = \frac12 \int_X
c_3(V) \ ,
\eeq
where we have used the index theorem.
We are interested in three-family models, that is theories for which 
\beq\label{Ngen2}
N_{gen} = 3 \ .
\eeq
It then follows from \eref{chernV}, \eref{Ngen1} and \eref{Ngen2}
that
\beq
\label{Ngen}
3 = \lambda \eta \cdot (\eta - n \ c_1(TB)) \ .
\eeq

We will, henceforth, require that this constraint be satisfied.
Note that \eref{Ngen} simplifies the expression for $a_F$ in
\eref{anom}. It follows that the condition on $a_F$ in \eref{anom-eff}
becomes
\beq\label{aFnew}
a_F = c_2(TB) + (11 + \frac{n^3-n}{24}) c_1(TB)^2 -
	\frac{3}{2\lambda}n(\lambda^2 - \frac14) \ge 0 \ .
\eeq
\subsection{Irreducibility of the Spectral Cover}
In order for the gauge connection to preserve $N=1$ supersymmetry,
it must satisfy the hermitian Yang-Mills equation. The theorems of
Uhlenbeck and Yau \cite{UY} and Donaldson \cite{Don} state that a rank
$n$ holomorphic vector bundle $V$ will admit such a connection if $V$ is
stable. We will, therefore, impose stability as a constraint on $V$.
It was shown in \cite{fmw1,don,fmw2} that if $V$ is
constructed via a Fourier-Mukai transformation from spectral data,
then $V$ will be stable if the spectral
cover is irreducible\footnote{More precisely, the notion of stability
	requires the choice of
	an ample class $H \in H^2(X, \IZ)$ and, in the
	situation that the cover is irreducible, we can find an ample
	class $H$ such that $V$ is stable.
}.
What are the conditions on the linear system $|\cC_V|$ such that it
contains an irreducible divisor?
This will be the case when the following
two conditions are met \cite{transition}.
\bea
(1) && |\pi^* \eta| \mbox{~contains an irreducible divisor in~} X \ , \nn \\
(2) && c_1(\pi^* K_B^{\otimes n} \otimes \cO_X(\pi^* \eta)) \mbox{~is
effective in~} H_4(X,\IZ) \ .
\eea
We will satisfy these two conditions if we impose
\bea
\label{irred}
(1) && \mbox{The linear system~} |\eta| \mbox{~is base-point free in~} 
	B \ , \nn \\
(2) && \eta + n c_1(K_B) \mbox{~is effective in~} B \ .
\eea
In order to preserve $N=1$ supersymmetry, the conditions 
in \eref{irred} must be imposed in addition to 
\eref{anom-eff} and \eref{Ngen}.

\subsection{Summary of the Constraints}
For clarity, we summarize here the physical constraints, namely
\eref{anom-eff}, \eref{Ngen}, and \eref{irred}, which we
impose on our vector bundle. These become the following conditions
on the effective base curve $\eta$ as
well as on the parameter $\lambda$.
\beq
\label{cons1}
\ba{llll}
(1)&W_B\mbox{~effective}& :
	&12 c_1(TB) - \eta \mbox{~~~~is effective}, \\
(2)&a_F>0&:&c_2(TB) + (11 + \frac{n^3-n}{24}) c_1(TB)^2 -
\frac{3}{2\lambda}n(\lambda^2 - \frac14)\ge 0, \\
(3)&\mbox{Three families}&:
	&\lambda \ \eta \cdot (\eta - n \ c_1(TB)) = 3, \qquad
	\lambda \in \left\{ \ba{ll}
	\IZ, & n\mbox{~even} \\
	\IZ/2, & n\mbox{~odd},
	\ea \right. \\
(4)&\mbox{Stability of~}V&:& 
|\eta| \mbox{~is base-point free}, \\
(5)&\mbox{Stability of~}V&:
	&\eta - n c_1(TB) \mbox{~~~~is effective} \ .
\ea
\eeq
In this paper, for specificity, we will 
restrict our attention to the case
\beq
n=5 \ .
\eeq
This corresponds to constructing an $SU(5)$ GUT model at low
energy.
Because $n=5$ is odd,
$\lambda$ has to be half integral by \eref{lambda}. 
Thus, the third condition in \eref{cons1}
implies, since $ \eta \cdot (\eta - n \ c_1(TB))$ is integral,
that the only possibilities for $\lambda$ are
\beq
\label{lambda2}
\lambda = \pm \frac12, \ \pm \frac32 \ .
\eeq
Therefore, \eref{cons1} simplifies to the constraints
\beq
\label{cons}
\ba{ll}
(1)
	&12 c_1(TB) - \eta \mbox{~~~~is effective}, \\
(2)
	&c_2(TB) + 16 c_1(TB)^2 -
\frac{15}{2\lambda}(\lambda^2 - \frac14)\ge 0, \\
(3)
	&\lambda \ \eta \cdot (\eta - 5 \ c_1(TB)) = 3, \qquad
	\lambda = \pm \frac12 \mbox{~or~} \pm \frac32 \ ,
\\
(4)
	& |\eta| \mbox{~is base-point free}, \\
(5)
	&\eta - 5 c_1(TB) \mbox{~~~~is effective} \ . \\
\ea
\eeq

%

%
%
\section{A Classification of $SU(5)$ GUT Theories 
from \\ Heterotic M-Theory Compactification}\label{s:class}
\subsection{Eliminating the Enriques Surfaces}
In \cite{holo} it was shown that Enriques surfaces will never satisfy
the first condition in constraint \eref{anom-eff}, that is, condition
(1) in \eref{cons1} and \eref{cons}.
We briefly present a simplified version of this argument. 
Recalling the expression for $W_B$ in \eref{anom},
we have
\beq
\label{WB}
W_{\IE} 
= \sigma \cdot \pi^* (12 c_1(T\IE) - \eta) \ .
\eeq
Furthermore, 
from \eref{torKE}, we know that
\beq
K_{\IE}^{\otimes 12} = \cO_{\IE} 
\eeq
because 12 is even. Therefore,
expression \eref{WB} then becomes
\beq
W_{\IE} = - \sigma \cdot \pi^* \eta \ .
\eeq
Since $\eta$ is an effective class, it follows
that $W_{\IE}$ can be effective only if $\eta$ is trivial. This,
of course, would violate the three-family condition (3) of
\eref{cons}.
We conclude that the Enriques surface is
not consistent with the anomaly cancellation and three-family
conditions.
\subsection{$d\IP_9$ Surfaces}
We now show that the generic
$d\mathbb{P}_9$ is ruled out as well. Recall from \eref{f-dP9}
that 
\beq
f = c_1(Td\IP_9)
\eeq
is the fiber class over $\IP^1$. It then follows from
condition (1) of \eref{cons} that
\beq
W_{d\IP_9} = 12 f - \eta
\eeq
must be effective.
Since we are considering generic $d\IP_9$, we can use the results in 
\tref{tab:dPr} and show that
\beq\label{effdP9-1}
12 f - \eta = \alpha f + \sum_i \beta_i y_i \ , \quad \mbox{for~some~}
\alpha, \beta_i \in \IZ_{\ge 0} \ ,
\eeq
where $y_i$ are such that
\beq\label{effdP9}
y_i^2 = -1, \ y_i \cdot f = 1 \ .
\eeq
We remark that, for a non-generic $d\IP_9$, expressions
\eref{effdP9-1} and \eref{effdP9}
need not be valid.
Now, by \eref{f2dP9} and \eref{effdP9},
\beq\label{etaf1}
(12 f - \eta) \cdot f = - \eta \cdot f = (\sum_i \beta_i y_i) 
\cdot f = \sum_i \beta_i \ge 0
\ .
\eeq
On the other hand, $\eta$ must
be effective, so we can write
\beq
\eta = \alpha' f + \sum_j \beta'_j y_j \ , \quad \mbox{for~some~}
\alpha', \beta'_j \in \IZ_{\ge 0} \ .
\eeq
It follows that
\beq\label{etaf2}
\eta \cdot f = \sum_j \beta'_j \ge 0 \ .
\eeq
Combining \eref{etaf1} and \eref{etaf2}, we have
\beq
\eta \cdot f = - \sum_i \beta_i = \sum_j \beta'_j \ .
\eeq
However, all $\beta_i$ and $\beta_j'$ are non-negative. Therefore, we
must have
\beq
\beta_i = \beta_j' = 0 \ ,
\eeq
which implies that
\beq\label{eta3}
\eta = \alpha' f \ .
\eeq
That is, $\eta$ is proportional to the fiber class.
Finally, the three-family condition (3) of \eref{cons} requires that
\beq\label{eta4}
\lambda \eta \cdot (\eta - n \ f) = 3 \ .
\eeq
However, the left hand side of \eref{eta4}
vanishes due to \eref{f-dP9} and
\eref{eta3}. It follows that on a generic $d\IP_9$ surface the
effectiveness condition for five-branes and the three family
constraint are in contradiction.
Therefore, to obtain phenomenologically acceptable theories of this
type, one must consider special non-generic $d\IP_9$ base
surfaces. See, for example, \cite{dopw-i,opr-i}.
%
%
\subsection{Hirzebruch Surfaces}
The remaining possibilities for the base surfaces are then the
Hirzebruch surfaces $\IF_r$, certain blowups of these surfaces
and the del Pezzo surfaces $d\mathbb{P}_r$
for $r = 1, \ldots, 8$. In this paper, we will not discuss the blowups
of $\IF_r$.
Let us first consider the case of $\IF_r$.
Using \eref{eff-Fr},
we can write the effective class of the
base curve in the spectral cover
\eref{coverX} as
\beq
\eta = a S + b \cE, \qquad a,b \in \IZ_{\ge 0} \ .
\eeq

Next, we must satisfy the five requirements in \eref{cons}.
Recalling the Chern classes from \eref{chernFr} and the intersection
numbers from \eref{interFr}, these translate into
the following conditions for $a,b,r \in \IZ_{\ge 0}$ and 
$\lambda = \pm \frac12, \ \pm \frac32$.
\bea
\label{consFr}
(1)&&24 - a \ge 0 \ , \quad 12(r+2) - b \ge 0, \nn\\
(2)&& 
44 - \frac{5}{2 \lambda} (\lambda^2 - \frac14) \ge 0, \nn \\
(3)&&\lambda(2\,a\,b - 10\,a - 10\,b - a^2\,r + 5 a\,r) = 3, \nn\\
(4)&&b \ge a \ r,  \nn\\ 
(5)&&a - 10 \ge 0 \ , \quad b - 5(r+2) \ge 0 \ .
\eea
Note that in the fourth condition in \eref{consFr} 
we have used \eref{Frbpf}.
The five expressions in \eref{consFr} constitute a system of 
Diophantine inequalities.
We have studied these inequalities for the 
four allowed values of $\lambda$ and
found that only
\beq\label{lam-F1}
\lambda = \frac12
\eeq
permits solutions. Subject to this constraint, 
the only solutions are
\beq\label{Frsol}
(a,b,r) = (12,15,1), \ (13,15,1) \ .
\eeq
That is,
\beq\label{F1sol}
B = \IF_1, \quad \eta = 12 S + 15 \cE, \ 13 S + 15 \cE \ .
\eeq
We see that our physical conditions are so stringent that they
restrict the Hirzebruch surfaces to $\IF_1$ and the possible spectral
covers on it to those specified by the two curves in \eref{F1sol}.

\subsection{The $d\IP_2$ Surface}
Let us now consider the del Pezzo surfaces. Since, as we remarked at
the end of Subsection \ref{sec:dP}, $d\IP_1 \simeq \IF_1$, we can
start with the next surface in the del Pezzo series, namely, $d\IP_2$.
Referring to \tref{tab:dPr}, we write the effective base curve $\eta$
as
\beq
\label{etadP2}
\eta = a E_1 + b(\ell - E_1 - E_2) + c E_2 \ , \qquad
a,b,c \in \IZ_{\ge 0} \ .
\eeq
Next, we must satisfy the constraints \eref{cons}. The difficult
constraint to satisfy is
condition (4), which requires that the linear system
$|\eta|$ be base-point free.  
Using the theorem stated in Subsection \ref{sec:dP}
we have that
the base-point-free condition (4) of \eref{cons} becomes, for
$d\IP_2$,
\beq
\eta \cdot E_1 \ge 0 \ , \quad
\eta \cdot (\ell - E_1 - E_2) \ge 0 \ , \quad
\eta \cdot E_2 \ge 0 \ .
\eeq
Substituting expression \eref{etadP2} for $\eta$ and using the
intersection relations \eref{interdPr}, the condition becomes
\beq\label{dP2free}
b-a \ge 0 \ , \quad a-b+c \ge 0 \ , \quad b-c \ge 0 \ .
\eeq
Using \eref{cherndPr} and \eref{dP2free}, the full set of
constraints in \eref{cons}
explicitly becomes a system of Diophantine inequalities for the
parameters $a,b,c \in \IZ_{\ge 0}$ in \eref{etadP2} and $\lambda =
\pm \frac12, \ \pm \frac32$. They are
\bea
(1)&&24 \ge a, \ 36 \ge b, \ 24 \ge c, \nn \\
(2)&&39-\frac{5}{2\lambda}(\lambda^2 - \frac14)\ge 0, \nn \\ 
(3)&&\lambda(-a^2 + 2ab - b^2 + 2bc - c^2 - 5a  - 5b  - 5c) = 3, 
	\nn \\
(4)&& b-a \ge 0, \ a-b+c \ge 0, \ b-c \ge 0, \nn \\
(5)&&a \ge 10, \ b \ge 15, \ c \ge 10.
\label{consdP2}
\eea
We have studied these equations numerically for the four allowed values of
$\lambda$. Once again, we find
that only $\lambda = \frac12$ is allowed.
For this $\lambda$, the allowed values of $a,b,c$ are found to be
\beq\label{soldP2-1}
(a,b,c) \ = \ (10, 15, 12), (10, 15, 13), (10, 17, 10), (10, 18, 10), 
(12, 15, 10), (13, 15, 10).
\eeq
The six solutions in \eref{soldP2-1} correspond to the following classes
\bea\label{soldP2}
\eta &=& 15\ell - 5E_1 - 3E_2, \quad 
15\ell - 5E_1 - 2E_2, \quad
17\ell - 7E_1 - 7E_2, \nn \\
&& 18\ell - 8E_1 - 8E_2, \quad 
15\ell - 3E_1 - 5E_2, \quad
15\ell - 2E_1 - 5E_2.
\eea
We conclude that for $B=d\IP_2$, exactly six types of vector bundles
corresponding to the curves in \eref{soldP2} will satisfy all of the
physical constraints.

\subsection{$d\IP_3$ Surface}
We move on to the third del Pezzo surface. Again,
referring to \tref{tab:dPr}, we write the effective base class $\eta$ as
\beq
\label{etadP3}
\eta = 
n_1E_1 + n_2(\ell - E_1 - E_2) + n_3E_2 + n_4(\ell - E_1 - E_3) + 
 n_5(\ell - E_2 - E_3) + n_6E_3 \ , 
\qquad
n_i \in \IZ_{\ge 0} \ .
\eeq
As above, we can use the theorem in Subsection \ref{sec:dP}
to obtain the following conditions for the linear system $|\eta|$ to
be base-point free. They are
\bea\label{bpfreedp3}
-n_1 + n_2 + n_4 \ge 0, \quad n_1 - n_2 + n_3\ge 0, \quad n_2 - n_3 + 
n_5 \ge 0,
\\ \nn 
n_1 - n_4 + n_6 \ge 0, \quad n_3 - n_5 + n_6 \ge 0, 
\quad n_4 + n_5 - n_6 \ge 0 \ .
\eea
The conditions (1) and (5) for effectiveness in \eref{cons} are now
more complicated than previously. For example, condition (1) becomes
\bea
12 c_1(Td\IP_3) - \eta &=& 
(36 - n_2 - n_4 - n_5) \ell + (-12 - n_1 + n_2 + n_4) E_1 + \nn \\
&&(-12 + n_2 - n_3 + n_5) E_2 + (-12 + n_4 + n_5 - n_6) E_3 \ .
\eea
This can be written as
\beq
12 c_1(Td\IP_3) - \eta =
\smat{36 - a_3 - a_6 - n_1 - n_3 - n_6\cr 
24 - a_6 - n_2 - n_6\cr a_3\cr 
24 - a_3 - n_3 - n_4 \cr
-12 + a_3 + a_6 + n_3 - n_5 + n_6\cr 
a_6}
\cdot v \ ,
\eeq
where
\beq
v = \{E_1, \ell - E_1 - E_2, E_2, \ell - E_1 - E_3, \ell - E_2 - E_3,
E_3\}
\eeq
is the basis of the Mori cone of $d\IP_3$ and
$a_3, a_6 \in \IZ$ are arbitrary parameters.
Similarly, condition (2) becomes
\beq
\eta - 5 c_1(Td\IP_3) =
\smat{-10 - b_5 + n_1 + n_5\cr -10 - b_6 + n_2 + n_6\cr 
-5 + b_5 - b_6 + n_3 - n_5 + n_6\cr -5 - b_5 + b_6 + n_4 + n_5 -
n_6\cr b_5\cr b_6} \cdot v
\eeq
for arbitrary parameters $b_3, b_6 \in \IZ$.
Effectiveness of the two classes $\eta - 5 c_1(Td\IP_3)$ and $12
c_1(Td\IP_3) - \eta$ means that the components of the two vectors 
in brackets must
be non-negative for at least one choice of the parameters $a_3$, $a_6$
and $b_3$, $b_6$ respectively.
Using these results and \eref{bpfreedp3}, the five conditions in
\eref{cons} for $d\IP_3$ translates into the following
system of Diophantine inequalities for $n_{i=1,\ldots6} \in \IZ_{\ge
0}$, $a_3, a_6, b_5, b_6 \in \IZ$ and $\lambda = \pm \frac12$ or  
$\pm \frac32$. They are
\bea\label{consdP3}
(1)&&
\ba{l}
36 - a_3 - a_6 - n_1 - n_3 - n_6\ge 0, \
24 - a_6 - n_2 - n_6\ge 0, \ a_3\ge 0, \ \\
24 - a_3 - n_3 - n_4 \ge 0, \ 
-12 + a_3 + a_6 + n_3 - n_5 + n_6\ge 0, \ 
a_6\ge 0,
\ea \nn \\
(2)&&
34 - \frac{5}{2\lambda}(\lambda^2 - \frac14)\ge 0, \nn \\
(3)&&
\ba{c}
\lambda \left(-5n_1 - n_1^2 - 5n_2 + 2n_1n_2 - n_2^2 - 5n_3 + 2n_2n_3 - 
 n_3^2 - 5n_4 + 2n_1n_4 \right. \\
\left. - n_4^2 - 5n_5 + 2n_3n_5 - n_5^2 - 
 5n_6 + 2n_4n_6 + 2n_5n_6 - n_6^2 \right) = 3,
\ea
\nn \\
(4)&& 
\ba{l}
-n_1 + n_2 + n_4\ge 0, \ n_1 - n_2 + n_3\ge 0, \ n_2 - n_3 + n_5\ge 0,\\
n_1 - n_4 + n_6\ge 0, \ n_3 - n_5 + n_6\ge 0, \ n_4 + n_5 - n_6\ge 0,
\ea
\nn \\
(5)&&
\ba{l}
-10 - b_5 + n_1 + n_5\ge 0, \ -10 - b_6 + n_2 + n_6\ge 0, \ 
-5 + b_5 - b_6 + n_3 - n_5 + n_6\ge 0,\\ 
-5 - b_5 + b_6 + n_4 + n_5 - n_6\ge 0, \  b_5\ge 0, \ b_6\ge 0.
\ea
\nn \\
\eea
We can find all solution to \eref{consdP3} numerically
by testing all
lattice points in the polytope defined by the above
inequalities. We find precisely
6930 solutions for $\lambda = \frac32$ and 6990 solutions for
$\lambda = \frac12$, giving a total of 13,920 solutions.
Presenting all these solutions is obviously un-illustrative. 
A few examples are the following. For
\beq
\lambda  = \frac12,
\eeq
we find one solution to be
\beq
(n_1, n_2, n_3, n_4, n_5, n_6) = (0, 0, 5, 5, 10, 12).
\eeq
This corresponds to the vacuum defined by
\beq
\eta = 15 \ell - 5 E_1 - 5E_2 - 3E_3 \ .
\eeq
For
\beq
\lambda  = \frac32,
\eeq
we find one solution to be
\beq
(n_1, n_2, n_3, n_4, n_5, n_6) = (2,4,8,2,10,1),
\eeq
corresponding to the vacuum
\beq
\eta = 22 \ell - 10 E_1 - 14E_2 - 17E_3 \ .
\eeq
We will not present any further solutions for the
higher del Pezzo surfaces because the exercise is not
enlightening.
In general, we see that because the inequality signs in constraints
(1) and (5) in \eref{cons} run in opposite directions, the solutions
will always be within some finite polytope. Additionally,
the solutions are constrained to be special lattice
points within the polytope which also obey (2), (3) and (4). 
In other words, for 
each generic base surface, there will always be a finite number of
solutions. A very crude upper-bound to the number of solutions is, of
course, the size of the polytope. For the del Pezzo surfaces, this is
roughly $16^N$, where $N$ is the number of generators of the Mori cone
in \tref{tab:dPr}.

\section{The Particle Spectrum in Heterotic
\\Compactifications}\label{s:spec-comp}
We now address the key
issue of this paper, namely, computing the particle
spectra of grand unified theories in four dimensions
arising from the compactification of heterotic theory
on a Calabi-Yau threefold $X$ endowed with a vector bundle $V$.
We briefly review the requisite quantities in such a calculation.

Consider the $E_8 \times
E_8$ gauge group of heterotic theory. 
In heterotic M-theory, one $E_8$ lives on the ``observable''
nine-brane while the other $E_8$ is restricted to the ``hidden''
brane. We will focus on the observable brane only and, hence, consider
a single $E_8$ gauge group. Now compactify on a Calabi-Yau
threefold $X$.
A vector bundle $V$ on $X$ breaks this $E_8$ group down to some GUT
group in the low energy theory.
The canonical example is to take
\beq
V = TX \ ,
\eeq 
where $TX$ is the tangent bundle of $X$. See, for example,
\cite{GSW}. Since $X$ has $SU(3)$ holonomy, it follows that $V$
has the structure group $SU(3)$. This is known as the 
``standard embedding.'' The
gauge connection on $V$ is then identified with the spin connection of
$X$.  The low-energy GUT group is the commutant of $SU(3)$ in $E_8$,
which is $E_6$. In other words, we have the breaking pattern
\beq\label{su3e6}
E_8 \rightarrow SU(3) \times E_6 \ .
\eeq
The relevant fermionic fields in the low-energy
four dimensional theory arise from the
decomposition of the gauginos in the vector supermultiplet of the ten
dimensional theory which transforms as 
the 248 of $E_8$.
Under the decomposition \eref{su3e6}, one finds that
\beq\label{248split}
248 \rightarrow (1,78) \oplus (3,27) \oplus 
(\overline{3},\overline{27}) \oplus(8,1) \ .
\eeq
To be observable at low energy,
the fermion fields transforming under the $E_6$ must be
massless modes of the Dirac operator on $X$ \cite{GSW,Witten}.
It was shown in \cite{Witten} that the number of massless modes for a
given representation equals the dimension of a certain
cohomology group. Let us first consider the representation $(1,78)$
in \eref{248split}. In this case, we note that $h^1(X, \cO_X)$
vanishes while
\beq
n_{78} = h^0(X, \cO_X) = 1 \ .
\eeq
These are the gauginos of a vector supermultiplet transforming in the
78 representation of $E_6$.
For the other representations, the zeroth cohomology groups vanish and
we have the following.
\beq
n_{27} = h^1(X, TX), \quad n_{\overline{27}} = h^1(X, TX^*),
\eeq
and
\beq
n_{1} = h^1(X, \mbox{End(TX)}) = h^1(X, TX \otimes TX^*) \ .
\eeq
Now
\beq
H^1(X, TX) \simeq H^{2,1}_{\overline{\partial}}(X), \
H^1(X, TX^*) \simeq H^{1,1}_{\overline{\partial}}(X),
\eeq
where $H^{p,q}_{\bar{\partial}}(X)$ are the Dolbeault cohomology groups of
$X$. 
It follows that
\beq
n_{27} = h^{2,1}, \quad n_{\overline{27}} = h^{1,1}
\eeq
where $h^{1,1}$ and $h^{2,1}$ are the Betti numbers of the Calabi-Yau
threefold $X$. Each such multiplet is the fermionic component of a
chiral superfield transforming in the 27 or $\overline{27}$
representation of $E_6$. The remaining quantity, $h^1(X, TX \otimes
TX^*)$, 
is a familiar object in deformation theory. It corresponds to the
number of moduli of infinitesimal deformations of the tangent bundle
$TX$. Therefore, the complex scalar superpartners of the fermions
transforming as singlets under $E_6$ are the bundle moduli of
$V$. These form $n_1 = h^1(X, TX \otimes TX^*)$ chiral superfields,
each an $E_6$ singlet. We
remark that the $8$ of $SU(3)$ is actually in the traceless part $Ad(TX)$ 
of $TX\otimes TX^*$. Note, however, that
$TX\otimes TX^* = \cO_X \oplus Ad(TX)$, where $\cO_X$ is the trivial
bundle on $X$, and that $h^1(X, \cO_X)$ vanishes. Therefore
$h^1(X,Ad(X)) =  h^1(X, TX \otimes TX^*)$. 

Since the Dolbeault
cohomology groups for Calabi-Yau threefolds are known, one can compute
the 27 and $\overline{27}$ part of the
$E_6$ particle spectrum. Furthermore, the number of moduli of $V$ can
be found in a straight-forward manner. This has been discussed, 
for example, in \cite{GSW}.
It is important to note, however, that this can only be accomplished because
one has chosen the standard embedding $V=TX$.

Having discussed the standard embedding, let us move on to
so-called ``non-standard'' embeddings. It was realized in
\cite{dlow,holo,Witten}
that using such vector bundles one could get, in addition to $E_6$, 
more appealing GUT groups such as $SU(5)$ and $SO(10)$. 
This is done by taking $V$ not to
be the tangent bundle $TX$ as was done above, but some more general
holomorphic vector bundle $V$ with structure group $G$. Since $V$ is no
longer $TX$, $G$ need not be $SU(3)$. Then,
the low-energy effective theory has gauge group $H$, where $H$ is the
commutant of $G$ in $E_8$. 
For example, if we take $V$ to be an $SU(4)$ bundle, then the
low-energy GUT group is $SO(10)$. If $V$ has structure group $SU(5)$, 
the low-energy GUT group is $SU(5)$. The decomposition
of the 248 of $E_8$ under these groups is as follows.
\beq
\label{decomp}
\ba{|l|ccl|}\hline
E_8 \rightarrow G \times H & && 
\\ \hline
SU(3) \times E_6 & 
248 & \rightarrow& (1,78) \oplus (3,27) \oplus 
	(\overline{3},\overline{27}) \oplus (8,1) \\
SU(4) \times SO(10) &
248 & \rightarrow& (1,45) \oplus (4,16) \oplus 
	(\overline{4},\overline{16}) \oplus (6,10) \oplus  (15,1) \\
SU(5) \times SU(5) &
248 & \rightarrow& (1,24) \oplus (5,\ol{10}) \oplus 
	(\overline{5},10) \oplus
	(10, 5) \oplus (\overline{10},\ol{5}) \oplus  (24,1) \\
\hline
\ea
\eeq
For a non-standard vector bundle $V$, the zero mode spectrum continues
to depend on the dimensions of certain cohomology groups.
In this paper, for specificity,
we will be primarily interested in $SU(5)$ GUTS. To differentiate the
two $SU(5)$ groups in $SU(5) \times SU(5)$, denote the structure group
of $V$, the first factor, by $SU(5)_G$ and the low energy GUT group,
the second factor, by $SU(5)_H$. From \eref{decomp}, we see that
the 1 , 5, $\overline{5}$, 10, $\overline{10}$ and 24 representations
of $SU(5)_G$  are paired with 
the $24$ ,$\ol{10}$, $10$, $5$, $\ol{5}$, and 1
respectively of $SU(5)_H$. Furthermore, note that the 5,
$\overline{5}$, 10, $\overline{10}$ and 24 representations of
$SU(5)_G$ are associated with the vector bundles $V$, $V^*$, $\av$,
$\av^*$ and $\vv$. Then, the spectrum of zero mass fields transforming
as the $24$ ,$\ol{10}$, $10$, $5$, $\ol{5}$, and 1 of
$SU(5)_H$ are the following. First, as previously, 
\beq
n_{24} = 1 \ ,
\eeq
indicating that there is a single vector supermultiplet transforming
in the adjoint 24 representation of $SU(5)_H$. The remaining
representations all occur as chiral superfields. Their
spectrum is given by
\bea
\label{VVdaV-1}
&&n_{\ol{10}} = h^1(X,V), \quad n_{10} = h^1(X,V^*), \nn \\ 
&&n_{5} = h^1(X,\av), \quad n_{\ol{5}} = h^1(X, \av^*)
\eea\
and
\beq\label{VVdaV-2}
n_{1} = h^1(X,\vv) \ . \nn \\
\eeq
This is a straightforward generalization of the formula for the
spectrum in the standard embedding case. Now, however, these
cohomology groups are unrelated to the Dolbeault cohomology of $X$ and,
hence, far more difficult to calculate. It will be the task of the
remainder of this paper to present a general method for computing the
quantities in \eref{VVdaV-1} and \eref{VVdaV-2}. Finally, note that
even though we will work within the context of $SU(5)$ GUTs, our
formalism and results generalize in a straight-forward manner to any
vector bundle $V$.

\subsection{Constraints from the Index Theorem}
Before computing the cohomology groups in \eref{VVdaV-1} and
\eref{VVdaV-2}, let us
see what simplifications can be achieved using the index theorem.
First, we apply Serre duality. This states that
\beq
\label{Serre}
H^i(X,V) \simeq H^{3-i}(X, V^* \otimes K_X)  \simeq 
H^{3-i}(X, V^*),
\eeq
where we have used the fact that $K_X$ is trivial on a
Calabi-Yau threefold $X$.
Therefore, we have
\beq\label{h0-h3}
H^0(X,V) \simeq H^3(X,V^*), \quad H^0(X,V^*) \simeq H^3(X,V)
\eeq
and
\beq\label{h1-h2}
H^1(X,V) \simeq H^2(X,V^*), \quad H^1(X,V^*) \simeq H^2(X,V).
\eeq
It can be shown that for a stable vector bundle $V$ of rank greater
than one and with vanishing
first Chern class,
\beq
\label{sd1}
H^0(X,V) = H^0(X,V^*) = 0 \ .
\eeq
It then follows from \eref{h0-h3} that
\beq
\label{sd2}
H^3(X,V^*) = H^3(X,V) = 0 \ .
\eeq
The Atiyah-Singer index theorem states that
\beq\label{AS}
\sum\limits_{i=0}^3 (-1)^i h^i(X,V)
= \int_X \ch(V) \td(X) = \frac12 \int_X c_3(V),
\eeq
where, in deriving the final term,
we have used the facts from \eref{c1TX} and \eref{chernV}
that $c_1(TX)$ and $c_1(V)$ both vanish. Expressions \eref{h0-h3},
\eref{h1-h2}, \eref{sd1} and \eref{sd2} 
allow us to simplify \eref{AS} to 
\beq
-h^1(X,V)+h^1(X,V^*) = \frac12 \int_X c_3(V).
\eeq
Now, recall the physical condition \eref{Ngen2} that our theory have
three quark/lepton generations.
It then follows from \eref{Ngen1} that
\beq\label{Ngens2}
\frac12 \int_X c_3(V) = 3.
\eeq
Therefore, the index theorem becomes
\beq\label{v-vd-1}
-h^1(X,V) + h^1(X,V^*) = 3.
\eeq
This will be an important constraint for us. It implies that, of
the two terms $h^1(X,V)$ and  $h^1(X,V^*)$ that we need to compute the
spectrum, 
it suffices to calculate only one of them and the other will differ from
it by $\pm 3$.

Similarly, we have that
\beq
H^0(X, \av) = H^0(X, \av^*) = 0,
\eeq
since $\av$ and $\av^*$ are both stable bundles
and have vanishing first Chern
class. Therefore, the
above arguments lead to the index theorem
\beq\label{AS-av}
- h^1(X,\av) + h^1(X,\av^*)
= \frac12 \int_X c_3(\av) \ .
\eeq
We have computed the Chern classes of the antisymmetrized
products of $V$ in Appendix B. Using \eref{chernV},
\eref{chernaV} and the fact that we have chosen the structure group of
$V$ to be $SU(5)$, it follows that
\beq
c_3(\av) = c_3(V).
\eeq
Therefore, the physical constraint \eref{Ngens2} simplifies the index
relation \eref{AS-av} to
\beq\label{v-vd-2}
- h^1(X,\av) + h^1(X,\av^*)
= 3 \ .
\eeq
As above, this constraint says that we need to compute only one of 
$h^1(X,\av)$ and $h^1(X,\av^*)$. The other will be determined by
adding or subtracting 3 according to \eref{v-vd-2}.

Finally, the above index theorem is inert when applied to
$\vv$. 
This is because Serre duality \eref{Serre} implies
\beq
H^0(X,\vv) \simeq H^3(X,(\vv)^*) = H^3(X, \vv)
\eeq
and
\beq
H^1(X,\vv) \simeq H^2(X,\vv).
\eeq
Therefore, application of the index theorem gives
\beq
0 = \frac12 \int_X c_3(\vv).
\eeq
This is consistent with the fact that
\beq
c_3(\vv) 
= 0,
\eeq
which holds since for any vector bundle $W$, $c_3(W^*) = -c_3(W)$, and
for $W = \vv$ we have $W = W^*$.
Hence, we must compute $h^1(X,\vv)$ directly.

\subsection{Determining The Spectral Data}\label{s:spec}

\def\n2{\cN_{V}^{\otimes 2}}
\def\nav{\cN_{\av}}
\def\navd{\cN_{\av^*}}

It follows from the previous discussion that the vector bundles
over $X$ that we will need to determine the spectrum are of five
types
\beq\label{U}
U = V, \ V^*, \ \av, \ \av^*, \ V \otimes V^* \ .
\eeq
For the first four bundles, 
it will be necessary to extract the relevant
cohomological data using the associated spectral data. 
The relevant quantity for the
bundle $V \otimes V^*$, namely $h^1(X,V \otimes V^*)$, can be computed
using a different technique and will be discussed separately in a
later section.

Recall, from the
discussion in Section \ref{s:FM}, that the Fourier-Mukai
transformation relates $U$ to its spectral data as
\beq
(\cC_U, \cN_U) \stackrel{FM}{\longleftrightarrow} U
\eeq
where $\cC_U$ and $\cN_U$ are a divisor on $X$ and a line bundle on
$\cC_U$ respectively. 
As we will see, for $U=V, \ V^*$ the line bundle $\cN_U$ on $\cC_U$
is the
restriction of a line bundle on $X$. We will extend the notation
discussed previously and not distinguish between $\cN_U$ on $X$ and
$\cN_U$ restricted to $\cC_U$, denoting both as $\cN_U$. In fact, to
simplify our notation, if $L$ is any line bundle on $X$ we will denote
both $L$ and $L|_{\cC_U}$ as $L$.
Note, however, that not all line bundles on $\cC_U$ are restrictions
of line bundles on $X$. Specifically, we will show that this is the
case for $U = \av, \ \av^*$.
We turn, therefore, to computing the spectral
data for each of $U= V,\  V^*, \ \av, \ \av^*$. Before computing, we note
that the curve defined by the intersection of the spectral cover with
the zero section, that is,
\beq\label{supportC-1}
\overline{c}_U = \cC_U \cap \sigma \ ,
\eeq
will play an important role in our analysis. We
will refer to this as the
``support curve.'' Note that in \eref{supportC-1} we use the
actual intersection ``$\cap$'', as opposed to the intersection
``$\cdot$'' in cohomology, since $\cC_U$ and $\sigma$ are specific
surfaces. Henceforth, writing $\cC_U \cdot \sigma$, for example, will
refer to the intersection of the class $[\cC_U]$ with the class
$[\sigma]$.
As defined, $\bar{c}_U$ is a curve in $\sigma$, $\cC_U$ and
$X$. Now, consider $\pi_{\sigma}(\bar{c}_U)$ in $B$. Since 
$\pi_{\sigma} : \sigma \rightarrow B$ is an isomorphism we will,
henceforth, not distinguish between $\pi_{\sigma}(\bar{c}_U)$ and
$\bar{c}_U$, denoting them both by $\bar{c}_U$.

\subsubsection{Spectral Data for $V$}
The spectral data for $U = V$ was presented in Section \ref{s:V}. For
the readers' convenience, we repeat those expressions here. The
spectral cover for a rank $n$ vector bundle $V$ has the form
\beq\label{coverX-2}
\cC_V \in |n\sigma + \pi^*\eta| \ ,
\eeq
where $\eta$ is some effective curve class 
in $B$. The spectral line bundle $\cN_V$ is specified by
expression \eref{c1N}
\begin{equation}
c_{1}(\cN_V)=n(\frac{1}{2}+\lambda)\sigma+(\frac{1}{2}-\lambda)
\pi^{*}\eta+(\frac{1}{2}+n\lambda)\pi^{*}c_{1}(TB) \ ,
\end{equation}
where $\lambda$ is a rational number satisfying conditions
\eref{lambda}.

\subsubsection{Spectral Data for $V^*$}
Let us now proceed to determine the necessary
data for $U = V^*$. 
We first recall from \eref{tau} that there is a natural involution
$\tau$ on $X$. 
The spectral cover $\cC_{V^*}$ is given by $\tau \cC_{V}$. Since
the linear system $|\cC_{V}|$ is invariant under $\tau$, we
have that $|\cC_{V^*}| = |\cC_V|$.
Now, note that the Chern classes have the property that
\beq\label{cstar}
c_i(V^*) = (-1)^i c_i(V)
\eeq
for any vector bundle $V$. Using this relation and expressions 
\eref{chernV} we can compute the Chern classes $V^*$. We see that,
since $|\cC_{V^*}| = |\cC_V|$, 
these Chern classes will be consistent with
choosing $\cN_{V^*}$ to be $\cN_V$ in \eref{c1N} with
\beq\label{cstar2}
\lambda \rightarrow -\lambda \ .
\eeq
In summary, we have
\bea
|\cC_{V^*}| &=& |\cC_V| \nn \\
c_{1}(\cN_{V^*})&=&n(\frac{1}{2}-\lambda)\sigma+(\frac{1}{2}+\lambda)
\pi^{*}\eta+(\frac{1}{2}-n\lambda)\pi^{*}c_{1}(TB) \ .
\label{c1Ndual}
\eea

\subsubsection{Spectral Data for $\av$}\label{s:specav}
Next, let us construct the spectral data for $U=\av$. The linear
system of the spectral cover
is easily determined, again, by considering the Chern classes. In
Appendix B, we compute the Chern classes of the
antisymmetric products of
a vector bundle. The result in \eref{chernaV} confirms that the vector
bundle $\av$ must have rank $\frac12n(n-1)$, as expected from
antisymmetry. Furthermore, since for our vector bundle $c_1(V)=0$, it
follows from \eref{chernaV} that
\beq
c_2(\av) = (n-2) c_2(V),
\eeq
where $c_2(V)$ is given in \eref{chernV}. We can now try to recast
$c_2(\av)$ in the same form as \eref{chernV}, but with $n$ replaced by
$\frac12 n(n-1)$ and possibly new values for $\eta$ and $\lambda$.
We see that the first term in $c_2(\av)$, the horizontal component of
the Chern class, can be put in the same form as \eref{chernV} by
choosing a new base curve $\eta'$ defined as $\eta' =
(n-2)\eta$. Since one can show that this horizontal term only depends
on the spectral cover, we can conclude that the spectral cover for
$\av$ is given by
\beq
\label{cXav}
\cC_{\av} \in \left|
\frac{n(n-1)}{2} \sigma + (n-2) \pi^* \eta \right|
\ .
\eeq
However, the remaining terms in $c_2(\av)$ do not arise from any line
bundle on $X$ of the form specified in \eref{c1N}.
It follows that $\cN_{\av}$ is a line bundle on $\cC_{\av}$
which is not the restriction of any bundle on
$X$. That is, $c_1(\cN_{\av})$ is represented by some curve in
$\cC_{\av}$, but this curve is not the complete intersection of
$\cC_{\av}$ with any divisor in $X$~\footnote{A 
theorem of Lefschetz shows that
any curve class on $\cC_V$ comes from a divisor in $X$, when $\cC_V$
is smooth and very ample. But $\cC_{\av}$ has no reason to satisfy
these conditions, so the Lefschetz theorem does not apply.}.
What
then is $c_1(\cN_{\av})$? This is a daunting problem. Happily, as we
will show in the next section, to compute the cohomology of $\av$ it
will be necessary to find not all of
$\cN_{\av}$ but, rather, its restriction
$\cN_{\av}|_{\overline{c}_{\av}}$, where ${\overline{c}_{\av}}$ is the
support curve defined by
\beq\label{caV}
{\overline{c}_{\av}} = \cC_{\av} \cap \sigma.
\eeq
This we can accomplish, as we now show.

Consider the action of the involution $\tau$, defined in
\eref{tau}, on the spectral cover surface
$\cC_V$. Denote the transformed surface by $\tau\cC_V$ and intersect
it with $\cC_V$ to obtain the curve $\tau\cC_V \cap \cC_V$. It is not
hard to show that this curve, which has multiple components, 
decomposes as
\beq\label{tCC}
\tau\cC_V \cap \cC_V = (\cC_V \cap \sigma) \cup
(\cC_V \cap \sigma_2) \cup D \ ,
\eeq
where $\sigma$, we recall, is the zero section and
$\sigma_2$ is the trisection that intersects each fiber at the
three non-trivial points of order 2. 
It is given by
\beq\label{sig2}
\sigma_2 \in |3 \sigma + 3 \pi^*(c_1(TB))| .
\eeq
Note, however, that there is a third component curve in \eref{tCC}
which we denote by $D$. Now, the linear system $|\cC_V|$ is invariant
under $\tau$. Therefore,
using expressions \eref{coverX} and \eref{sig2}, we can solve for
$D$. We find that it is a representative of the class
\beq
D 
\in
\left[
\sigma \cdot
\pi^*((2n-4)\eta - (n^2-n)c_1(TB)) + (\eta^2-3 \eta \cdot
c_1(TB))F \right] \ .
\label{D}
\eeq
Note that $D$ is contained in
$X$. Let $\bar{c}_{\av}$ be the curve
in the base associated with $\bar{c}_{\av}$ defined in \eref{caV}. We
remind the reader that, notationally, we are not distinguishing these
curves. Then, one
can show that $D$ is actually the double cover of the support
curve $\overline{c}_{\av}$, with covering map
$\pi_D: D \rightarrow \overline{c}_{\av}$. There are, in principle, a
number of branch points and the associated ramification points of this
mapping. 
The branch divisor in $\overline{c}_{\av}$ will be denoted by $Br$,
whereas the ramification divisor in $D$ is written as $R$. The numbers
of branch points and ramification points are given by $\deg(Br)$ and
$\deg(R)$ respectively. Note that
\beq
\deg(Br) = \deg(R) \ .
\eeq
In the following, we may, for simplicity, denote both the divisor
and its degree by the same symbol. For example, we will write $R$ for
both the ramification divisor and its degree.
This
divisor can be obtained as the intersection of $D$ with the
zero section and the points of order two. 
That is,
\beq\label{ram-1}
R = D \cap (\sigma + \sigma_2) \ .
\eeq
Numerically, $R$ can be computed using 
\beq\label{ram}
R = D \cdot (4\sigma + 3\pi^*(c_1(TB)) \ ,
\eeq
where $D$ is given in \eref{D}.
While the integer $R$ is always defined by formula \eref{ram},
the divisor $R$ is not always determined by formula \eref{ram-1}
unless the intersection is proper. In a crucial situation which we
will encounter, a component of $D$ is actually contained in
$\sigma$. In this situation, the divisor cannot be uniquely
determined. To our rescue comes intersection theory \cite{fulton},
which says that the line bundle $\cO_D(R)$ is nevertheless well
defined. It is given by a refinement of \eref{ram}, namely, $\cO_D(R)$
is the restriction to $D$ of the line bundle 
$\cO_X(R) = \cO_X(4\sigma + 3 \pi^* c_1(TB))$ on $X$. That is,
\beq\label{ram-2}
\cO_D(R) = \cO_X(4\sigma + 3 \pi^* c_1(TB))|_D \ .
\eeq


The definition of $D$ as a double cover of $\overline{c}_{\av}$ allows us
to relate the spectral line bundle $\nav|_{\overline{c}_{\av}}$ to
$\n2|_D$. After a detailed analysis, we can prove that
\beq
\pi_D^* \nav|_{\overline{c}_{\av}} = \n2|_D \otimes \cO_D(-R),
\eeq
or, equivalently,
\beq\label{pushN2}
\n2|_D 
= \pi_D^* \nav|_{\overline{c}_{\av}} \otimes \cO_D(R) 
\eeq
where $\cO_D(R)$ is given in \eref{ram-2}. 
Here and henceforth, we use the following notation. As discussed
earlier, $\cN_U$ denotes a line bundle both on $X$ and restricted to
$\cC_U$. In either case $\cN_U |_{\bar{c}_U}$ is well-defined. Now
consider $\pi_\sigma|_{\bar{c}_U *}(\cN_U |_{\bar{c}_U})$, which is a line
bundle on the curve $\bar{c}_U$ in the base $B$. Since 
$\pi_{\sigma} : \sigma \rightarrow B$ is an isomorphism, we will not
distinguish between $\pi_\sigma|_{\bar{c}_U *}(\cN_U |_{\bar{c}_U})$ and 
$\cN_U |_{\bar{c}_U}$, denoting them both by $\cN_U |_{\bar{c}_U}$.
It then follows that
\beq\label{c1ND-1}
c_1(\n2|_D) =
2c_1( \nav|_{\overline{c}_{\av}}) + R,
\eeq
where we have used the fact that $\pi_D$ is of degree two and that
$c_1(\cO_D(R)) = R$. Now, note that
\beq\label{c1ND-2}
c_1(\n2|_D) = 2 c_1(\cN_V) \cdot D,
\eeq
where $c_1(\cN_V)$ is given in \eref{c1N}. Combining \eref{c1ND-1} and
\eref{c1ND-2} yields the desired result
\beq\label{c1ND-3}
c_1(\nav|_{\overline{c}_{\av}}) = c_1(\cN_V) \cdot D 
- \frac{R}{2}.
\eeq
In summary, from \eref{cXav} and \eref{c1ND-3},
we conclude that the required part of the
spectral data for $\av$ is given by
\bea\label{N2Nav}
\cC_{\av} &\in& \left|\frac{n(n-1)}{2} \sigma + (n-2) \pi^* \eta
\right| \ , \nn \\
c_1(\nav)|_{\overline{c}_{\av}}
&=&\left(n(\frac{1}{2}+\lambda)\sigma+(\frac{1}{2}-\lambda)
\pi^{*}\eta+(\frac{1}{2}+n\lambda)\pi^{*}c_{1}(TB) \right)\cdot D -
\frac{R}{2} \ ,
\eea
with $D$ and $R$ are defined in \eref{D} and \eref{ram} respectively.

We can be even more specific about the structure of
$\nav|_{\overline{c}_{\av}}$. Pushing equation \eref{pushN2} down onto
$\overline{c}_{\av}$, one finds that
\beq\label{projODR}
\pi_{D*}(\n2|_D) =  \nav|_{\overline{c}_{\av}} \otimes
	\pi_{D*}(\cO_D(R)) \ .
\eeq
However, $\cO_D(R)$ pushed down onto $\overline{c}_{\av}$ is a rank two
vector bundle which is the direct sum of two line bundles,
\beq\label{pushODR}
\pi_{D*}(\cO_D(R)) = \cO_{\overline{c}_{\av}} \oplus 
	\cO_{\overline{c}_{\av}}(\frac{Br}{2}) \ ,
\eeq
where 
$Br = \pi_D(R)$ is the divisor of branch points in
$\overline{c}_{\av}$. Substituting 
\eref{pushODR} into \eref{projODR} gives
\beq\label{N2split}
\pi_{D*}(\n2|_D) =
\nav|_{\overline{c}_{\av}}
\oplus 
\left( \nav|_{\overline{c}_{\av}} \otimes 
	\cO_{\overline{c}_{\av}}(\frac{Br}{2}) \right) \ .
\eeq
This implicitly contains an expression for $\nav|_{\overline{c}_{\av}}$,
not just its first Chern class, that we will find useful later in the
paper.

\subsubsection{Spectral Data for $\av^*$}\label{s:specavd}
First, using the fact that
\beq
\wedge^2 V^* = (\wedge^2 V)^* \ ,
\eeq
we conclude that
\beq
|\cC_{\av^*}| = |\cC_{\av}| .
\eeq
Furthermore, we know from \eref{cstar} and \eref{cstar2}
that the Chern classes of $V^*$
are obtained from those of $V$ by letting $\lambda \rightarrow
-\lambda$. Using this and expression \eref{N2Nav} we can compute
$c_1(\navd)|_{\overline{c}_{\av^*}}$. In summary, we find, using
\eref{caV} and \eref{N2Nav}, that
\bea\label{N2Navstar}
\cC_{\av^*} &\in& \left|\frac{n(n-1)}{2} \sigma + (n-2) \pi^* \eta
\right| \ , \nn \\
c_1(\navd)|_{\overline{c}_{\av^*}}
&=&\left(n(\frac{1}{2}-\lambda)\sigma+(\frac{1}{2}+\lambda)
\pi^{*}\eta+(\frac{1}{2}-n\lambda)\pi^{*}c_{1}(TB) \right)\cdot D -
\frac{R}{2} \ .
\eea
Similarly, for the bundle $\av^*$, expression \eref{N2split} becomes
\beq\label{N2split2}
\pi_{D*}(\cN^{\otimes 2}_{V^*}|_D) =
\cN_{\av^*}|_{\overline{c}_{\av^*}}
\oplus 
\left( \cN_{\av^*}|_{\overline{c}_{\av^*}} \otimes 
	\cO_{\overline{c}_{\av^*}}(\frac{Br}{2}) \right) \ .
\eeq
where $Br$ is the divisor of branch points in
$\overline{c}_{\av^*}$. It is not hard to show that 
$\overline{c}_{\av^*} = \overline{c}_{\av}$.

\subsection{Computing the Particle Spectrum}\label{s:spectrum}
We know from the discussions at the beginning of this section that to
determine the particle spectrum one must compute 
$H^1(X, U)$ for $U = V,V^*,\av,\av^*$ and $V\otimes
V^*$. Now, recall that any such $U$ and $X$ have the following
structure
\beq\ba{ccc}
U & & \\
\downarrow & & \\
X & \stackrel{\pi}{\rightarrow} & B \ .
\ea\eeq
The standard technique for finding $H^1(X,U)$ on such a
structure is to evoke the Leray spectral sequence. This
reduces the cohomology of $U$ over $X$ to that of derived functors
$R^i \pi_*U$ over the base $B$. 
Since the fibers of the projection map
$\pi: X \rightarrow B$ are one-dimensional, we see that for
any vector bundle $U$ on $X$, the Leray
spectral sequence reduces to a single long exact sequence
\beq
\label{leray}
0 \rightarrow H^1(B, \pi_* U) \rightarrow 
H^1(X, U) \rightarrow 
H^0(B, R^1\pi_* U) \rightarrow  
H^2(B,\pi_* U) \rightarrow  
\ldots
\eeq
where $R^i \pi_*$ is the $i$-th right derived functor for the
push-forward map $\pi_*$. The reader is referred to 
\cite{hart,griffith} for a discussion of Leray sequences.
We first recall some key facts from \cite{redumod} concerning the
properties of $\pi_* V$ and $R^1 \pi_* V$.
On the base surface $B$ of our elliptic
fibration, we have
\beq
\label{piV}
\pi_* V = 0, \quad \rk_B(R^1 \pi_* V) = 0 \ .
\eeq
This follows from the fact that $V$ is a vector bundle corresponding
to an irreducible spectral cover.
Similarly,
this holds for $U=V^*,\av$ and $\av^*$. That is
\beq\label{piU}
\pi_* U = 0, \quad \rk_B(R^1 \pi_* U) = 0
\eeq
for $U = V, V^*, \av, \av^*$.
For $U=V \otimes V^*$, however,
it is not true that $\pi_* U$ vanishes. For this reason, as
mentioned earlier, we will compute $H^1(X,V \otimes V^*)$ separately,
using a different formalism. Here, we will restrict $U$ to be $V, V^*,
\av, \av^*$ only.

It follows from the first equation in \eref{piU} that
\beq
H^1(B, \pi_* U) = H^2(B, \pi_* U) = 0.
\eeq
The sequence \eref{leray} then implies
\beq\label{reduB}
H^1(X,U) \simeq H^0(B, R^1\pi_* U).
\eeq
The first direct image $R^1 \pi_* U$
does not vanish identically on $B$ but, rather, is 
a torsion sheaf. It is supported on the curve
\beq
\label{supportC}
\overline{c}_U = \cC_U \cap \sigma
\eeq
in $B$.
The genus of $\sC$ can be obtained from the adjunction formula
\beq
\label{adjunct}
2 g - 2 = \overline{c}_U \cdot (\overline{c}_U + c_1(K_B)) \ .
\eeq
To recapitulate, our spectrum calculation simplifies to finding the global 
holomorphic sections of $R^1\pi_* U$ on the support curve
$\overline{c}_U \subset B$. That is,
\beq\label{X2c}
H^1(X,U) \simeq H^0(\overline{c}_U, R^1\pi_* U |_{\overline{c}_U}) \ .
\eeq
It is important to note, however, that even though $\pi_*U$ and
$\rk_B(R^1\pi_*U)$ vanish, $R^1\pi_*U$ need not be zero.
\subsection{The First Image $R^1\pi_* U$}
Let us determine, using the Fourier-Mukai techniques presented in
Section \ref{s:FM}, the torsion sheaf
$R^1\pi_* U$.
The commutativity of the diagram in \eref{commU} allows us to write
\beq
\pi' \circ \pi_1 = \pi_{\cC} \circ \pi_2 \ .
\eeq
This implies that, in the derived category, the functors 
$R$ of these projection maps obey
\beq\label{Rpi}
R \pi'_* \circ R \pi_{1*} = R  \pi_{\cC *} \circ R \pi_{2 *} \ .
\eeq
However, $\pi_{1}$ and $\pi_{\cC}$ are finite covering maps 
so their higher direct images vanish.  This means that
\beq
R \pi_{1 *} = \pi_{1 *}, \ \ R \pi_{\cC *} =  \pi_{\cC *} \ .
\eeq
Subsequently, \eref{Rpi} becomes
\beq
\label{comm}
R^i \pi'_* \circ \pi_{1*} =  \pi_{\cC *} \circ R^i \pi_{2 *} \ , 
\quad i \ge 0.
\eeq
Applying the $i=1$ case of \eref{comm} to the sheaf 
$\pi_2^*\cN_U \otimes \cP$, we have that
\beq
\label{comm2}
R^1 \pi'_* \circ \pi_{1*}(\pi^*_2\cN_U \otimes \cP) 
= 
 \pi_{\cC *} \circ R^1 \pi_{2*}(\pi_2^*\cN_U \otimes \cP) \ .
\eeq
Now we recall the definition of $U$ from \eref{FM} and use the
projection formula
\beq
R^1 \pi_{2*}(\pi_2^*\cN_U \otimes \cP) =
\cN_U \otimes R^1 \pi_{2*} \cP.
\eeq
Then \eref{comm2} simplifies to
\beq
\label{comm3}
R^1 \pi'_* U =  \pi_{\cC *} ( \cN_U \otimes R^1 \pi_{2*} \cP) \ .
\eeq
From now on, using the fact that $X \simeq X'$, we will omit the prime
and replace $\pi'$ by $\pi$.
The left hand side is precisely the term we desire, while the
right hand side can be further simplified using the following
commutative diagram
\beq
\begin{diagram}
\cC_U \times_B X' & \rTo{i_{\cC\times_B X'}}{}
	& X \times_B X' & \leftarrow \cP \\
\pi_2 \dTo & & \dTo \pi_X &\\
\cC_U & \rTo{}{i_\cC} & X &
\end{diagram}
\ .
\eeq
Therefore,
\beq
R^1\pi_{2*} \cP = i_\cC^* (R^1 \pi_{X*} \cP) \ .
\eeq
Finally, we know that
\beq
R^1 \pi_{X*} \cP = \sigma_* K_B \ ,
\eeq
and thus
\beq
R^1 \pi_{2 *} \cP = i_\cC^* \sigma_* K_B \ ,
\eeq
where $\sigma : B \rightarrow X$ is the zero section map discussed in
Section \ref{s:CY}.
Substituting this into \eref{comm3} gives us
\beq\label{R1pU1}
R^1\pi_* U =  \pi_{\cC*}( \cN_U \otimes (i_\cC^* \sigma_* K_B ) ) \ .
\eeq
It is clear that $i_\cC^* \sigma_* K_B$ is a sheaf on $\cC_U$ with
support on the curve $\bar{c}_U = \cC_U \cap \sigma$. Note that
this can be thought of as $\pi_{\cC}^*K_B |_{\bar{c}_U}$. Then,
for any line bundle $\cN_U$, we can identify
\beq\label{NU-id}
\cN_U \otimes (i_\cC^* \sigma_* K_B) = 
(\cN_U \otimes \pi_{\cC}^* K_B)|_{\bar{c}_U} \ .
\eeq
For $U = V, \ V^*$, when $\cN_U$ is defined globally on $X$, we can
replace $\pi_{\cC}^*$ by $\pi^*$ in \eref{NU-id} and write
\beq
\cN_U \otimes (i_\cC^* \sigma_* K_B) = 
(\cN_U \otimes \pi^* K_B)|_{\bar{c}_U} \ .
\eeq
Note that this latter equation does not apply to $U = \av, \ \av^*$
since, in these cases, $\cN_U$ is defined on $\cC_U$ only. Be that as
it may, to avoid having to give separate discussions we will,
henceforth, always denote this sheaf by $(\cN_U \otimes \pi^*
K_B)|_{\bar{c}_U}$. Whether the lift is by $\pi^*$ or $\pi^*_{\cC}$
will be clear from the context.
Since, as
discussed previously, we will not distinguish a line bundle on
$\bar{c}_U \subset \cC_U$ from the associated line bundle on 
$\bar{c}_U \subset B$, we can write
\beq\label{R1pU2}
 \pi_{\cC*}( \cN_U \otimes (i_\cC^* \sigma_* K_B ) )
= ( \cN_U \otimes \pi^* K_B )|_{\overline{c}_U} \ .
\eeq
Combining \eref{R1pU1} and \eref{R1pU2} 
with expression \eref{X2c}, we have \beq\label{h1V-1}
H^1(X,U) \simeq H^0(\overline{c}_U, ( \cN_U \otimes \pi^*K_B
)|_{\overline{c}_U}). 
\eeq
Since we know $\sC$, $\cN_U|_{\sC}$ and $K_B|_{\sC}$, this
expression allows us, in principle, to compute $H^1(X,U)$ for
$U = V, V^*, \av, \av^*$. In practice, this calculation will depend on
the properties of the support curve $\sC$.

\subsection{Riemann-Roch Theorem on a Smooth Support Curve}\label{s:sC}
Generically, the support curve $\sC$ 
will be smooth and irreducible. However, our three-family
constraint may make it reducible and even non-reduced and, therefore,
much more difficult to analyze. 
Even in these cases, however, $\sC$ may contain a smooth, irreducible
component. It is of interest, therefore, to discuss
the Riemann-Roch
theorem for smooth curves. As we will see, this theorem is very
helpful in computing either all, or part, of $H^0(\sC, (\cN_U \otimes
\pi^* K_B)|_{\sC})$.

The Riemann-Roch theorem states that for a smooth
curve $C$ and a line bundle $\cF$ on $C$, we have
\beq\label{RR-C}
h^0(C, \cF) - h^1(C, \cF) = \deg(\cF) - g(C) + 1,
\eeq
where $g(C)$ is the genus of the curve and
\beq
\deg(\cF) = c_1(\cF)
\eeq
is the degree of $\cF$. In our problem,
we need to calculate the term $h^0(C, \cF)$.
If $\deg(\cF)<0$, the term $h^0(C, \cF)$ simply vanishes because
there are no global holomorphic sections to a line bundle of negative
degree. Now, we assume that $\deg(\cF) \ge 0$.
Using Serre duality, we have
\beq
h^1(C, \cF) = h^0(C, \cF^* \otimes K_C),
\eeq
where $K_C$ is the canonical bundle of $C$. Then,
\eref{RR-C} becomes
\beq\label{RR-C-1}
h^0(C, \cF) - h^0(C, \cF^* \otimes K_C) = \deg(\cF) - g(C) + 1.
\eeq
Now,
\beq\label{degF}
\deg( \cF^* \otimes K_C ) = -\deg(\cF) + \deg(K_C) =  -\deg(\cF) +
2g(C) - 2,
\eeq
where we have used the fact that
\beq
\deg(K_C) = 2g(C) - 2.
\eeq
When $\deg( \cF^* \otimes K_C ) < 0$, $h^0(C, \cF^* \otimes K_C)$
vanishes and \eref{RR-C-1} becomes
\beq\label{RR-C-stab}
h^0(C, \cF) = \deg(\cF) - g(C) + 1.
\eeq
Thus, for line bundles $\cF$ on $C$ for which $\deg(\cF^* \otimes K_C ) <
0$, the so-called ``stable range,'' we can compute $h^0(C, \cF)$
explicitly using \eref{RR-C-stab}. Note, however, that outside this
range the Riemann-Roch theorem is not sufficient in determining
$h^0(C, \cF)$.

As a simple example, let us assume that the support curve $\sC$ 
is smooth and consider the line bundle $(\cN_U \otimes
\pi^*K_B)|_{\sC}$.
That is, take $C = \sC$ and $\cF = (\cN_U \otimes \pi^*K_B)|_{\sC}$.
Let us denote
\beq
\label{deg-i}
d = \deg( (\cN_U \otimes \pi^* K_B)|_{\sC} ) = 
\int_{\sC} c_1(\cN_U \otimes \pi^*K_B).
\eeq
If $d < 0$, 
then $h^0(\sC, (\cN_U \otimes \pi^* K_B)|_{\sC})$ vanishes. Now assume that
$d \ge 0$. It follows from \eref{degF} that
\beq
\label{h1piece}
\deg((\cN_U \otimes \pi^*K_B)|_{\sC}^* \otimes K_{\sC})
= - d + 2g(\sC) - 2 \ .
\eeq
If expression \eref{h1piece} is non-negative, the Riemann-Roch theorem
is insufficient to determine
$h^0(\sC, (\cN_U \otimes \pi^*K_B)|_{\sC})$. However, if
\beq\label{sCrange}
-d + 2g(\sC) - 2 < 0,
\eeq
that is, if $(\cN_U \otimes \pi^*K_B)^* \otimes K_C$ is in the stable range,
\eref{RR-C-stab} implies that
\beq
\label{h1V_2}
h^0(\sC, (\cN_U \otimes \pi^*K_B)|_{\sC}) = d - g(\sC) + 1 \ .
\eeq
To complete this computation, note that
\beq
c_1((\cN_U \otimes \pi^*K_B)|_{\sC}) =
c_1(\cN_U|_{\sC}) +  c_1(K_B|_{\sC}).
\eeq
Then, expression \eref{h1V_2} becomes
\beq
h^0(\sC,(\cN_U \otimes \pi^*K_B)|_{\sC}) = 
(-c_1(TB) + c_1(\cN_U)) \cdot \sC - g(\sC) + 1.
\eeq
Recalling that the genus $g(\sC)$ can be computed by adjunction
\eref{adjunct}, and using \eref{h1V-1}, the above 
expression simplifies to
\beq
h^1(X,U) =  \left( c_1(\cN_U) - \frac12 c_1(TB) - \frac12 \sC
\right) \cdot \sC \ .
\eeq
Note that the first intersection is to be carried out in $X$ where the
remaining two intersections occur in the base $B$.
This result is the final expression for $h^1(X,U)$
in the simple case that the
support curve $\sC$ is smooth and the degree of 
$\cN_U \otimes \pi^*K_B$ 
falls in the stable range \eref{sCrange}. Unfortunately, as we will
see below, this simplified example is not realised in the realistic
three-family constrained models of interest. Be that as it may, the
general analysis of the Riemann-Roch theorem presented here will play
an important role in determining part of $h^1(X,U)$ on smooth
components of $\sC$.
%
%
\section{An Explicit Calculation}
The calculation of $H^1(X,U)$ for $U = V,V^*,\av$ and $\av^*$ on the
complicated curves $\sC$ that are encountered in phenomenologically
realistic theories is rather intricate. We find it expedient, and more
enlightening, to present an explicit example that will illustrate all
of the techniques necessary to compute $H^1(X,U)$. These techniques
can be applied to most other examples that could arise. With this in
mind, we choose the following example from the $SU(5)$ GUT vacua
classified in Section \ref{s:class}. Recall from \eref{lam-F1} and
\eref{F1sol} that
\beq\label{eg}
B = \IF_1, \ n=5, \ \eta = 12 S + 15 \cE, \ \lambda = \frac12
\eeq
is an explicit solution which satisfies all of the physical
constraints. We proceed to calculate $H^1(X,U)$ for $U = V,V^*,\av$
and $\av^*$ as well as $\vv$ within the context of this example.

\subsection{Calculation of $h^1(X,V)$ and $h^1(X,V^*)$}
It follows from \eref{coverX-2}, \eref{c1N} and \eref{eg} that the
spectral data for $V$ is given by
\beq\label{cVeg}
\cC_V \in | 5\sigma + \pi^*(12 S + 15 \cE) |
\eeq
and
\beq\label{NVeg}
c_1(\cN_V) = 5 \sigma + \pi^*(3 c_1(T\IF_1)),
\eeq
where, using \eref{chernFr},
\beq\label{c1F1}
c_1(T\IF_1) = -c_1(K_{\IF_1}) = 2S + 3 \cE.
\eeq
It follows from \eref{h1V-1} that
\beq\label{h1h0Veg}
h^1(X,V) = h^0(\overline{c}_V, 
	( \cN_V \otimes \pi^*K_{\IF_1} )|_{\overline{c}_V}),
\eeq
so we need to find expressions for both $\overline{c}_V$ and $( \cN_V
\otimes \pi^*K_{\IF_1} )|_{\overline{c}_V}$. 
The support curve $\overline{c}_V$ is
easily calculated from \eref{intersig}, \eref{interFr},
\eref{supportC-1}, \eref{cVeg} and \eref{c1F1} to be
\beq\label{sceg}
\overline{c}_V = 2 S.
\eeq 
Furthermore, it follows from \eref{NVeg} and \eref{c1F1} that 
\beq\label{c1egA}
c_1(\cN_V \otimes \pi^*K_{\IF_1}) = 5 \sigma + \pi^*(4S + 6\cE)
\eeq
and, hence, using \eref{intersig}, \eref{interFr} and \eref{sceg} that
\beq
c_1(\cN_V \otimes \pi^*K_{\IF_1}) \cdot {\overline{c}_V} = -6.
\eeq
Unfortunately, the support curve \eref{sceg} is non-reduced, being
scheme-theoretically twice the sphere $S$. The linear system $|2S|$
contains only one curve, this element is given by the unique global
section of $\cO_{\IF_1}(2S)$, which vanishes along $S$ to second
order.
Since such a curve is not smooth, we can not apply the above
Riemann-Roch analysis. How then, can we proceed to evaluate
$h^1(X,V)$? To do this, we first note that
\beq
S \simeq \IP^1.
\eeq
Now, for simplicity, denote the line bundle $\cN_V \otimes \pi^*
K_{\IF_1}$
on $X$ by
\beq
L = \cN_V \otimes \pi^*K_{\IF_1}.
\eeq
When restricted to $S$, this bundle has degree
\beq
c_1(L)|_S 
= -3,
\eeq
where we have used \eref{intersig}, \eref{interFr} and
\eref{c1egA}. However, since $S$ is a $\IP^1$, it follows that 
$L|_S$ is none other than the line bundle
\beq\label{LegF}
L|_S = \cO_{\IP^1}(-3).
\eeq
Let us now invoke the following short exact sequence for the
non-reduced scheme $2S$, 
\beq\label{S-2S-1}
0 \rightarrow \cO_S(-S) \rightarrow \cO_{2S} \rightarrow \cO_S
\rightarrow 0 \ ,
\eeq
which, since $S$ is a $\IP^1$ in $\IF_1$, becomes
\beq\label{S-2S}
0 \rightarrow \cO_{\IP^1}(1) \rightarrow \cO_{2S} \rightarrow
\cO_{\IP^1} \rightarrow 0 \ .
\eeq
Tensoring this sequence with $L|_S$ in \eref{LegF} gives us
\beq
\label{hVLseq}
0 \rightarrow \cO_{\IP^1}(-2) \rightarrow L|_{2S} \rightarrow
\cO_{\IP^1}(-3) \rightarrow 0 \ .
\eeq
Now neither $ \cO_{\IP^1}(-2)$ nor $\cO_{\IP^1}(-3)$ has global
holomorphic sections, being of negative degree.
This implies that $L|_{2S}$ also has no global sections. But
\beq
L|_{2S} = (\cN_V \otimes \pi^*K_{\IF_1})|_{\overline{c}_V}.
\eeq
It then follows from \eref{h1h0Veg} that
\beq\label{h1Veg}
h^1(X,V) = h^0(\bar{c}_V,L|_{\overline{c}_V}) = 0 \ .
\eeq
Thus by exploiting the exact sequence \eref{S-2S-1} on $S \simeq
\IP^1$, we have succeeded in computing $h^1(X,V)$.

As discussed previously, the dimensions of the first cohomology group
of the dual bundle, $h^1(X,V^*)$, can be immediately computed from
the index theorem result \eref{v-vd-1}
\beq\label{v-vd-1-1}
-h^1(X,V) + h^1(X,V^*) = 3.
\eeq
It follows, using \eref{h1Veg}, that
\beq\label{h1Vdeg}
h^1(X,V^*) = 3.
\eeq
It is reassuring to calculate $h^1(X,V^*)$ directly
using the method employed to compute $h^1(X,V)$.
Hopefully, this will reproduce the result in
\eref{h1Vdeg}. It follows from \eref{c1Ndual}, \eref{eg} and
\eref{c1F1} that the
spectral data for $V^*$ is given by
\beq\label{cVegd}
\cC_{V^*} \in | 5\sigma + \pi^*(12 S + 15 \cE) |
\eeq
and
\beq\label{NVegd}
c_1(\cN_{V^*}) = \pi^*(8 S + 9 \cE).
\eeq
Note that since $\cC_{V^*}$ is in the same linear system as $\cC_V$, 
it follows that
\beq\label{scegd}
\overline{c}_{V^*} = \overline{c}_V = 2S.
\eeq
Defining
\beq
L' = \cN_{V^*} \otimes \pi^* K_{\IF_1},
\eeq
we see, using \eref{c1F1} and \eref{NVegd}, that
\beq
c_1(L') = \pi^*(6S+6\cE)
\eeq
and, hence, using \eref{interFr}
\beq
c_1(L')|_S = 0. 
\eeq
That is,
\beq\label{c1Lp0}
L'|_S = \cO_{\IP^1}.
\eeq
Tensoring the short exact sequence in \eref{S-2S} with $L'|_S$ in
\eref{c1Lp0} gives
\beq\label{Lpseq}
0 \rightarrow \cO_{\IP^1}(1) \rightarrow L'|_{2S} \rightarrow
\cO_{\IP^1} \rightarrow 0 \ .
\eeq
Now, recall that
\beq\label{degP1}
h^0(\IP^1, \cO_{\IP^1}(n)) = \left\{\ba{ll}
0, & n < 0 \nn \\
n+1, & n \ge 0,
\ea\right.
\eeq
for integer $n$.
It then follows from \eref{Lpseq} and
\eref{degP1} that
the number of global holomorphic sections of $L'|_{2S}$ is simply
\beq
h^0(S, \cO_{\IP^1}(1)) + h^0(S, \cO_{\IP^1}) 
= 3.
\eeq
In deriving this result, we have used the fact that $h^1(\IP^1,
\cO_{\IP^1}(1)) = 0$. 
Noting that
\beq
L'|_{2S} = (\cN_{V^*} \otimes \pi^* K_{\IF_1})|_{\overline{c}_{V^*}} 
\ ,
\eeq
it follows from \eref{h1h0Veg} that
\beq\label{h1Vstareg}
h^1(X,V^*) = h^0(\overline{c}_{V^*}, (\cN_{V^*} \otimes
\pi^* K_{\IF_1})|_{\overline{c}_{V^*}}) = 3 \ ,
\eeq
which is consistent with the index theorem result presented
in \eref{h1Vdeg}.

\subsection{Calculation of $h^1(X,\av)$ and $h^1(X,\av^*)$}\label{s:avstar}

\def\n2d{\cN_{V^*}^{\otimes 2}}
\def\sCav{\overline{c}_{\av^*}}

We will calculate the term $h^1(X,\av^*)$ first because it will turn
out to be computationally easier. We then use the index theorem
\eref{v-vd-2} to compute $h^1(X,\av)$.

It follows from the first equation in \eref{N2Navstar} and \eref{eg}
that the spectral cover for $\av^*$ is
\beq\label{caveg}
\cC_{\av^*} \in |10 \sigma + \pi^* 3 (12 S + 15 \cE)| \ .
\eeq
Using \eref{intersig}, \eref{supportC-1} and \eref{c1F1}, we find that
the associated support curve is given by
\beq\label{sCwedge}
\overline{c}_{\av^*} \in [ 16 S + 15 \cE ] \ .
\eeq
Note that every curve in this linear system
is reducible, and decomposes into two generically
non-intersecting components as follows.
\beq\label{C1C2}
\overline{c}_{\av^*} = C_1 \cup C_2,
\eeq
where
\beq\label{defC1C2}
C_1 \in [S], \quad C_2 \in [15(S + \cE)].
\eeq
Using \eref{interFr} we see that
\beq
C_1 \cdot C_2 = 0.
\eeq
So either $C_1$ and $C_2$ are disjoint, or $C_2$ must decompose
further, with $S$ a component of $C_2$. 
Since $\sCav$ splits into $C_1$ and $C_2$, 
$D$ also splits into two generically disjoint components,
\beq
D = D_1 \cup D_2,
\eeq
where $D_1$ and $D_2$ are double covers of $C_1$ and $C_2$
respectively.  It can be shown that
\beq
D_1 \in [2S],
\eeq
and
\beq\label{defD2}
D_2 \in [ 30(S+\cE) + 90 F ] \ .
\eeq
The restriction of $\pi_D: D \rightarrow C$ gives a double cover 
$\pi_{D_2}: D_2 \rightarrow C_2$ with branch divisor $Br_2$ and
ramification $R_2$. However, $D_1$, as mentioned earlier, is actually
contained in $\sigma$, and is, in fact,
non-reduced so there is no divisor candidate for $Br_1$ and $R_1$.

The term we wish to compute is $h^1(X, \av^*)$.
In principle, one could try to compute it in a manner
similar to the calculation of $h^1(X,U)$ for $U=V,V^*$ above. However,
this is not possible in the present case. Recall that in the previous
section $\overline{c}_V =\overline{c}_{V^*} = 2S$, where $S \simeq \IP^1$ is a
rigid curve in $\IF_1$. This allowed us to use a short exact sequence
to relate the spectral line bundle twisted with $K_{\IF_1}$ on $2S$ to
line bundles on $\IP^1$. The result then followed. Unfortunately, in
the present case $\overline{c}_{\av}$ contains a component, 
$C_2 \in [15(S +\cE)]$, 
that is not a multiple copy of an isolated $\IP^1$.
This greatly complicates the problem, and
makes a solution along the lines of the last section impossible. 
One might try to directly apply the Riemann-Roch theorem to the curve
$C_2$. However, one finds that the associated line bundle lies outside
the stable range. How then, can we proceed?

Let us first define the bundles
\beq\label{W-def}
W = \n2d \otimes \pi^*K_{\IF_1}
\eeq
and
\beq\label{Z-def}
Z = \navd \otimes \pi^*K_{\IF_1} \ .
\eeq
Then, the term we wish to compute is $h^0(\sCav,Z|_{\sCav})$.
Applying \eref{N2split2} to $\pi_2 : D_2 \rightarrow C_2$ gives
\beq
\pi_{D_{2*}}(\cN^{\otimes 2}_{V^*}|_{D_2}) =
\cN_{\av^*}|_{C_2}
\oplus 
\left( \cN_{\av^*}|_{C_2} \otimes 
	\cO_{C_2}(\frac{Br_2}{2}) \right) \ .
\eeq
Tensoring with $K_{\IF_1}$ and taking $h^0$, this becomes
\beq
h^0(D_2, W|_{D_2}) = 
h^0(C_2, Z|_{C_2}) + 
h^0(C_2, (Z \otimes \cO_{C_2}(\frac{Br_2}{2}))|_{C_2}) \ .
\eeq
Putting this together with \eref{h1V-1} and using the definitions
\eref{W-def} and \eref{Z-def}, we obtain
\bea\label{F14terms}
h^1(X, \av^*) &=& h^0(\sCav, Z|_{\sCav}) \nn \\
	&=& h^0(C_1, Z|_{C_1}) + h^0(C_2, Z|_{C_2}) \nn \\
	&=& h^0(C_1, Z|_{C_1}) + h^0(D_2, W|_{D_2}) - 
	h^0(C_2, (Z \otimes \cO_{C_2}(\frac{Br_2}{2}))|_{C_2})\nn \\
	&=& 
	- h^0(D_1, W|_{D_1}) 
	+ h^0(C_1, Z|_{C_1})
	- h^0(C_2, (Z \otimes \cO_{C_2}(\frac{Br_2}{2}))|_{C_2})
	+ h^0(D, W|_D)\nn \\
\eea
Therefore, we need to compute four terms, 
$h^0(D_1, W|_{D_1})$,
$h^0(C_1, Z|_{C_1})$
$h^0(C_2, (Z \otimes \cO_{C_2}(\frac{Br_2}{2}) )|_{C_2})$
and $h^0(D, W|_{D})$ in \eref{F14terms} to finish our calculation.

We begin by calculating the first term $h^0(D_1, W|_{D_1})$.
Since $D_1 = 2S$, we see that this term can be computed exactly as in
the previous section. First, we note that on 
$S$, $W$ is simply
\beq\label{LS1}
W|_S = \cO_{\IP^1}(1).
\eeq
Next, recall from \eref{S-2S} that, for
$S \simeq \IP^1$, we have the exact sequence
\beq\label{C1seq}
0 \rightarrow \cO_{\IP^1}(1) \rightarrow \cO_{2S} \rightarrow
\cO_{\IP^1} \rightarrow 0 \ .
\eeq
Then,  tensoring by $W|_S$ in \eref{LS1} gives
\beq\label{S-2S-2}
0 \to \cO_{\IP^1}(2) \to W|_{2S} \to \cO_{\IP^1}(1) \to 0.
\eeq
It now follows that
\beq\label{h0D1-1}
h^0(D_1, W|_{D_1}) 
=h^0(\IP^1, \cO_{\IP^1}(2)) + h^0(\IP^1, \cO_{\IP^1}(1)) = 5 
\eeq
where we have used \eref{degP1}. 

Next, we compute the second of the four terms in \eref{F14terms},
namely, $h^0(C_1, Z|_{C_1})$.
Recalling that $C_1=S$, this can also be readily computed.
Note that
\beq
Z|_{C_1} = 
(\navd \otimes \pi^*K_{\IF_1})|_{C_1} \simeq W_S(-S) = \cO_{\IP^1}(2).
\eeq
Therefore, using \eref{degP1}, we find
\beq\label{h0D1-2}
h^0(C_1, Z|_{C_1}) = 3.
\eeq

Now, we move on to the third requisite term 
$h^0(C_2, (Z \otimes \cO_{C_2}(\frac{Br_2}{2}) )|_{C_2})$
in \eref{F14terms}.
One way to do this
is to try to use the Riemann-Roch theorem on the support curve
as outlined in Subsection \ref{s:sC}.
In this regard, define the line bundle
\beq
\cF = ( Z \otimes \cO_{C_2}(\frac{Br_2}{2}) )|_{C_2}
\eeq
and recall that
\beq
\deg(\cF) = c_1(\cF) \ .  
\eeq
The restriction of \eref{N2Navstar} to $C_2$ implies
\beq
c_1(\navd|_{C_2}) = c_1(\cN_{V^*}) \cdot D_2 - \frac{R_2}{2}
\eeq
where, by definition
\beq
c_1(\cO_{C_2}(\frac{Br_2}{2})) = \frac{R_2}{2}
\eeq
and $R_2$ are the ramification points on $D_2$.
It follows that
\beq
\deg(\cF) = c_1(\n2d) \cdot D_2 + c_1(K_{\IF_1}) \cdot C_2.
\eeq
Evaluating \eref{c1Ndual} for the vacuum given in \eref{eg} and using
\eref{c1F1} and the curves $C_2$ and $D_2$ presented in \eref{defC1C2}
and \eref{defD2} respectively, we find
\beq\label{F225}
\deg(\cF) = 225.
\eeq
Now we compute the genus of $C_2$ using \eref{adjunct}. Using
\eref{c1F1} and \eref{defC1C2}, we find that
\beq\label{gC2}
2g(C_2) - 2 = 180.
\eeq
We note that the quantity
\beq
-\deg(\cF) + 2g(C_2) - 2 = -45 <0.
\eeq
Thus, the line bundle $\cF$ is in the stable range discussed in
Subsection \ref{s:sC}. Therefore, we can use \eref{RR-C-stab} to
determine $h^0(C_2, \cF)$. It follows from \eref{RR-C-stab}, 
\eref{F225} and \eref{gC2} that
\beq\label{h0D2}
h^0(C_2, (Z \otimes \cO_{C_2}(\frac{Br_2}{2}) )|_{C_2}) =
h^0(C_2, \cF) = 
225 - g(C_2) + 1 = 135.
\eeq
This completes the calculation of the third of the four required
terms.

Finally, we proceed to compute the remaining term 
$h^0(D, W|_D)$ in \eref{F14terms}
to complete our computation. Unfortunately,
evaluating this quantity is considerably more difficult.
The term we need to compute is $h^0(D, W|_D)$. 
We will use the fact that $W|_D$ is a restriction of the global line
bundle $W$ on $X$. For such cases we have the technology to count
global sections. In our particular
example,
\beq\label{W-val}
W = \pi^*\cO_B( 2 \eta - 5 c_1(T\IF_1)) 
= \pi^*\cO_B(14 S + 15 \cE) \ ,
\eeq
%
%
where we have used \eref{c1Ndual} and \eref{eg}. We proceed by
considering the short exact sequence on $\cC_V$
\beq
0 \rightarrow W|_{\cC_V}(-D) \rightarrow W|_{\cC_V} \rightarrow W|_D
\rightarrow 0 \ ,
\eeq
where, using standard notation, we denote $W|_{\cC_V} \otimes
\cO_{\cC_V}(-D)$ by $W|_{\cC_V}(-D)$.
This induces a long exact sequence in cohomology
\beq\label{seq1}
\ba{ccccccccc}
0 & \to & H^0(\cC_V,W|_{\cC_V}(-D)) & \to & H^0(\cC_V,W|_{\cC_V}) & \to &
	\fbox{$H^0(D,W|_D)$} & \to &\\
& \to & H^1(\cC_V,W|_{\cC_V}(-D)) & \stackrel{M_3}{\longrightarrow} 
	& H^1(\cC_V,W|_{\cC_V}) & \to &
	H^1(D,W|_D) & \to &\\
& \to & H^2(\cC_V,W|_{\cC_V}(-D)) & \to & H^2(\cC_V,W|_{\cC_V}) & \to &
	H^2(D,W|_D) & \to &\\
& \to & H^3(\cC_V,W|_{\cC_V}(-D)) & \to & H^3(\cC_V,W|_{\cC_V}) & \to &
	H^3(D, W|_D) & \to & 0 \ ,
\ea
\eeq
where, in the third column, 
we have used the fact that for all $i \ge 0$,
\beq
H^i(\cC_V, W|_D) \simeq H^i(D, W|_D)
\eeq
because we are restricting $W$ to $D$.
Note that the cohomology group we are interested in,
$H^0(D, W|_D)$, occurs in this sequence. For emphasis, we
have boxed this term and indicated 
an explicit map, which we call
$M_3$ and which will be essential in our calculation.

In general, for an exact sequence
\beq
\ldots 
\to A_1 \stackrel{d_1}{\longrightarrow}
A_2 \stackrel{d_2}{\longrightarrow}
A_3 \stackrel{d_3}{\longrightarrow}
A_4 \stackrel{d_4}{\longrightarrow}
A_5 \longrightarrow
\ldots \ ,
\eeq
we have
\beq\label{seqdim}
\dim A_3 = \dim A_2 + \dim A_4 - \rk d_1 - \rk d_4 \ .
\eeq
Therefore, \eref{seq1} would give us
\beq\label{dim}
h^0(D, W|_D) = h^0(\cC_V,W|_{\cC_V}) +
h^1(\cC_V,W|_{\cC_V}(-D)) - h^0(\cC_V,W|_{\cC_V}(-D)) - \rk M_3 \ .
\eeq
We have used the fact that the rank of the map between
$H^0(\cC_V,W|_{\cC_V}(-D))$ and $ H^0(\cC_V,W|_{\cC_V})$ is simply equal to
$h^0(\cC_V,W|_{\cC_V}(-D))$ because this map, being the first in an exact
sequence, is injective.
We subsequently need to compute
the cohomologies of $W|_{\cC_V}(-D)$ and
$W|_{\cC_V}$ for which there are two more short exact sequences, both
on $X$. These are 
\beq
0 \rightarrow W(-\cC_V) \rightarrow W \rightarrow W|_{\cC_V}
\rightarrow 0 \ ,
\eeq
inducing the long exact sequence
\beq\label{seq2}
\ba{ccccccccc}
0 & \to & H^0(X,W(-\cC_V)) & \to & H^0(X,W) & \to &
	\fbox{$H^0(\cC_V,W|_{\cC_V})$} & \to &\\
& \to & H^1(X,W(-\cC_V)) & \stackrel{M_2}{\longrightarrow} 
	& H^1(X,W) & \to &
	H^1(\cC_V,W|_{\cC_V}) & \to &\\
& \to & H^2(X,W(-\cC_V)) & \to & H^2(X,W) & \to &
	H^2(\cC_V,W|_{\cC_V}) & \to &\\
& \to & H^3(X,W(-\cC_V)) & \to & H^3(X,W) & \to &
	H^3(\cC_V,W|_{\cC_V}) & \to & 0 \ ,
\ea
\eeq
as well as the sequence
\beq\label{W-C-D}
0 \rightarrow W(-\cC_V-D) \rightarrow W(-D) \rightarrow W|_{\cC_V}(-D)
\rightarrow 0 \ ,
\eeq
which induces the long exact sequence
\beq\label{seq3}
\ba{ccccccccc}
0 & \to & H^0(X,W(-\cC_V-D)) & \to & H^0(X,W(-D)) & \to &
	\fbox{$H^0(\cC_V,W|_{\cC_V}(-D))$} & \to &\\
& \to & H^1(X,W(-\cC_V-D)) & \stackrel{M_1}{\longrightarrow} &
H^1(X,W(-D)) & \to &
	\fbox{$H^1(\cC_V,W|_{\cC_V}(-D))$} & \to &\\
& \to & H^2(X,W(-\cC_V-D)) & \to & H^2(X,W(-D)) & \to &
	H^2(\cC_V,W|_{\cC_V}(-D)) & \to &\\
& \to & H^3(X,W(-\cC_V-D)) & \to & H^3(X,W(-D)) & \to &
	H^3(\cC_V,W|_{\cC_V}(-D)) & \to & 0 \ .
\ea
\eeq
Note that $D$, as defined in \eref{tCC} and given in \eref{D}, is a
curve in $\cC_V$. However, it is not hard to show from \eref{tCC} that
$D$ is the intersection of a divisor $\cC_V - \sigma -\sigma_2$ on $X$
with $\cC_V$. Somewhat abusing notation, we will also denote this
divisor by $D$. That is, let
\beq
D = \cC_V - \sigma -\sigma_2 \ .
\eeq
It is this divisor of $X$ that occurs in the terms $W(-\cC_V-D)$ and
$W(-D)$ of \eref{W-C-D} and \eref{seq3}, whereas $D$ in
$W|_{\cC_V}(-D)$ is the curve \eref{D}. Which $D$ we are referring to
will be clear by context.
As before, in the third column of \eref{seq2} and \eref{seq3}, 
we have used the fact that for all $i \ge 0$,
\beq
H^i(X,W|_{\cC_V}) \simeq H^i(\cC_V,W|_{\cC_V})
\eeq
and
\beq
H^i(X,W|_{\cC_V}(-D)) \simeq H^i(\cC_V,W|_{\cC_V}(-D))
\eeq
because of the restriction to $\cC_V$. Again,
we have boxed the requisite terms in \eref{seq2} and
\eref{seq3} that we need in \eref{dim}. 
We have also labeled two more maps, $M_1$ and
$M_2$, which will be required in our calculation.

Now, the dimensions of the cohomology groups in the 
first two columns of the
exact sequences \eref{seq2} and \eref{seq3} can be determined. We
show how this is done in Appendix C. Using the techniques therein,
we can fill in these dimensions as subscripts in the two sequences. 
We find that
\beq\label{seq2-2}
\ba{ccccccccc}
0 & \to & H^0(X,W(-\cC_V))_0 & \to & H^0(X,W)_{135} & \to &
	\fbox{$H^0(\cC_V,W|_{\cC_V})$} & \to &\\
& \to & H^1(X,W(-\cC_V))_{180} & \stackrel{M_2}{\longrightarrow} 
	& H^1(X,W)_{91} & \to &
	H^1(\cC_V,W|_{\cC_V}) & \to &\\
& \to & H^2(X,W(-\cC_V))_2 & \to & H^2(X,W)_0 & \to &
	H^2(\cC_V,W|_{\cC_V}) & \to &\\
& \to & H^3(X,W(-\cC_V))_0 & \to & H^3(X,W)_0 & \to &
	H^3(\cC_V,W|_{\cC_V}) & \to & 0
\ea
\eeq
and
\beq\label{seq3-2}
\ba{ccccccccc}
0 & \to & H^0(X,W(-\cC_V-D))_0 & \to & H^0(X,W(-D))_0 & \to &
	\fbox{$H^0(\cC_V,W|_{\cC_V}(-D))$} & \to &\\
& \to & H^1(X,W(-\cC_V-D))_{84} & \stackrel{M_1}{\longrightarrow} &
	H^1(X,W(-D))_{28}  & \to &
	\fbox{$H^1(\cC_V,W|_{\cC_V}(-D))$} & \to &\\
& \to & H^2(X,W(-\cC_V-D))_0 & \to & H^2(X,W(-D))_0 & \to &
	H^2(\cC_V,W|_{\cC_V}(-D)) & \to &\\
& \to & H^3(X,W(-\cC_V-D))_{26} & \to & H^3(X,W(-D))_0 & \to &
	H^3(\cC_V,W|_{\cC_V}(-D)) & \to & 0 \ .
\ea
\eeq
Applying \eref{seqdim} to \eref{seq2-2}, we have
\beq\label{h0WcV}
h^0(\cC_V,W|_{\cC_V}) = 135 +180 - \rk\left(M_2\right)  
= 315 - \rk(M_2) \ .
\eeq
Note that the matrix $M_2$ has dimensions 
\beq\label{dimM2}
(M_2)_{91\times 180} \ .
\eeq
Similarly, we can use \eref{seqdim} for \eref{seq3-2} to obtain
\beq\label{h0WcVD-A}
h^0(\cC_V,W|_{\cC_V}(-D)) 
= 84 - \rk(M_1) \ ,
\eeq
and
\beq\label{h0WcVD-B}
h^1(\cC_V,W|_{\cC_V}(-D)) 
= 28 -  \rk(M_1) \ .
\eeq
The matrix $M_1$ has the dimensions
\beq\label{dimM1}
(M_1)_{28 \times 84} \ .
\eeq
Substituting \eref{h0WcV}, \eref{h0WcVD-A} and \eref{h0WcVD-B}
into \eref{dim} gives
\beq\label{259M2M3}
h^0(D,W|_D) = 259 - \rk(M_2) - \rk(M_3) \ .
\eeq

Let us study some limiting cases. It follows from \eref{dimM1} that
$M_1$ has maximal rank 28 while its minimal
rank is 0. Let us first assume that
\beq
\rk(M_1) = 28.
\eeq
Then, it follows from \eref{h0WcVD-B} that
\beq
H^1(\cC_V,W|_{\cC_V}(-D)) = 0 
\eeq
and, hence, from the sequence \eref{seq1} that $M_3$ is the
zero map. That is,
\beq
\rk(M_3) = 0 \ .
\eeq
Also, we see from \eref{dimM2} that the rank of $M_2$
is in the range
\beq\label{rkM2}
\rk(M_2) \in [0, 91] \ .
\eeq
Throughout the remainder of this paper, we will use the symbol $[m,n]$
to indicate the range of integers from $m$ to $n$. It does not imply
that the quantity in question must assume all values in this range.
Therefore expression \eref{259M2M3} becomes a bound on the dimension
of $h^0(D,W|_D)$ given by
\beq\label{M1-max}
h^0(D,W|_D) = 259 - \rk(M_2) \in [168, 259] \ .
\eeq

On the other hand, let us assume that $M_1$ has its minimal rank. That is,
\beq
\rk(M_1) = 0.
\eeq
In this case, it follows from \eref{h0WcVD-A} and \eref{h0WcVD-B} that
\beq
h^0(\cC_V,W|_{\cC_V}(-D)) = 84, \quad 
h^1(\cC_V,W|_{\cC_V}(-D)) = 28.
\eeq
Since $H^1(\cC_V,W|_{\cC_V}(-D))$ is not trivial, the mapping
$M_3$ in sequence \eref{seq1} is no longer the zero map. From
\eref{rkM2}, 
we know that $\rk(M_2) \in [0, 91]$. What are the possible
ranks for $M_3$? First, assume $M_2$ has its maximal rank, that is,
\beq
\rk(M_2) = 91.
\eeq
By inspecting \eref{seq2-2}, we see that, in this case,
\beq
h^0(\cC_V,W|_{\cC_V}) = 224, \ h^1(\cC_V,W|_{\cC_V}) = 2.
\eeq
It then follows from \eref{seq1} that the rank of $M_3$ is in the
range
\beq
\rk(M_3) \in [0,2]  \ .
\eeq
On the other hand, if we assume
\beq
\rk(M_2)=0 \ , 
\eeq
then \eref{seq2-2} implies that
\beq
h^0(\cC_V,W|_{\cC_V}) = 315, \ h^1(\cC_V,W|_{\cC_V}) = 93 \ . 
\eeq
Then, from \eref{seq1} and \eref{h0WcVD-B} we find
\beq
\rk(M_3) \in [0,28]. 
\eeq
Putting this information together, we can conclude the following. When
$M_1$ has its minimal rank, that is $\rk(M_1) = 0$, then
\beq
\rk(M_2) = 91 \Rightarrow
\rk(M_3) \in [0,2], \ \  h^0(D,W|_D) = 259 - \rk(M_2) -
	\rk(M_3) \in [166,168]
\eeq
and
\beq
\rk(M_2) = 0  \Rightarrow
\rk(M_3) \in [0,28],  \ \ h^0(D,W|_D) = 259 - \rk(M_2) -
	\rk(M_3) \in [231,259].
\eeq
Having discussed these limiting ranges, we now need to explicitly
compute the ranks of the maps $M_{i=1,2,3}$ to finish our
calculation. This is a
rather technical exercise and we leave the exposition of the method to
Appendix D.
We see from our discussion in the Appendix that the maps $M_i$ depend
on various complex parameters. These are the moduli associated with the
vector bundle $V$. As we will show in the next section, there are
$223$ such moduli and, hence, the calculation of the ranks is
rather complicated.

It will be very helpful if one can choose $M_1$ to have 
its maximal rank of 28. 
Then, by the above discussion, $M_3$ is the
zero map and the requisite calculations are greatly
simplified. 
To show that $M_1$ can, in fact, have maximal rank,
at least for some subclass of vector bundle moduli,
we explicitly compute the matrix $M_1$ given in \eref{M1},
setting all moduli appearing in it to zero except those in the
$m_{(3)0}$, $m_{(2)0}$ and $m_{(1)0}$ sub-blocks. The sub-blocks
$m_{(R)i}$ of $M_1$ are defined in \eref{blockM}.
If, in addition, we
identify the moduli in these sub-blocks with a single modulus $\phi$,
the matrix $M_1$ can be explicitly written as
\beq\label{M1-set}
M_1 = \phi
\left(
\ba{ll}
\left[
\matrix{
\II_{1,1} & 0 & 0 & 0 & 0 & 0 & 0 & 0 & 0 \cr 0 & \II_{2,2} & 0 & 0 & 0 & 0 & 0 & 0 & 0 \cr 0 & 0 & \II_{3,3} & 0 & \II_{3,
   3} & 0 & 0 & 0 & 0 \cr 0 & 0 & 0 & \II_{4,4} & 0 & \II_{4,4} & 0 & 0 & 0 \cr 0 & 0 & 0 & 0 & 0 & 0 & \II_{5,
   5} & 0 & 0 \cr 0 & 0 & 0 & 0 & 0 & 0 & 0 & \II_{6,6} & 0 \cr 0 & 0 & 0 & 0 & 0 & 0 & 0 & 0 & \II_{7,7} \cr
}
\right]_{28 \times 35}
&
\left[\mbox{{\Huge 0}}\right]_{28 \times 49}
\ea\right) \ ,
\eeq
where $\II_{n,n}$ is the identity matrix of size $n$.
For $\phi \ne 0$, the rank of this matrix is clearly 28, as
required. Now turn on all other moduli in $M_1$. In general, it is not
hard to show that, for generic values of these moduli, the rank of
$M_1$ remains 28, decreasing only on specific loci of co-dimension
one, or greater, in the moduli space. That is,
\beq
\rk(M_1) = 28 \quad \mbox{generically.}
\eeq
Therefore, not only can we choose $M_1$ to have rank 28 but it is the
generic value.
Then, as discussed above, the matrix $M_3$ is the zero matrix
and we have
\beq\label{h0WD-res}
h^0(D,W|_D) = 259 - \rk(M_2) \ .
\eeq

In Appendix D, we explicitly construct $M_2$ and
find that it is a $91 \times 180$ matrix depending on 139 complex
moduli
\beq
\left\{
\phi^{[(4)i]}_p, \ \phi^{[(6)i]}_q
\right\}, \qquad
i=1,\ldots,9, \ p=1,\ldots,i+1, \quad  j=3,\ldots,12, \ q=1,\ldots,j+1 \ .
\eeq
The quantity $\rk(M_2)$ is discussed in Appendix E. It is found
to be extremely
sensitive to the choice of these parameters. \fref{f:M2jump},
the table in
\eref{tabrk} and \fref{f:moduli} show us that, depending on the
values one chooses for these moduli,
\beq\label{rkM2-1}
\rk(M_2) \in [28, 85]
\eeq
and is, in fact, expected to attain all integer values between these
bounds. The generic value is 85.
It follows from this and \eref{h0WD-res} that
\beq\label{h0D3}
h^0(D,W|_D) = [174, 231] \ ,
\eeq
where 174 is the generic value.

We have now computed the last term $h^0(D, W|_D)$ 
of the four requisite terms in \eref{F14terms}.
Substituting \eref{h0D1-1}, \eref{h0D1-2}, \eref{h0D2}
and \eref{h0WD-res} into \eref{F14terms}, we obtain
\beq\label{F13-terms2}
h^1(X,\av^*) = 122 - \rk(M_2) \ .
\eeq
Therefore, substituting \eref{rkM2-1} into \eref{F13-terms2}, 
we find at last that
\beq\label{h1aveg}
h^1(X,\av^*) = [37,94] \ ,
\eeq
where
\beq\label{h1aveg-gen}
h^1(X,\av^*) = 37 \mbox{~generically} \ .
\eeq
Let us comment on what the result \eref{h1aveg} means. One can make a
plethora of choices when computing the ranks of the explicit maps
$M_1$, $M_2$ and $M_3$. These correspond to the choice of the
moduli on which each of these matrices depend.
The final answer for $h^1(X,\av^*)$ will also depend, rather
dramatically, on the values of these parameters. 
The generic result of 37 in \eref{h1aveg-gen} occurs for
$\rk(M_1)=28$ and $\rk(M_2)=85$, which are their respective generic
ranks. This means that as we move in the moduli space of the
vector bundle $V$,
we generically expect $h^1(X,\av^*)$ to be 37. However,
as we hit special loci of co-dimension one or higher, as will be shown
in \fref{f:specjump}, the value of $h^1(X,\av^*)$ 
can jump to higher integers lying in the range \eref{h1aveg}.
We conclude that the 
particle spectrum of the low-energy effective theory in
heterotic compactifications depends crucially
on the choice of the moduli of the vector
bundle and can change as the values of the moduli change.

We have gone to great lengths to determine $h^1(X,\av^*)$ and
to elucidate the fact that
its value depends on the vector bundle
moduli.
To obtain $h^1(X,\av)$, we simply use the index relation \eref{v-vd-2}
\beq
h^1(X,\av)  = h^1(X,\av^*) - 3 \ .
\eeq
Given a value for $h^1(X,\av^*)$, this relation uniquely fixes
$h^1(X,\av)$. It then follows from \eref{h1aveg} and \eref{h1aveg-gen}
that
\beq
h^1(X,\av) = [34,91] \ ,
\eeq
where
\beq
h^1(X,\av) = 34 \mbox{~generically} \ .
\eeq
It is important to note that even though the values of
$h^1(X,\av)$ and $h^1(X,\av^*)$ depend
dramatically upon the choice of moduli, their difference is
constrained by the index theorem to be 3.

%
%
\subsection{Calculation of $h^1(X,\vv)$}\label{s:vv}
As discussed earlier, we will compute the term 
$h^1(X, \vv)$ using a different
technique.
We know that $h^1(X,\vv)$ are the number of
moduli associated with the
vector bundle $V$. In Section 4 of \cite{Evgeny}, it was shown 
that this is equal to
\beq\label{h1vv}
h^1(X, \vv) = (h^0(X, \cO_X(\cC_V)) - 1) + h^1(\cC_V, \cO_{\cC_V}).
\eeq
Furthermore, $h^0(X, \cO_X(\cC_V))$ was computed and $h^1(\cC_V,
\cO_{\cC_V})$ was shown to vanish in the case when
the spectral cover $\cC_V$ is positive. 
The conditions for positivity of a spectral cover
\beq
\cC_V \in |n\sigma + \pi^*(a S + b \cE)|
\eeq
over a base surface $B = \IF_r$ were shown in \cite{Evgeny} to be
\beq\label{ample}
b > a\ r - n(r-2), \quad a>2n.
\eeq
Under these circumstances, it was found that 
\beq\label{resEvgeny}
h^1(X, \vv) = \frac{n}{3}(4n^2-1) + nab - (n^2-2)(a+b) +
ar(\frac{n^2}{2}-1) - \frac{n}{2} r a^2 - a.
\eeq

Can we  use this expression to
compute $h^1(X,\vv)$ in the explicit example \eref{eg} being
considered? Recall that, in this case,
\beq
\cC_V \in |5\sigma + \pi^*(12 S + 15 \cE)|
\eeq
is the spectral cover over $B = \IF_1$. Putting the data
$r=1$, $n=5$ and $(a,b)=(12,15)$ into \eref{ample}, we see that the
requisite inequalities are violated. That is, in our specific example
$\cC_V$ is not a positive divisor and, hence, we can not use
\eref{resEvgeny} to compute $h^1(X,\vv)$. Unfortunately, for the $SU(5)$ GUT
theories classified in the paper, this will often be the case. We,
therefore, must use a different technique to compute $h^0(X,
\cO_X(\cC_V))$ and $h^1(\cC_V, \cO_{\cC_V})$.

\subsubsection{Moduli for the Spectral Cover}
First, let us compute the term $h^0(X, \cO_X(\cC_V))$. These are the
moduli associated with the spectral cover $\cC_V$.
To do this, first use \eref{H-push} and \eref{pi-nsig} in Appendix
C to push $H^0(X, \cO_X(5 \sigma + \pi^*(12 S + 15 \cE)))$  
onto the base $\IF_1$. We obtain
\bea
H^0(X, \cO_X(5 \sigma + \pi^*(12 S + 15 \cE))) &=& 
H^0(\IF_1, \pi_* \cO_X(5 \sigma) \otimes \cO_{\IF_1}(12 S + 15 \cE))\\
&=& H^0(\IF_1, (\cO_{\IF_1} \oplus \bigoplus_{i=2}^5
	\cO_{\IF_1}(-i c_1(T\IF_1))) 
	\otimes \cO_{\IF_1}(12 S + 15 \cE)) \nn
\eea
Since $\IF_1$ is itself a $\IP^1$-fibration
over $\IP^1$, one can use \eref{H-push2} and \eref{beta-push} to push
this result down further to $\IP^1$. The answer is
\bea\label{h0cV}
&&H^0(X, \cO_X(5 \sigma + \pi^*(12 S + 15 \cE))) \nn \\
&=&
\bigoplus_{i=3}^{15} H^0(\IP^1, \cO_{\IP^1}(i)) \oplus
\bigoplus_{i=1}^{9} H^0(\IP^1, \cO_{\IP^1}(i)) \oplus
\bigoplus_{i=0}^{6} H^0(\IP^1, \cO_{\IP^1}(i)) \nn \\
&& \quad
\oplus \bigoplus_{i=-1}^{3} H^0(\IP^1, \cO_{\IP^1}(i)) \oplus
\bigoplus_{i=-2}^{0} H^0(\IP^1, \cO_{\IP^1}(i)).
\eea
Using \eref{degP1}, expression \eref{h0cV} gives
\beq\label{cover-mod}
h^0(X, \cO_X(5 \sigma + \pi^*(12 S + 15 \cE))) = 223.
\eeq

\subsubsection{Moduli for the Spectral Line Bundle}
Next, we compute the term $h^1(\cC_V, \cO_{\cC_V})$. These are the
moduli associated with the spectral line bundle, that is, the
continuous moduli of the Picard group $H^1(\cC_V, \cO_{\cC_V}^*)$
of line bundles on $\cC_V$.
To show this, consider the exact sequence
\beq
\rightarrow H^1(\cC_V, \IZ)
\stackrel{}{\rightarrow} 
H^1(\cC_V, \cO_{\cC_V}) \rightarrow
H^1(\cC_V, \cO_{\cC_V}^*) \rightarrow
H^2(\cC_V, \IZ) \stackrel{}{\rightarrow} \ .
\eeq
Now note that 
$H^1(\cC_V, \IZ)$ and  $H^2(\cC_V, \IZ)$ are rigid lattices.
Therefore,
the only continuous moduli of $H^1(\cC_V,
\cO_{\cC_V}^*)$ come from $H^1(\cC_V, \cO_{\cC_V})$. Hence, we
need  only to compute $h^1(\cC_V, \cO_{\cC_V})$.
To do this, we use the
short exact sequence
\beq
0 \to \cO_X(-\cC_V) \to \cO_X \to \cO_{\cC_V} \to 0 \ ,
\eeq
which implies the long exact sequence
\beq\label{bundle-mod2-1}
\to H^1(X,\cO_X) \to H^1(\cC_V, \cO_{\cC_V}) \to H^2(X, \cO_X(-\cC_V))
\to H^2(X, \cO_X) \to \ .
\eeq
Now,
\beq
H^1(X,\cO_X) = H^{0,1}_{\overline{\partial}}(X, \IC), \quad
H^2(X,\cO_X) = H^{0,2}_{\overline{\partial}}(X, \IC)
\eeq
for the Dolbeault cohomolgy groups $H^{0,1}_{\overline{\partial}}(X, \IC)$ and
$H^{0,1}_{\overline{\partial}}(X, \IC)$, both of which vanish for a
Calabi-Yau threefold $X$. Therefore, \eref{bundle-mod2-1} implies that
\beq\label{bundle-mod2}
H^1(\cC_V, \cO_{\cC_V}) \simeq H^2(X, \cO_X(-\cC_V)) \ .
\eeq
We can simplify this expression further by using Serre duality
\eref{Serre}, which dictates that
\beq\label{bundle-mod3}
H^2(X, \cO_X(-\cC_V)) \simeq H^1(X, \cO_X(\cC_V))
\eeq
on a Calabi-Yau threefold $X$.
In summary, \eref{bundle-mod2}
and \eref{bundle-mod3} together imply that the number of moduli
associated with the spectral line bundle is
\beq\label{bundle-mod4}
h^1(\cC_V, \cO_{\cC_V}) =  h^1(X, \cO_X(\cC_V)) \ .
\eeq
Recalling from \eref{cVeg}
that $\cC_V \in | 5 \sigma + \pi^*(12 S + 15 \cE)|$, 
we see that the
computation of \eref{bundle-mod4} can be carried out using the
techniques presented 
in Appendix C, in complete
analogy with the above calculation for the moduli associated with the
spectral cover. We find, after pushing everything onto the base
$\IP^1$, 
that we have
\beq\label{bun-mod}
h^1(\cC_V, \cO_{\cC_V}) = 1.
\eeq

Substituting \eref{cover-mod} and \eref{bun-mod} into \eref{h1vv}, 
we finally obtain
\beq
h^1(X, \vv) = 223.
\eeq
That is, in our explicit example, there are 223 vector bundle moduli.

\subsubsection{Checking Against the Case of Positive Spectral Cover}
As a check on our method, let us derive an expression for
$h^1(X, \vv)$ in the case of $B=\IF_1$ and 
$\eta = a S + b \cE$ with $a,b \in \IZ_{\ge 0}$ where 
$n,a,b$ satisfy \eref{ample}, that is, when $\cC_V$ is a positive 
spectral cover.
In the case of positive spectral cover, 
it was shown in \cite{Evgeny} that
$h^1(\cC_V, \cO_{\cC_V})$ vanishes. Therefore, we need only
compute $h^0(X, \cO_X(\cC_V))$. Then,
\beq\label{h1vv-pos}
h^1(X, \vv) = h^0(X, \cO_X(\cC_V) - 1 \ .
\eeq

First, recalling that $n$ is always positive, we have
\bea\label{vvF1-1}
h^0(X, \cO_X(n \sigma + \pi^* \eta)) &=&
h^0(B, (\cO_B \oplus \bigoplus_{i=2}^n 
\cO_B(-i c_1(T\IF_1))) \otimes \cO_B(\eta)) \nn \\
&=& h^0(B, \cO_B(aS+b\cE)) + \sum_{i=2}^n h^0\left(B, 
\cO_B((a-2i) S + (b-3i) \cE) \right), \nn \\
\eea
where we have used the expression for $c_1(T\IF_1)$ in \eref{chernFr}.
Now, \eref{ample} clearly requires that $a>0$, so the first term in
\eref{vvF1-1} becomes
\beq\label{vvF1-2}
h^0(B, \cO_B(aS+b\cE)) = h^0(\IP^1, (
\cO_{\IP^1} \oplus \bigoplus_{i=1}^a \cO_{\IP^1}(-i)
) \otimes \cO_{\IP^1}(b)) \ ,
\eeq
where we have used \eref{beta-push} to push down onto the
base $\IP^1$ of $\IF_1$.
Similarly, the second term in \eref{vvF1-1} becomes
\beq\label{vvF1-3}
\sum_{i=2}^n (
h^0(\IP^1, \cO_{\IP^1}(b-3i)) + \sum_{j=1}^{a-2i} 
h^0(\IP^1, \cO_{\IP^1}((b-3i)-j))
) \ .
\eeq
Note that the bound $a-2i$ in the second sum is always positive since
the positivity conditions for $\cC_V$ in \eref{ample} require that $a
> 2n$.
Furthermore, the degrees $(b-3i)-j$ in the
second term of \eref{vvF1-3} are always positive for $j \le a-2i$
because \eref{ample} requires that $b > a + n$.
Substituting \eref{vvF1-2} and \eref{vvF1-3} into
\eref{vvF1-1} and using \eref{degP1}, we have at last
\bea
h^0(X,\cO_X(\cC_V)) &=& (b+1) + \sum_{i=1}^a(b-i+1) + \sum_{i=2}^n (
(b-3i+1) + \sum_{j=1}^{a-2i} (b-3i-j+1) ) \nn \\
&=& a + 2\,b - \frac{n}{3} - \frac{a^2\,n}{2} + a\,b\,n - 
  \frac{a\,n^2}{2} - b\,n^2 + \frac{4\,n^3}{3},
\eea
which implies, using \eref{h1vv-pos}, that
\beq
h^1(X, \vv) = 
\frac{n}{3}(4n^2-1) + nab - (n^2-2)(a+b) + a (\frac{n^2}{2}-1)
- \frac{n}{2}a^2 - 1 \ .
\eeq
This result agrees completely with \eref{resEvgeny} and
Eq.4.39 of \cite{Evgeny} which were computed by other methods for
positive spectral cover. By this we are thus much assured.

\subsection{Summary of the Particle Content}
It is useful to summarize here
the results of our calculation.
We have compactified heterotic M-theory on an elliptic Calabi-Yau threefold
whose base surface is $\IF_1$ and on which there is a stable
holomorphic vector bundle $V$ 
with a structure group $G=SU(5)$
and spectral data
\beq
\cC_V \in |5\sigma + \pi^*(12 S + 15 \cE)|
\eeq
and
\beq
c_1(\cN_V) = 5 \sigma + \pi^*(3 c_1(T\IF_1)) \ .
\eeq
This compactification satisfies the three physical constraints
discussed in Section \ref{s:cons}, that is, it is anomaly free, has
three families of quarks and leptons and admits a gauge connection
satisfying the hermitian Yang-Mills equation. The low energy GUT group
is $H=SU(5)$. To distinguish the structure group of $V$ from the GUT
group, we will denote them by $SU(5)_G$ and $SU(5)_H$
respectively. The 5, $\overline{5}$, 10, $\overline{10}$ and 24
representations of $SU(5)_G$  are associated with the bundles
$V$, $V^*$, $\av$, $\av^*$ and $\vv$ respectively. The dimensions of
the relevant cohomologies of these bundles were computed in the
previous two sections and found to be
\beq
\ba{|c|c|c|}\hline
SU(5)_G & \mbox{cohomology} & \mbox{spectrum} \\ \hline
5	& h^1(X, V)	& 0 \\ \hline
\overline{5}& h^1(X, V^*)	& 3 \\ \hline
10	& h^1(X, \av)	& [34,91], \mbox{~generically~}34
\\ \hline
\overline{10}	& h^1(X, \av^*)	& [37,94], \mbox{~generically~}37
\\ \hline
24	& h^1(X, \vv)	& 223 \\ \hline
\ea\eeq
The low energy theory is a four-dimensional $N=1$ supersymmetric
GUT theory with gauge group
$SU(5)_H$. The expressions for the spectrum of chiral superfields
transforming as the $\ol{10}$, $10$, $5$, $\ol{5}$, and 1
of $SU(5)_H$ were discussed in Section \ref{s:spec-comp} 
and given in \eref{VVdaV-1} and
\eref{VVdaV-2}. Combining these expressions, we have the following.
Trivially, 
there is one vector supermultiplet transforming as the adjoint 78
representation of $SU(5)_H$. That is,
\beq
n_{78} = 1 \ .
\eeq
The number of chiral supermultiplets in the $\ol{10}$ and $10$
representations of $SU(5)_H$ are
\beq
n_{\ol{10}} = 0, \quad n_{10} = 3
\eeq
respectively. The number of chiral supermultiplets in the $\ol{5}$
representation of $SU(5)_H$ depends on the values of the vector bundle
moduli. Generically, we find that
\beq
n_{\ol{5}} = 37.
\eeq
However, on loci of co-dimension one or higher in the moduli space
this value can abruptly jump, spanning the range
\beq\label{n5-range}
n_{\ol{5}} \in [37, 94] \ .
\eeq
We expect that each integer value in this range is realized on some
subset of moduli space. 
As a graphic example of this phenomenon, we show in \fref{f:specjump}
a nine-dimensional region of vector bundle moduli space discussed in
Appendix E. Note that for a generic point in this space,
$n_{\ol{5}}=37$. 
However, on various sub-planes of co-dimension one or higher
$n_{\ol{5}}$ jumps, taking the values $n_{\ol{5}}=37,40,43$ and $52$. 
These numbers
are obtained using \eref{F13-terms2} and the results in table 
\eref{tabrk}.

\begin{figure}
\centerline{\input{n5spec.pstex_t}}
\caption{A subspace of the moduli space $\cM$ of $\phi$'s spanned
by $\phi^{[(4)1]}_{p=1,2}$, $\phi^{[(4)2]}_{q=1,2,3}$ and 
$\phi^{[(4)3]}_{r=1,2,3,4}$. Generically, in the bulk, $n_{\ol{5}} =
37$,
its minimal value. As we restrict to various planes and
intersections thereof, we are confining ourselves to special
sub-spaces of co-dimension one or higher. In these subspaces, the
value of $n_{\ol{5}}$ can increase dramatically.
\label{f:specjump}}
\end{figure}

The index theorem tells us that the number of
chiral supermultiplets transforming in the $5$ representation of
$SU(5)_H$ is given by
\beq\label{n5bar}
n_{5} = n_{\ol{5}} - 3.
\eeq
Therefore, generically
\beq
n_{5} = 34.
\eeq
However, it follow from \eref{n5-range} and \eref{n5bar} that this
number can jump, spanning the range
\beq
n_{5} \in [34, 91] \ .
\eeq
Note, however, that the index theorem guarantees that at every point
in moduli space
\beq
n_{\ol{5}} - n_{5} = 3 \ .
\eeq
Finally, the number of chiral superfields transforming as singlets
under $SU(5)_H$, that is, the number of vector bundle moduli, is given
by
\beq
n_1 = 223 \ .
\eeq

We have succeeded, therefore, in computing the exact particle spectrum
of our $SU(5)$ GUT theory. Rather remarkably, we find that, although
the difference $n_{\ol{5}} - n_{5}$ is fixed by the three family
condition to be 3, the individual values of $n_5$ and
$n_{\overline{5}}$ depend on the location in vector bundle moduli
space at which they are evaluated.

\section{Conclusions}

We have shown, for general heterotic vacua, that
the calculation of the particle spectrum consists of computing
the sheaf cohomology of five 
vector bundles:  $V,V^*,\wedge^2 V,\wedge^2 V^*,V \otimes V^*$. Among
these, the cohomology group $H^1(X,V \otimes V^*)$ has a topological
interpretation as the number of deformation moduli of the vector bundle
$V$ and, therefore, is always deformation invariant. Hence,
it cannot jump when the bundle moduli are varied continuously. 
However, no
such topological
interpretation is available for the cohomologies of the remaining
bundles. In fact, we have shown that, as $V$ varies continuously, the
cohomologies of $\wedge^2 V$ and $\wedge^2 V^*$ do jump in the particular
theory under consideration in this paper. 

The novelty of these results is seen when contrasted with the standard
embedding. In this latter case, there are two distinct ways to deform
the vector bundle. The first is to deform the Calabi-Yau threefold $X$
while keeping $V=TX$. In this case, jumps in the spectrum
can never occur since all the cohomologies in
question do have a topological interpretation. By definition, in the
standard embedding we have
$V=\wedge^2 V^*=TX=\Omega^2_X$ and $V^*=\wedge^2 V=T^*X=\Omega^1_X$.
Hence,
the cohomologies are all of the form $H^i(\Omega^j_X)$,
$j=1,2$ and $i=0,\ldots,3$ where $\Omega^j_X$ is the sheaf of
holomorphic $j$-forms on $X$.
On our Calabi-Yau threefold $X$ these are topological invariants since
their Hodge numbers can be related to the Betti numbers as 
\beq
h^{1,1}=b^2, \qquad 2h^{2,1}+2=b^3.
\eeq
This
explains why jumps in the spectrum 
as the moduli of the Calabi-Yau threefold are varied 
were never observed in heterotic
compactifications based on the standard embedding. The second way to
deform the vector bundle is to start with the standard embedding
$V=TX$ on a fixed Calabi-Yau threefold and to deform its tangent
bundle, as in \cite{PR}. In this case, it is not excluded, and is in
fact quite likely, that jumps in the particle 
spectrum will occur.

\vspace{2cm}

\section*{Acknowlegements}
We are grateful to V.~Braun, E.~Buchbinder and 
T.~Pantev for many insightful comments and conversations.
This Research was supported in part by
the Dept.~of Physics and the Maths/Physics Research Group
at the University of Pennsylvania
under cooperative research agreement \#DE-FG02-95ER40893
with the U.~S.~Department of Energy and an NSF Focused Research Grant
DMS0139799 for ``The Geometry of Superstrings.'' R.~D.~further
acknowledges an NSF grant DMS 0104354.
R.~R.~would like to thank Dmitriy Boyarchenko for interesting 
discussions and is also supported
by the Department of Physics and Astronomy of Rutgers University under
grant DOE-DE-FG02-96ER40959.

%
%
%


\newpage

\noindent
{\bf {\huge Appendices}}

\appendix

\section{Chern and Todd Classes}
For convenience, we remind the reader of the
expansion of the Chern and Todd classes for a vector bundle $U$ on a
complex manifold $X$.
\bea\label{ChTd}
\td(U) &=& 1 + \td_1(U) + \td_2(U) + \td_3(U) + \ldots \nn \\
\ch(U) &=& \ch_0(U) + \ch_1(U) + \ch_2(U) + \ch_3(U) + \ldots 
\eea
with
\beq
\label{Todd}
\td_1(U) = \frac12 c_1(U), \quad
\td_2(U) = \frac{1}{12} (c_2(U)+c_1(U)^2), \quad
\td_3(U) = \frac{1}{24} (c_1(U)c_2(U)) \ ,
\ldots
\eeq
and
\bea
\label{Chern}
\ch_0(U) &=& \mbox{rk}(U), \quad
\ch_1(U) = c_1(U), \quad
\ch_2(U) = \frac12(c_1(U)^2-2c_2(U)), \nn \\
\ch_3(U) &=& \frac16(c_1(U)^3 - 3 c_1(U)c_2(U) + 3 c_3(U)) \ . \ldots
\eea
We will make frequent use of these formulas.

%
\section{Chern Classes of Antisymmetric Products}
In this Appendix, we obtain the expressions for the Chern classes of
the antisymmetric product $\av$ of a rank $n$ vector bundle $V$ on a
threefold $X$.
Using the splitting principle,
let us first decompose $V$ into $n$ line bundles $L_i$ as
\beq
V = \bigoplus_{i=1}^n L_i \ .
\eeq
Then, we have that
\beq
\label{expVL}
\ch(V) = \sum_{i=1}^n e^{x_i} \,
\eeq
where
\beq
x_i = c_1(L_i) \ .
\eeq
Since $X$ is a threefold, \eref{expVL} can be expanded as
\bea
\label{chVx}
\ch(V) &=& \sum_{i=1}^n 1 + x_i + \frac12 x_i^2 + \frac16x_i^3  \ .
\eea
Therefore, we can read off from \eref{chVx} that
\beq
\ch_0(V) = \rk(V) = n, \quad
\ch_1(V) = \sum_{i=1}^n x_i, \quad
\ch_2(V) = \frac12 \sum_{i=1}^n x_i^2, \quad
\ch_3(V) = \frac16 \sum_{i=1}^n x_i^3 \ .
\eeq
The result we are after is
\beq
\ch(\av) = \sum_{i<j}^n e^{x_i}e^{x_j} \ ,
\eeq
which, using \eref{chVx}, becomes
\bea
\label{chernaV-1}
\ch(\av) &=& \sum_{i<j}^n 1 + (x_i + x_j) + \frac12 (x_i +x_j)^2
	+ \frac16(x_i + x_j)^3 \nn \\
&=& \frac{n(n-1)}{2} + (n-1)\sum_{i=1}^n x_i\frac{}{}
+
\frac14
\left(
2n\sum_{i=1}^n x_i^2 + \frac{}{} 2\sum_{i=1}^n x_i \sum_{j=1}^n x_j 
- 4 \sum_{i=1}^n x_i^2
\right)
+ \nn \\
&&\qquad \frac1{12}
\left(
2n\sum_{i=1}^n x_i^3 +
6\sum_{i=1}^n x_i^2 \sum_{j=1}^n x_j - 8\sum_{i=1}^n x_i^3 
\right)\nn \\
&=&
\left[\frac{n(n-1)}{2}\right]
+
\left[(n-1)\frac{}{}\ch_1(V)\right]
+
\left[(n-2)\ch_2(V) + \frac12 \ch_1(V)^2 \right]
+ \nn \\
&&\quad\left[
(n-4)\ch_3(V) +\frac{}{} \ch_1(V)\ch_2(V)
\right] \ .
\eea
Using the relations given in \eref{ChTd} and \eref{Chern}, we conclude
	that the Chern classes of $\av$ are
\bea
\label{chernaV}
\rk(\av) &=& \frac{n(n-1)}{2}, \nn \\
c_1(\av) &=& (n-1) c_1(V), \nn \\
c_2(\av) &=& \frac{(n-1)(n-2)}{2} c_1(V)^2 + (n-2) c_2(V), \nn \\
c_3(\av) &=& \frac{(n-1)(n-2)(n-3)}{6} c_1(V)^3 + (n-2)^2 c_1(V)
c_2(V) + (n-4)c_3(V).
\eea
%
%
\section{Determining $H^i(X, \cO_X(n \sigma) \otimes \pi^* L)$}
In this Appendix, we determine, using the Leray spectral sequence, the
cohomology groups $H^i(X,T)$ 
for line bundles of the form $T = \cO_X(n \sigma) \otimes \pi^* L$
where $L$ is some line bundle on the base $B$. In particular, we will be
interested in the dimensions of these groups and the
explicit maps between
them. We will use the Leray spectral sequence to reduce the
calculation of the cohomology on $X$ to that on the base $B$. Then, we
specialize to the case when $B=\IF_1$, which is itself a $\IP^1$
fibration over $\IP^1$. In this case, we use the Leray spectral
sequence again to reduce the cohomology on $B$ to that on $\IP^1$
with which we are familiar. In all, for $B=\IF_1$, we can compute the
cohomology groups $H^i(X, \cO_X(n \sigma) \otimes \pi^* L)$ in general
by reducing them to direct sums of cohomologies over $\IP^1$.

For $\pi: X \rightarrow B$, with $B$ being a surface and $T$ a line
bundle, the Leray spectral sequence becomes the long exact sequence
\beq\label{leray2}
\ba{ccccccccc}
0 & \to & H^1(B,\pi_*T) & \to & H^1(X,T) & \to &
	H^0(B,R^1\pi_*T) & \to &\\
& \to & H^2(B,\pi_*T) & \to & H^2(X,T) & \to &
	H^1(B,R^1\pi_*T) & \to &\\
& \to & H^3(B,\pi_*T) & \to & H^3(X,T) & \to &
	H^2(B,R^1\pi_*T) & \to & 0. \\
\ea
\eeq
For $T$ of the form $T=\cO_X(n \sigma) \otimes \pi^* L$, 
\eref{leray2} gives us
\beq\label{H-push}
\ba{l|l|l}
n>0 & R^1\pi_* T = 0 & \left\{\ba{l}
	H^i(X, T) = H^i(B, \pi_* \cO_X(n \sigma) \otimes L), \ 
	i=0,1,2 \\ 
	H^3(X, T) = 0
	\ea\right. \\ \hline
n<0 & \pi_*T=0 & \left\{
	\ba{rcl}
	H^0(X,T) &=& 0 \\
	H^i(X,T) &=& H^{i-1}(B, R^1\pi_*T)\\ 
	&=&  H^{i-1}(B, R^1\pi_*(\cO_X(n\sigma))\otimes L ), \ i=1,2,3 
	\ea\right. \\ \hline
n=0 & 
\ba{l}
\pi_*T = L,\\ 
R^1\pi_* T = K_B \otimes L 
\ea
& 
\left\{\ba{l}
	H^0(X, \pi^*L) = H^0(B, L) \\
	H^1(X, \pi^*L) =
	H^1(B,L) \oplus H^0(B, K_B \otimes L) \mbox{~~if~~} 
	H^2(B,L) = 0
\ea \right.
\ea
\eeq
where, for $n > 0$ and $n < 0$, we have used the fact that
\bea\label{pi-nsig}
\pi_*(\cO_X(n \sigma)) &=& \left\{
\ba{lll}
\cO_B \oplus \cO_B(-2 c_1(TB)) \oplus \ldots \oplus \cO_B(-n c_1(TB))
& \mbox{for} & n > 0 \\
0 & \mbox{for} & n < 0
\ea
\right. \nn \\
R^1\pi_*(\cO_X(n \sigma)) &=& \left\{
\ba{lll}
0 & \mbox{for} & n > 0 \\
\cO_B((-n-1)c_1(TB)) \oplus \ldots \oplus \cO_B(c_1(TB)) 
\oplus \cO_B(- c_1(TB))
& \mbox{for} & n < 0 \ . \\
\ea
\right. \nn \\
\eea
In the last case of $n=0$ in \eref{H-push}, 
we have used the relation
\beq
R^1 \pi_* \cO_X = K_B
\eeq
and the fact that \eref{leray2} reduces, for $n=0$, to
\beq
0 \to H^1(B,L) \to H^1(X, \pi^*L) \to H^0(B, R^1\pi_*\cO_X \otimes L)
\to H^2(B,L) \to \ldots \ .
\eeq

Now the Hirzebruch surface $B = \IF_1$
is a $\IP^1$ fibration over $\IP^1$.
Therefore, there is a projection map
\beq
\beta : B \rightarrow \IP^1 \ .
\eeq
Any line bundle $L$ on $B=\IF_1$ can be written, by \eref{eff-Fr}, as
\beq
L = \cO_{B}(a S + b \cE) \ , \quad a,b \in \IZ \ .
\eeq
A similar application as \eref{leray2}, replacing $X$, $B$ and $T$ by
$B$, $\IP^1$ and $\cO_B(aS + b \cE)$ respectively, gives us
\bea\label{H-push2}
H^0(B, \cO_B(a S + b\cE)) &=& H^0(\IP^1, \beta_* \cO_B(aS) \otimes
\cO_{\IP^1}(b)),
\nn \\
H^1(B, \cO_B(a S + b\cE)) &=& H^0(\IP^1, R^1 \beta_* \cO_B(aS) \otimes
\cO_{\IP^1}(b)) \oplus H^1(\IP^1, \beta_* \cO_B(a S) \otimes
\cO_{\IP^1}(b)),
\nn \\
H^2(B, \cO_B(a S + b\cE)) &=& H^1(\IP^1, R^1\beta_* \cO_B(aS) \otimes
\cO_{\IP^1}(b)) \ .
\eea
where we have, for $\beta_* \cO(a S)$ and $R^1\beta_* \cO(a S)$,
\beq\label{beta-push}\ba{c|c|c}
& \beta_* \cO(a S) & R^1\beta_* \cO(a S) \\ \hline
a \ge 0 & \cO_{\IP^1} \oplus  \cO_{\IP^1}(-1) \oplus
\ldots \oplus  \cO_{\IP^1}(-a) & 0 \\ \hline
a < 0 &  0 &  \cO_{\IP^1}(1) \oplus \ldots \oplus
\cO_{\IP^1}(-a-1) \\
\ea\eeq
Combining the results in \eref{H-push} and \eref{H-push2} gives us a
method of expressing $H^i(X, \cO_X(n \sigma) \otimes \pi^* L)$ in
terms of a much more familiar object which can be handled with ease,
namely $H^1(\IP^1, \cO_{\IP^1}(a))$.

Let us demonstrate this technique by computing $H^1(X,W)$ for a
specific example.
Choose $n=0$ and
\beq
L = \cO_B(14 S + 15 \cE).
\eeq
Note this corresponds to choosing
\beq
T = W \ ,
\eeq
where the line bundle $W$ is defined in \eref{W-val}.
That is, the sheaf cohomology group we wish to consider is
$H^1(X, \pi^* L)$.
First, we note that
\beq
H^2(B, L) = H^1(\IP^1, R^1\beta_* \cO_B(14S) \otimes \cO_{\IP^1}(15))
= 0,
\eeq
since $R^1\beta_* \cO_B(14S)$ vanishes by
\eref{beta-push}.
Therefore, using \eref{H-push}, we conclude that
\beq
H^1(X, \pi^*\cO_B(14S + 15 \cE)) = H^1(B, \cO_B(14S + 15 \cE)) 
	\oplus H^0(B, \cO_B(-c_1(\IF_1)+14S + 15 \cE)).
\eeq
Using \eref{beta-push}, this becomes
\beq\label{H1W-1}
H^0(\IP^1, R^1\beta_*\cO_B(14S) \otimes \cO_{\IP^1}(15)) \oplus 
H^1(\IP^1, \beta_*\cO_B(14S) \otimes \cO_{\IP^1}(15)) \oplus
H^0(B, \beta_*\cO_B(12S) \otimes \cO_{\IP^1}(12)).
\eeq
Upon simplifying \eref{H1W-1}, we at last have
\beq
H^1(X,W) \simeq \bigoplus\limits_{i=0}^{12} 
	H^0(\IP^1, \cO_{\IP^1}(i)).
\eeq

We can similarly determine the other cohomology groups for $W$,
in summary, one finds that
\bea
H^0(X,W) &\simeq& \bigoplus\limits_{i=0}^{15} 
	H^0(\IP^1, \cO_{\IP^1}(i)) \ ,
\nn \\
H^1(X,W) &\simeq& \bigoplus\limits_{i=0}^{12} 
	H^0(\IP^1, \cO_{\IP^1}(i)) \ , 
\nn \\
H^2(X,W) &\simeq& H^3(X,W) = 0 \ .
\eea
Using the dimensions of the
cohomology groups on $\IP^1$ given in \eref{degP1},
we readily find that
\beq
h^0(X,W) = 135, \ h^1(X,W) = 91, \  h^2(X,W) = h^3(X,W) = 0 \ .
\eeq

%
%
\section{Constructing the Maps $M_i$ Explicitly}
In order to construct the linear maps $M_i$ in
\eref{seq1}, \eref{seq2} and \eref{seq3}, we first need
to determine the relevant cohomology groups $H^i(X,T)$ as vector
spaces. 
This was done in Appendix C. Next, we will construct explicit bases for these
vector spaces. Finally, we write the matrix representative for $M_i$ by
finding the multiplication rules which transform a basis of the
domain to a basis for the range.

We illustrate our technique with $M_1$ which, we recall from
\eref{seq3-2}, is the following mapping
\beq
H^1(X,W(-\cC_V-D))_{84} \stackrel{M_1}{\longrightarrow}
H^1(X,W(-D))_{28} \ .
\eeq
The method we use is similar to that of \cite{Evgeny2}.
We will need the bases for three vector spaces, the domain
$H^1(X,W(-\cC_V-D))_{84}$, the range $H^1(X,W(-D))_{28}$ 
and $H^0(X,\cO_X(\cC_V))$. This last space classifies the mapping
$M_1$. To see this, note that our map on cohomology
\beq
M_1 \in \Hom(H^1(X,W(-\cC_V-D)), H^1(X,W(-D)))
\eeq
is induced by a sheaf map
\beq
\tilde{M}_1: W(-\cC_V-D) \rightarrow W(-D)
\eeq
and both are given by the cup product with a class in
\beq
H^0(X,W(-\cC_V-D)^* \otimes W(-D)) = H^0(X, \cO_X(\cC_V)) \ .
\eeq
We will illustrate our technique
with the space $H^1(X,W(-D))_{28}$, the range of
the map $M_1$.
From \eref{W-val}, we have
\beq
W(-D) = \cO_X(-\sigma + \pi^*(8S + 9 \cE)).
\eeq
The method of expressing a cohomology group of this form
in terms of those on the
base $\IF_1$ and then on the base $\IP^1$ of $\IF_1$ has
already been presented in Appendix C
using the Leray spectral sequence. By
\eref{H-push}, \eref{pi-nsig}, \eref{H-push2} and \eref{chernFr}, 
we have
\beq\label{w-d1}\begin{diagram}
H^1(X, W(-D)) = H^1(X, \cO_X(-\sigma + \pi^*(8S + 9 \cE))) \\
\dTo{\pi_*} \\
H^0(B, \cO_B(6S + 6 \cE)) \\
\dTo{\beta_*} \\
H^0(\IP^1, \cO_{\IP^1}(6) \oplus \cO_{\IP^1}(5) \oplus \cO_{\IP^1}(4)
\oplus \cO_{\IP^1}(3) \oplus \cO_{\IP^1}(2) \oplus \cO_{\IP^1}(1)
\oplus \cO_{\IP^1}) \ .
\end{diagram}\eeq
We now describe the vector space on $\IP^1$ explicitly by defining
\beq
B_{(3)k} = H^0(\IP^1, \cO_{\IP^1}(k))
\eeq
for $k=0,1,\ldots, 6$ where
\beq
\dim B_{(3)k} = k+1.
\eeq
The vector space at the bottom of \eref{w-d1},
the cohomology group on $\IP^1$, can now be written as
\beq
\bigoplus_{k=1}^6 B_{(3)k}.
\eeq
Next, we pull back this space to $\IF_1$ using $\beta^*$ and define
\beq
b_{(3)k} = \beta^* B_{(3)k} \ .
\eeq
Note that $b_{(3)k}$ is the space of sections of $\cO_B(k \cE)$. That
is,
\beq
b_{(3)k} = H^0(B,\cO_B(k \cE)).
\eeq
In order to embed this as a 
subspace of $H^0(B, \cO_B(6S+6\cE))$, each element of
$H^0(B, \cO_B(k \cE))$ must be multiplied by
a fixed section of $\cO_B(6S+6\cE-k\cE)$. Defining
\beq\label{defsi}
s_{(3)k} = 3c_1(T\IF_1) - k \cE \ ,
\eeq
the reader can
readily verify, using \eref{chernFr}, 
that
\beq
s_{(3)k} - 3 \cE = 6S+6\cE-k\cE \ .
\eeq
Let us denote the fixed section by
\beq\label{def-s3k}
\tilde{s}_{(3)k} \in H^0(B, \cO_B(s_{(3)k} - 3 \cE))
\eeq
and the space of sections of $\cO_B(k \cE)$ multiplied by 
$\tilde{s}_{(3)k}$ as
$b_{(3)k}\tilde{s}_{(3)k}$.
Then, we can write the middle term in \eref{w-d1} as
\beq\label{pull-B-1}
H^0(B, \cO_B(6S + 6 \cE)) = 
\bigoplus\limits_{k=0}^6 b_{(3)k} \tilde{s}_{(3)k} \ .
\eeq

Finally, we pull back this space
to $X$ using $\pi^*$. We find that
\beq\label{h1XW-D}
H^1(X, W(-D)) = \bigoplus\limits_{k=0}^6 \hat{b}_{(3)k}
\hat{\tilde{s}}_{(3)k} \tilde{a},
\eeq
where
\beq
a = -\sigma + \pi^*(c_1(\IF_1)), \ \
\hat{b}_{(3)k} = \pi^*(b_{(3)k}), \ \
\hat{\tilde{s}}_{(3)k} = \pi^*(\tilde{s}_{(3)k})
\eeq
and
\beq
\tilde{a} \in H^0(X, \cO_X(a)) \ .
\eeq
The factor $\tilde{a}$ arises for the same reason as the 
$\tilde{s}_{(3)k}$ factor and we use similar notation. 
That is, $\tilde{a}$ is required so that each term on the right hand
side of \eref{h1XW-D} is a subspace of
$H^1(X, W(-D))$.
Specifically, the notation in \eref{h1XW-D} indicates that one should
take the element $\tilde{a}$ 
of $H^1(X, \cO_X(a))$ and multiply each element of
$\bigoplus\limits_{k=0}^6 \hat{b}_{(3)k} \hat{\tilde{s}}_{(3)k}$ by
it.
In summary, for the term $H^1(X, W(-D))$, we have
\beq\label{basis1}\begin{diagram}
H^1(X, W(-D)) = \bigoplus\limits_{k=0}^6 \hat{b}_{(3)k}
\hat{\tilde{s}}_{(3)k} \tilde{a} \\
\uTo{\pi^*} \\
H^0(B, \cO_B(6S + 6 \cE)) = 
\bigoplus\limits_{k=0}^6 b_{(3)k} \tilde{s}_{(3)k} \\
\uTo{\beta^*} \\
\bigoplus\limits_{k=0}^6 B_{(3)k}.
\end{diagram}\eeq

The above procedure 
can be repeated for the other two terms, the
domain $H^1(X,W(-\cC_V-D))$ and the mapping 
$M_1 \in H^0(X,\cO_X(\cC_V))$.
For $H^1(X,W(-\cC_V-D))$, we have the following decomposition.
\beq\label{basis2}\begin{diagram}
H^1(X, W(-\cC_V-D))
= \bigoplus\limits_{Q=0}^3
\bigoplus\limits_{j=Q}^{3Q} \hat{c}_{(Q)j} \hat{\tilde{s}}_{(Q)j}
\tilde{a}_{(Q)},
& \qquad \qquad
\ba{l}\hat{c}_{(Q)j}=\pi^*c_{(Q)j}, \\
	\hat{\tilde{s}}_{(Q)j} = \pi^*\tilde{s}_{(Q)j}, \\ 
	a_{(Q)} = -6\sigma-\pi^*((Q+2)c_1(T\IF_1)), \\
	\tilde{a}_{(Q)} \in H^0(X, \cO_X(a(Q)))\\~\\
\ea
\\
\uTo{\pi^*} \\
\bigoplus\limits_{Q=0}^3 \bigoplus\limits_{j=Q}^{3Q} c_{(Q)j}
\tilde{s}_{(Q)j}, & 
\ba{l} 
c_{(Q)j}=\beta^*C_{(Q)j}, \\
s_{(Q)j}=Qc_1(T\IF_1)-j\cE, \\
\tilde{s}_{(Q)j} \in H^0(B, \cO_B(s_{(Q)j}))
\ea \\
\uTo{\beta^*} \\
\bigoplus\limits_{Q=0}^3 \bigoplus\limits_{j=Q}^{3Q}
C_{(Q)j}, & C_{(Q)j}= H^0(\IP^1, \cO_{\IP^1}(j)).
\end{diagram}\eeq
Finally, the space $H^0(X,\cO_X(\cC_V))$, in which the map
$M_1$ lives, has the
basis
\beq\label{basis3}\begin{diagram}
H^0(X, \cO_X(\cC_V))
= \bigoplus\limits_{R=1,2,3,4,6}
\bigoplus\limits_{i=R-3}^{3R-3} \hat{m}_{(R)i} 
	\hat{\tilde{s}}_{(R)i} \tilde{A}_{(R)},
& \qquad \qquad
\ba{l}
\hat{m}_{(R)i} = \pi^*m_{(R)i}, \\
\hat{\tilde{s}}_{(R)i} = \pi^*s_{(R)i}, \\ 
A_{(R)i} = 5\sigma+\pi^*((6-R)c_1(T\IF_1)), \\
\tilde{A}_{(R)i} \in H^0(X, \cO_X(A_{(R)i}))\\~\\
\ea
\\
\uTo{\pi^*} \\
\bigoplus\limits_{R=1,2,3,4,6} \bigoplus\limits_{i=R-3}^{3R-3} m_{(R)i}
\tilde{s}_{(R)i} \ , & 
\ba{l} 
m_{(R)i}=\beta^*M_{(R)i}, \\
s_{(R)i}=R c_1(T\IF_1)- i \cE, \\
\tilde{s}_{(R)i} \in H^0(B, \cO_B(s_{(R)i}))
 \ea \\
\uTo{\beta^*} \\
\bigoplus\limits_{R=1,2,3,4,6} \bigoplus\limits_{i=R-3}^{3R-3}
M_{(R)i}, & M_{(R)i}= H^0(\IP^1, \cO_{\IP^1}(i)).
\end{diagram}\eeq

We have now explicitly constructed the bases for the vector spaces of
concern. In particular, $M_1 \in H^0(X,\cO_X(\cC_V))$ given in \eref{basis3}
must map $H^1(X,W(-\cC_V-D))_{84}$ given in \eref{basis2} linearly to
$H^1(X,W(-D))_{28}$ given in \eref{basis1}. The matrix for $M_1$ is
straight-forward to
construct. We multiply the basis in \eref{basis2} with the matrix
elements in \eref{basis3} and demand that the result be in
\eref{basis1}.
More specifically, we must guarantee that
\beq
\left( \hat{m}_{(R)i} \hat{\tilde{s}}_{(R)i} \tilde{A}_{(R)} \right)
\cdot 
\left( \hat{c}_{(Q)j} \hat{\tilde{s}}_{(Q)j} \tilde{a}_{(Q)} \right) =
\hat{b}_{(3)k} \hat{\tilde{s}}_{(3)k} \tilde{a} \ .
\eeq
This requires that
\beq\label{Aa}
\tilde{A}_{(R)} \cdot \tilde{a}_{(Q)} = \tilde{a} \ ,
\eeq
\beq\label{ss}
\hat{\tilde{s}}_{(R)i} \cdot \hat{\tilde{s}}_{(Q)j} 
= \hat{\tilde{s}}_{(3)k}
\eeq
and
\beq\label{bb}
\hat{m}_{(R)i} \cdot \hat{c}_{(Q)j} = \hat{b}_{(3)k}
\eeq
be satisfied individually. Using the definitions of these quantities
given in \eref{basis1}, \eref{basis2} and \eref{basis3}, we find from
\eref{Aa} that only terms with
\beq\label{MN-con}
R+Q=3
\eeq
can pair non-trivially.
Substituting this relation into \eref{ss} and \eref{bb} then
constrains $i,j$ and $k$ to satisfy
\beq\label{ijk-con}
i+j=k \ .
\eeq
We still must show that the sections $\tilde{A}_{(R)}$,
$\tilde{a}_{(S)}$, $\tilde{a}$ and $\hat{\tilde{s}}_{(R)i}$, 
$\hat{\tilde{s}}_{(Q)j}$, $\hat{\tilde{s}}_{(3)k}$ can be chosen to
satisfy \eref{Aa} and \eref{ss} respectively. This can be done as
follows. Fix once and for all a generic section $\tilde{\ell} \in
H^0(B, \cO_B(S + \cE))$. Now, each of our $\hat{\tilde{s}}$-type 
sections is
the lift of a section $\tilde{s} \in H^0(B, \cO_B(a S + b \cE))$ for
some integers $a \ge b \ge 0$. We will simply set
\beq\label{sSab}
\tilde{s} = \tilde{\ell}^b \tilde{S}^{a-b},
\eeq
where $\tilde{S}$ is a fixed section of $\cO_B(S)$. Note that
\eref{sSab} is a section of the desired bundle $\cO_B(a S + b
\cE)$. Furthermore, these sections clearly satisfy \eref{ss}. We can
make similar choices for the $a$-type sections, thus satisfying
\eref{Aa} as well.
Having established the above, we see that any element of 
$\hat{m}_{(R)i} \hat{\tilde{s}}_{(R)i} \tilde{A}_{(R)}$, $\hat{c}_{(Q)j}
\hat{\tilde{s}}_{(Q)j} \tilde{a}_{(Q)}$ or 
$\hat{b}_{(3)k} \hat{\tilde{s}}_{(3)k} \tilde{a}$ 
can be explicitly labeled by an element of $M_{(R)i}$,
$C_{(Q)j}$ or $B_{(3)k}$ respectively.

Now, consider the matrix $M_1 \in H^0(X,\cO_X(\cC_V))$.
Any matrix block in $M_1$ that does not satisfy
one or both of \eref{MN-con} and
\eref{ijk-con} must be a zero entry. On the other hand, any entry in
$M_1$ that satisfies both \eref{MN-con} and \eref{ijk-con} is a
potentially non-vanishing sub-matrix which we denote by
\beq\label{mat-m}
m_{(R)i} \in M_{(R)i} = H^0(\IP^1, \cO_{\IP^1}(i)) \ .
\eeq
Note that $m_{(R)i}$ may occur as many different sub-matrices within
$M_1$ with the $(R)i$ subscripts fixing its location.
This is most easily understood by simply looking at the result.
Using the data from \eref{basis1}, \eref{basis2} and \eref{basis3}
subject to the constraints \eref{MN-con} and \eref{ijk-con},
we find
\beq
\hspace{-1cm}
\smat{ m_{(3)0} & 0 & 0 & 0 & 0 & 0 & 0 & 0 & 0 & 0 & 0 & 0 & 0 & 0 & 0 & 0 \cr m_{(3)1} & m_{(2)
   0} & 0 & 0 & 0 & 0 & 0 & 0 & 0 & 0 & 0 & 0 & 0 & 0 & 0 & 0 \cr m_{(3)2} & m_{(2)1} & m_{(2)0} & 0 & m_{(1)
   0} & 0 & 0 & 0 & 0 & 0 & 0 & 0 & 0 & 0 & 0 & 0 \cr m_{(3)3} & m_{(2)2} & m_{(2)1} & m_{(2)0} & 0 & m_{(1)
   0} & 0 & 0 & 0 & 0 & 0 & 0 & 0 & 0 & 0 & 0 \cr m_{(3)4} & m_{(2)3} & m_{(2)2} & m_{(2)1} & 0 & 0 & m_{(1)
   0} & 0 & 0 & 0 & 0 & 0 & 0 & 0 & 0 & 0 \cr m_{(3)5} & 0 & m_{(2)3} & m_{(2)2} & 0 & 0 & 0 & m_{(1)
   0} & 0 & 0 & 0 & 0 & 0 & 0 & 0 & 0 \cr m_{(3)6} & 0 & 0 & m_{(2)3} & 0 & 0 & 0 & 0 & m_{(1)
   0} & 0 & 0 & 0 & 0 & 0 & 0 & 0 \cr  }_{28 \times 84}
\cdot \smat{
C_{(0)0}\cr
C_{(1)1}\cr
C_{(1)2}\cr
C_{(1)3}\cr
C_{(2)2}\cr
C_{(2)3}\cr
C_{(2)4}\cr
C_{(2)5}\cr
C_{(2)6}\cr
C_{(3)3}\cr
C_{(3)4}\cr
C_{(3)5}\cr
C_{(3)6}\cr
C_{(3)7}\cr
C_{(3)8}\cr
C_{(3)9}\cr
}_{84}=
\smat{
B_{(3)0}\cr
B_{(3)1}\cr
B_{(3)2}\cr
B_{(3)3}\cr
B_{(3)4}\cr
B_{(3)5}\cr
B_{(3)6}\cr
}_{28 \ .}
\eeq
Of course, each non-zero sub-matrix $m_{(R)i}$ maps a space
$B_{(Q)j} = H^0(\IP^1, \cO(j))$ linearly to a space $B_{(3)k} =
H^0(\IP^1, \cO_{\IP^1}(k))$ 
where $R+Q=3$ and $i+j=k$. That is, $m_{(R)i}$ is
a $(k+1) \times (j+1)$ matrix. We can emphasize this by extending our
notation and writing $m_{(R)i}$ as
\beq
m_{(R)i \{k+1, \ j+1\}} \ .
\eeq 
Using this notation, the matrix $M_1$ can be written as
\beq\label{M1}
\hspace{-1cm}
M_1 =
\tmat{ m_{(3)0\{ 1,1\} } & 0 & 0 & 0 & 0 & 0 & 0 & 0 & 0 & 0 & 0 & 0 & 0 & 0 & 0 & 0 \cr m_{(3)1\{ 2,1\} } & m_{(2)0\{ 2,2\} } & 0 & 0 & 0 & 0 & 0 & 0 & 0 & 0 & 0 & 0 & 0 & 0 & 0 & 0 \cr m_{(3)2\{ 3,1\} } & m_{(2)1\{ 3,2\} } & m_{(2)0\{ 3,3\} } & 0 & m_{(1)0
   \{ 3,3\} } & 0 & 0 & 0 & 0 & 0 & 0 & 0 & 0 & 0 & 0 & 0 \cr m_{(3)3\{ 4,1\} } & m_{(2)2\{ 4,2\} } & m_{(2)
   1\{ 4,3\} } & m_{(2)0\{ 4,4\} } & 0 & m_{(1)0
   \{ 4,4\} } & 0 & 0 & 0 & 0 & 0 & 0 & 0 & 0 & 0 & 0 \cr m_{(3)4\{ 5,1\} } & m_{(2)3\{ 5,2\} } & m_{(2)2,
   \{ 5,3\} } & m_{(2)1\{ 5,4\} } & 0 & 0 & m_{(1)0\{ 5,5\} } & 0 & 0 & 0 & 0 & 0 & 0 & 0 & 0 & 0 \cr m_{(3)
   5\{ 6,1\} } & 0 & m_{(2)3\{ 6,3\} } & m_{(2)2\{ 6,4\} } & 0 & 0 & 0 & m_{(1)0\{ 6,6\} } & 0 & 0 & 0 & 0 & 0 & 0 & 0 & 0 \cr m_{(3)6\{ 7,1\} } & 0 & 0 & m_{(2)3\{ 7,4\} } & 0 & 0 & 0 & 0 & m_{(1)0\{ 7,7\} } & 0 & 0 & 0 & 0 & 0
& 0 & 0 \cr  }_{28 \times 84 \ .}
\eeq

It remains to determine the block matrices
$m_{(R)i\{k+1,j+1\}}$ to finish constructing $M_1$. The method for
doing this was
presented in detail in Section 6 of \cite{Evgeny2}. We summarize the
results here.
A block of dimension $(k+1) \times (j+1)$ is a mapping
\beq\begin{diagram}
m_{\{k+1,j+1\}}:
H^0(\IP^1, \cO_{\IP^1}(j)) &\rTo{H^0(\IP^1, \cO_{\IP^1}(k-j))}&
H^0(\IP^1, \cO_{\IP^1}(k)) \ ,
\end{diagram}\eeq
where, for the moment, we have suppressed the subscript $(R)i$.
Now, we can write $H^0(\IP^1, \cO_{\IP^1}(j))$ in terms of a symmetrized
product of the vector space $\hat{V} = H^0(\IP^1, \cO_{\IP^1}(1))$.
$\hat{V}$ is a two-dimensional space whose basis we choose to be
$\{u,v\}$.
In other words,
\bea\label{sym-j}
H^0(\IP^1, \cO_{\IP^1}(j)) &=& \sym^j(H^0(\IP^1, \cO_{\IP^1}(1)))\nn\\
&=&\sym^j({\rm span}\{u,v\})\nn\\
&=&{\rm span}\{u^{j}, u^{j-1}v,\ldots,uv^{j-1}, v^{j}\}.
\eea
Similarly,
\beq\label{sym-k}
H^0(\IP^1, \cO_{\IP^1}(k)) = {\rm span}\{u^{k}, u^{k-1}v,
\ldots,uv^{k-1}, v^{k}\}.
\eeq
Finally, $m_{\{k+1,j+1\}}$ itself lives in $H^0(\IP^1,
\cO_{\IP^1}(k-j))$, which can be written as
\bea\label{symH0p}
H^0(\IP^1, \cO_{\IP^1}(k-j)) &=& \sym^{k-j}(H^0(\IP^1,
\cO_{\IP^1}(1)))\nn\\
&=&\sym^{k-j}({\rm span}\{u,v\})\nn\\
&=&{\rm span}\{u^{k-j}, u^{k-j-1}v,\ldots,uv^{k-j-1}, v^{k-j}\} . 
\eea
It is important to note that the sub-matrix is non-zero only when
\beq
k \ge j,
\eeq
for otherwise $H^0(\IP^1,\cO_{\IP^1}(k-j))=0$ 
and our matrix $m_{\{k+1,j+1\}}$ would vanish identically. This is
consistent with the constraint that $i+j=k$ given in \eref{ijk-con}.
Having determined the explicit bases, we can now construct the matrix
$m_{\{k+1,j+1\}}$.
First, note from \eref{symH0p} that any element of $H^0(\IP^1,
\cO_{\IP^1}(k-j))$ can be written
in terms of the basis $\{u,v\}$ as
\beq\label{mphi}
m_{\{k+1,j+1\}} = 
\phi_1 u^{k-j} + \phi_2 u^{k-j-1} v + \ldots + \phi_{k-j+1} v^{k-j},
\eeq
where $\phi_p \in \IC$ are $k-j+1$ complex moduli. 
Now, tensor expression \eref{mphi} into each basis element of 
$H^0(\IP^1,\cO_{\IP^1}(j))$ given in \eref{sym-j}. Expanding the
result into the basis \eref{sym-k} of $H^0(\IP^1, \cO_{\IP^1}(k))$
completely specifies the matrix. 
The final expression for $m_{\{k+1, j+1\}}$ is the following.
First of all, the matrix
$m_{\{k+1, j+1\}} = 0$ for $k < j$. For $k \ge j$, the $p,q$-th
matrix element is
\beq\label{blockM}
(m_{\{k+1, j+1\}})_{pq} = \left\{\ba{ll}
\phi_{p-q+1}, & q \le p \le q+k-j \ ,\\
0, & \mbox{otherwise}
\ea\right.
\eeq
for $p=1,2,\ldots,k+1$ and $q=1,2,\ldots,j+1$.
It is important to recall, however, that we have been suppressing the
$(R)i$ indices on $m_{(R)i\{k+1, j+1\}}$. Restoring these, the
expression \eref{mphi} becomes
\beq
m_{(R)i\{k+1,j+1\}} = 
\phi_1^{[(R)i]} u^{k-j} + \phi_2^{[(R)i]} u^{k-j-1} v + \ldots 
+ \phi_{k-j+1}^{[(R)i]} v^{k-j} \ .
\eeq
That is, the moduli labeling $m_{(R)i}$ in $M_{(R)i}$ are uniquely
determined by the $(R)i$ indices. Change either or both of these and
the set of moduli changes. Note, using the relation \eref{ijk-con},
that $m_{(R)i}$ is labeled by the moduli
\beq
\phi_{p}^{[(R)i]} \ , \qquad p = 1, \ldots, i+1 \ .
\eeq
Of course, expression \eref{blockM} for the $p,q$-th element of the
matrix $m_{(R)i\{k+1,j+1\}}$ remains the same, but with the moduli
replaced by $\phi_{p-q+1}^{[(R)i]}$. At this point, it would be
helpful to present some explicit examples. Let us consider the 
$m_{(1)0\{3,3\}}$ and $m_{(2)1\{3,2\}}$ sub-matrices of $M_1$ in
\eref{M1}. Then, it follow from \eref{blockM} that
\beq\label{blockeg}
m_{(1)0\{3,3\}} =
\mat{ 
\phi^{[(1)0]}_1 & 0 & 0 \cr 0 & \phi^{[(1)0]}_1 & 0 \cr 
0 & 0 & \phi^{[(1)0]}_1 \cr 
},
\qquad
m_{(2)1\{3,2\}} = 
\mat{
\phi^{[(2)1]}_1 & 0 \cr
\phi^{[(2)1]}_2 & \phi^{[(2)1]}_1 \cr
0 & \phi^{[(2)1]}_2
} \ .
\eeq
Indeed, expression \eref{blockM}, inserted into each sub-matrix 
$m_{(R)i\{k+1, j+1\}}$ in \eref{M1}, completes the construction
of the matrix $M_1$. 
The general result is unenlightening and will not be presented here. 

It is important, however, to
compute the generic value of the rank of $M_1$ To do this, we begin by
setting all moduli to zero except those in the $m_{(1)0}$, $m_{(2)0}$
and $m_{(3)0}$ sub-blocks of \eref{M1}. Note from \eref{blockeg} that
\beq
m_{(1)0\{3,3\}} = \phi^{[(1)0]}_1 \II_{3,3} \ ,
\eeq
where $\II_{3,3}$ is the $3 \times 3$ identity matrix. In fact, all 
$m_{(1)0}$, $m_{(2)0}$
and $m_{(3)0}$ sub-matrices are proportional to $\phi^{[(1)0]}_1$,
$\phi^{[(2)0]}_1$ and $\phi^{[(3)0]}_1$ moduli respectively times the
unit matrix of the appropriate size. If we then set
\beq
\phi^{[(1)0]}_1 = \phi^{[(2)0]}_1 = \phi^{[(3)0]}_1 \equiv \phi, 
\eeq
we find the matrix $M_1$ given in \eref{M1-set}. For any $\phi \ne 0$,
this matrix has
\beq\label{M1-max2}
\rk(M_1) = 28.
\eeq
One can also consider the complete expression for $M_1$, randomly
initialize the values of its moduli and  numerically compute the
rank. We have done this for a large number of random initial sets with
the caveat, to be explained below, that we insist that all moduli be
non-zero. Generically, we recover the result \eref{M1-max2}. This
justifies the important assumption in Subsection \ref{s:avstar} that
we can choose $\rk(M_1) = 28$, thus simplifying the calculation of
$h^0(D,W|_D)$ to expression \eref{h0WD-res}.

Similarly, we can construct the matrix $M_2$, which we recall from
\eref{seq2-2} is the mapping
\beq
H^1(X,W(-\cC_V))_{180} \stackrel{M_2}{\longrightarrow} H^1(X,W)_{91} \ .
\eeq
The result, in block-form, is
\beq\label{M2}
\hspace{-2cm}
{\tiny
\ba{l}
{\normalsize M_2 =}
\left(
\matrix{
0 & 0 & 0 & 0 & 0 & 0 & 0 & 0 & 0 & 0 & 0 & 0 & 0 & 0 & 0 & 0 & 0 & 0 & 0 & 0 \cr 0 & 0 & 0 & 0 & 0 & 0 & 0 & 
   0 & 0 & 0 & 0 & 0 & 0 & 0 & 0 & 0 & 0 & 0 & 0 & 0 \cr 0 & 0 & 0 & 0 & 0 & 0 & 0 & 0 & 0 & 0 & 0 & 0 & 0 & 0 & 
   0 & 0 & 0 & 0 & 0 & 0 \cr 0 & 0 & 0 & 0 & m_{(6)3
   \{ 4,1\} } & 0 & 0 & 0 & 0 & 0 & 0 & 0 & 0 & 0 & 0 & 0 & 0 & 0 & 0 & 0 \cr 0 & 0 & 0 & 0 & m_{(6)4
   \{ 5,1\} } & m_{(6)3
   \{ 5,2\} } & 0 & 0 & 0 & 0 & 0 & 0 & 0 & 0 & 0 & 0 & 0 & 0 & 0 & 0 \cr 0 & 0 & 0 & 0 & m_{(6)5
   \{ 6,1\} } & m_{(6)4\{ 6,2\} } & m_{(6)3
   \{ 6,3\} } & 0 & 0 & 0 & 0 & 0 & 0 & 0 & 0 & 0 & 0 & 0 & 0 & 0 \cr 0 & 0 & 0 & 0 & m_{(6)6\{ 7,1\} } & m_{(6)
   5\{ 7,2\} } & m_{(6)4\{ 7,3\} } & m_{(6)3
   \{ 7,4\} } & 0 & 0 & 0 & 0 & 0 & 0 & 0 & 0 & 0 & 0 & 0 & 0 \cr 0 & 0 & 0 & 0 & m_{(6)7\{ 8,1\} } & m_{(6)6
   \{ 8,2\} } & m_{(6)5\{ 8,3\} } & m_{(6)4\{ 8,4\} } & m_{(6)3
   \{ 8,5\} } & 0 & 0 & 0 & 0 & 0 & 0 & 0 & 0 & 0 & 0 & 0 \cr 0 & 0 & 0 & 0 & m_{(6)8\{ 9,1\} } & m_{(6)7
   \{ 9,2\} } & m_{(6)6\{ 9,3\} } & m_{(6)5\{ 9,4\} } & m_{(6)4\{ 9,5\} } & m_{(6)3
   \{ 9,6\} } & 0 & 0 & 0 & 0 & 0 & 0 & 0 & 0 & 0 & 0 \cr 0 & 0 & 0 & 0 & m_{(6)9\{ 10,1\} } & m_{(6)8
   \{ 10,2\} } & m_{(6)7\{ 10,3\} } & m_{(6)6\{ 10,4\} } & m_{(6)5\{ 10,5\} } & m_{(6)4\{ 10,6\} } & m_{(6)
   3\{ 10,7\} } & 0 & 0 & 0 & 0 & 0 & 0 & 0 & 0 & 0 \cr 0 & 0 & 0 & 0 & m_{(6)10\{ 11,1\} } & m_{(6)9
   \{ 11,2\} } & m_{(6)8\{ 11,3\} } & m_{(6)7\{ 11,4\} } & m_{(6)6\{ 11,5\} } & m_{(6)5\{ 11,6\} } & m_{(6)
   4\{ 11,7\} } & 0 & 0 & 0 & 0 & 0 & 0 & 0 & 0 & 0 \cr 0 & 0 & 0 & 0 & m_{(6)11\{ 12,1\} } & m_{(6)10
   \{ 12,2\} } & m_{(6)9\{ 12,3\} } & m_{(6)8\{ 12,4\} } & m_{(6)7\{ 12,5\} } & m_{(6)6\{ 12,6\} } & m_{(6)
   5\{ 12,7\} } & 0 & 0 & 0 & 0 & 0 & 0 & 0 & 0 & 0 \cr 0 & 0 & 0 & 0 & m_{(6)12\{ 13,1\} } & m_{(6)11
   \{ 13,2\} } & m_{(6)10\{ 13,3\} } & m_{(6)9\{ 13,4\} } & m_{(6)8\{ 13,5\} } & m_{(6)7
   \{ 13,6\} } & m_{(6)6\{ 13,7\} } & 0 & 0 & 0 & 0 & 0 & 0 & 0 & 0 & 0 \cr 
} \cdots  \ \ \right. \\ \\
\left.
\qquad \qquad \cdots \ 
\matrix{
0 & 0 & 0 & 0 & 0 & 0 & 0 & 0 & 0 & 0 & 0 \cr 0 & 0 & 0 & 0 & 0 & 0 & 0 & 0 & 0 & 0 & 0 \cr 0 & 0 & 0 & 0 & 
   0 & 0 & 0 & 0 & 0 & 0 & 0 \cr m_{(4)1\{ 4,3\} } & 0 & 0 & 0 & 0 & 0 & 0 & 0 & 0 & 0 & 0 \cr m_{(4)2
   \{ 5,3\} } & m_{(4)1\{ 5,4\} } & 0 & 0 & 0 & 0 & 0 & 0 & 0 & 0 & 0 \cr m_{(4)3\{ 6,3\} } & m_{(4)2
   \{ 6,4\} } & m_{(4)1\{ 6,5\} } & 0 & 0 & 0 & 0 & 0 & 0 & 0 & 0 \cr m_{(4)4\{ 7,3\} } & m_{(4)3
   \{ 7,4\} } & m_{(4)2\{ 7,5\} } & m_{(4)1\{ 7,6\} } & 0 & 0 & 0 & 0 & 0 & 0 & 0 \cr m_{(4)5\{ 8,3\} } & m_{(4)
   4\{ 8,4\} } & m_{(4)3\{ 8,5\} } & m_{(4)2\{ 8,6\} } & m_{(4)1\{ 8,7\} } & 0 & 0 & 0 & 0 & 0 & 0 \cr m_{(4)6\{ 9,3\} } & m_{(4)5\{ 9,4\} } & m_{(4)4\{ 9,5\} } & m_{(4)
   3\{ 9,6\} } & m_{(4)2\{ 9,7\} } & m_{(4)1\{ 9,8\} } & 0 & 0 & 0 & 0 & 0 \cr m_{(4)7\{ 10,3\} } & m_{(4)
   6\{ 10,4\} } & m_{(4)5\{ 10,5\} } & m_{(4)4\{ 10,6\} } & m_{(4)3\{ 10,7\} } & m_{(4)2
   \{ 10,8\} } & m_{(4)1\{ 10,9\} } & 0 & 0 & 0 & 0 \cr m_{(4)8\{ 11,3\} } & m_{(4)7\{ 11,4\} } & m_{(4)6
   \{ 11,5\} } & m_{(4)5\{ 11,6\} } & m_{(4)4\{ 11,7\} } & m_{(4)3\{ 11,8\} } & m_{(4)2\{ 11,9\} } & m_{(4)
   \{ 11,10\} } & 0 & 0 & 0 \cr m_{(4)9\{ 12,3\} } & m_{(4)8\{ 12,4\} } & m_{(4)7\{ 12,5\} } & m_{(4)6
   \{ 12,6\} } & m_{(4)5\{ 12,7\} } & m_{(4)4\{ 12,8\} } & m_{(4)3\{ 12,9\} } & m_{(4)2
   \{ 12,10\} } & m_{(4)1\{ 12,11\} } & 0 & 0 \cr 0 & m_{(4)9\{ 13,4\} } & m_{(4)8\{ 13,5\} } & m_{(4)7
   \{ 13,6\} } & m_{(4)6\{ 13,7\} } & m_{(4)5\{ 13,8\} } & m_{(4)4\{ 13,9\} } & m_{(4)3
   \{ 13,10\} } & m_{(4)2\{ 13,11\} } & m_{(4)1\{ 13,12\} } & 0 \cr
} \right)_{91 \times 180 \ .}\\
~\\
\ea}
\eeq
As explicit examples of sub-matrices in $M_2$,
we present the following two which we will use in our analyses below.
They are
\beq\label{M2eg}
m_{(4)1\{4,3\}} 
\mat{\phi^{[(4)1]}_1 & 0 & 0 \cr \phi^{[(4)1]}_2 & 
\phi^{[(4)1]}_1 & 0 \cr 0 & \phi^{[(4)1]}_2 & 
\phi^{[(4)1]}_1 \cr 0 & 0 & \phi^{[(4)1]}_2 \cr},
\qquad
m_{(4)2\{5,3\}} 
=
\mat{\phi^{[(4)2]}_1 & 0 & 0 \cr \phi^{[(4)2]}_2 & \phi^{[(4)2]}_1 & 0
\cr \phi^{[(4)2]}_3 & \phi^{[(4)2]}_2 & \phi^{[(4)2]}_1 \cr 0 &
\phi^{[(4)2]}_3 & \phi^{[(4)2]}_2 \cr 0 & 0 & \phi^{[(4)2]}_3 \cr} \ .
\eeq
We have computed all other $m_{(R)i\{k+1, j+1\}}$ sub-blocks in
\eref{M2} but, of course, will not present them here. This completes
the construction of the matrix $M_2$.

\section{The Rank of $M_2$}
\begin{figure}
\centerline{\psfig{figure=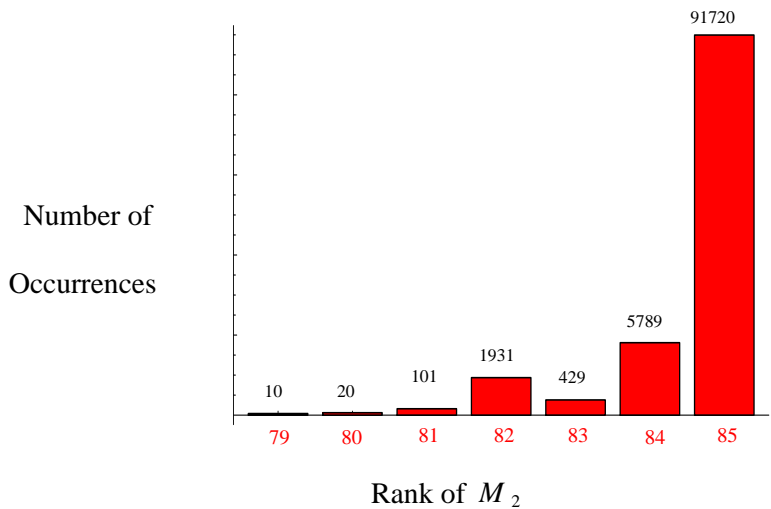,width=4in}}
\caption{ In 100,000 random initializations of the matrix
$M_2$ of integers valued between 1 and 3, 
the numbers of occurrences of the various values of $\rk(M_2)$
are plotted. We see that the generic value 85 dominates by far. 
\label{f:M2jump}}
\end{figure}

It is not enlightening to display $M_2$ in its full form in terms of
the moduli. However, it is important for
us to compute its rank. To do this, we begin by randomly selecting the
values of all moduli assuming, however, that each is non-zero. We then
numerically compute the rank of $M_2$. This process is continued for a
large number of different random, but non-zero,
initializations. The results of an explicit numerical calculation
involving 100,000 random integer initializations between 1 and 3
of the moduli are shown in
\fref{f:M2jump}. The horizontal axis indicates the ranks of $M_2$
found in the survey, while the vertical axis gives the number of
occurrences. We see that the rank of 85 by far dominates over any
other possibilities. It follows that at generic points in moduli space
\beq
\rk(M_2) = 85.
\eeq
This is, in fact, the maximal possible rank, as can be seen by
examining \eref{M2} and noting that 6 out of the 91 rows have all zero
entries.

Importantly, however, we notice that there are isolated
initializations of the moduli for which the rank of $M_2$ jumps to
values smaller than 85. This phenomenon is clearly seen in
\fref{f:M2jump} where $\rk(M_2)$ is shown to attain all integer values
between 79 and 84, in addition to its generic value of 85. It is clear
from the low statistics of these other values, that they occur at
non-generic points in the vector bundle moduli space.
In this set of 100,000 integer
randomizations, we have not seen any ranks lower than 79. However,
as we increase
the number of randomizations we expect to see smaller values of
$\rk(M_2)$. Statistically, this is expected.
For example, with a smaller set of only 1,000 randomizations,
only 2 values of $\rk(M_2)$ were observed.

Clearly, $\rk(M_2)$ jumps to values less than 85 on non-generic points
in the moduli space. It is, therefore, rather difficult and
unenlightening to search for such points numerically. We have
presented the analysis in \fref{f:M2jump} to clearly demonstrate that
the rank of $M_2$ can take different values in different regions of
the vector bundle moduli space.
Let us now take a more systematic and analytic approach to this
phenomenon of the jumping of the rank. We see from \eref{M2} that there are
two clusters of non-zero sub-matrices, namely, a triangle cluster
consisting of $m_{(6)i}$ sub-matrices and another consisting of 
$m_{(4)i}$ sub-matrices.
Let us leave the sub-blocks $m_{(6)i}$
untouched and proceed to consecutively set the 
sub-blocks $m_{(4)i}$ to zero.
Let us first set the block $m_{(4)1}$,
given explicitly in \eref{M2eg}, to
zero. That is, we set the two moduli 
\beq\label{phi4-12-0}
\phi^{[(4)1]}_{1} = \phi^{[(4)1]}_{2} = 0 \ .
\eeq
Now compute the rank of $M_2$ numerically, initializing the remaining
moduli to have random, but non-zero, values. Generically, we find
\beq
\rk(M_2) = 82 \ .
\eeq
Continuing in this manner, we can set both blocks 
$m_{(4)1}$ and $m_{(4)2}$ to zero. That is, in addition to 
\eref{phi4-12-0}, take
\beq
\phi^{[(4)2]}_{1} = 
\phi^{[(4)2]}_{2} = 
\phi^{[(4)2]}_{3} = 
0 \ .
\eeq
Again, numerically computing $\rk(M_2)$ for arbitrary, non-zero values
of the remaining moduli, we find that
\beq
\rk(M_2) = 79
\eeq
generically. And so on.
In the table below, we
present the generic rank of $M_2$ evaluated for specific 
blocks of moduli set to zero.
\beq\label{tabrk}
\ba{|l|c|}\hline
\mbox{Block $m_{(4)i}$ set to 0} & \mbox{Generic rank of~} M_2 \\
\hline
\mbox{none}	& 85 \\ \hline
i=1		& 82 \\ \hline
i=1,2		& 79 \\ \hline
i=1,2,3 	& 70 \\ \hline
i=1,2,3,4	& 61 \\ \hline
i=1,2,3,4,5	& 53 \\ \hline
i=1,2,3,4,5,6	& 46 \\ \hline
i=1,2,3,4,5,6,7	& 40 \\ \hline
i=1,2,3,4,5,6,7,8	&35 \\ \hline
i=1,2,3,4,5,6,7,8,9	&28 \\
\hline\ea
\eeq
We see that this procedure stops at rank 28 when we have set all of
the triangular cluster of $m_{(4)i}$ sub-matrices to zero while still
keeping the $m_{(6)i}$ generic. For reasons to be discussed at the end
of this section, we choose not to consider non-generic values of
$m_{(6)i}$. Nevertheless,
one sees from \eref{tabrk} that we have achieved a wide range of
values for $\rk(M_2)$. We have attempted to enlarge this range by
modifying our approach and setting only some, but not all, of the
moduli within each $m_{(4)i}$ block to zero. Unfortunately, we did not
achieve any new, intermediate values for $\rk(M_2)$ beyond those
already found.

Not surprisingly, these results differ from those obtained in the
purely numerical approach leading to \fref{f:M2jump}. They complement
each other in two ways. First, the results in \eref{tabrk} clearly
indicate that values of $\rk(M_2)$ much smaller than 79 are
attained. On the other hand, the results in \fref{f:M2jump} imply
that, given enough statistics, one would expect $\rk(M_2)$ to attain
all integer values from 85 all the way down to 28, not simply the
non-consecutive results given in \eref{tabrk}. However, we have not
definitively proven this.

\begin{figure}
\input{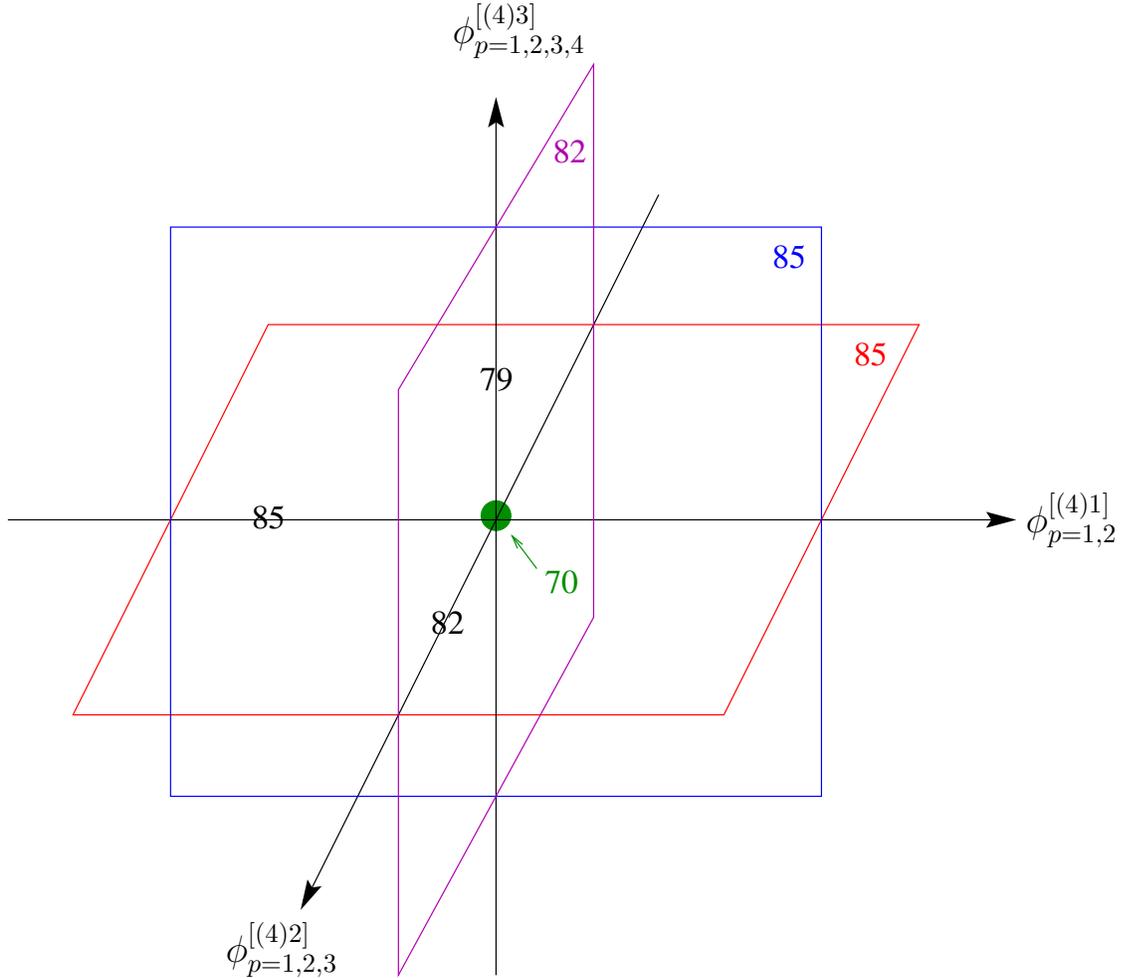}
\caption{{ A subspace of the moduli space $\cM$ of $\phi$'s spanned
by $\phi^{[(4)1]}_{p=1,2}$, $\phi^{[(4)2]}_{q=1,2,3}$ and 
$\phi^{[(4)3]}_{r=1,2,3,4}$. Generically, in the bulk, the rank of $M_2$
is 85, its maximal value. As we restrict to various planes and
intersections thereof, we are confining ourselves to special
sub-spaces of co-dimension one or higher. In these subspaces, the rank
of $M_2$ can drop dramatically.
}\label{f:moduli}}
\end{figure}

We can analyze this jumping phenomenon in the following way. 
There are, as shown in subsection \ref{s:vv}, 223 vector
bundle moduli in our specific example. These parametrize the full
moduli space. Of these, there are 139 moduli occurring in $M_2$,
constituting a 139 complex-dimensional subspace $\cM$. This number is
easily counted from the block form of $M_2$ given in \eref{M2} and
expression \eref{blockM}. As one
moves in $\cM$, the generic value of $\rk(M_2)$ is 85. However, as one
touches certain sub-spaces of $\cM$ of co-dimension one or
higher, the rank of $M_2$ drops. At the various 
intersections of these sub-spaces, that is, at sub-spaces of even
higher co-dimension, the rank may drop further. In the example
\eref{tabrk}, the sub-spaces, in particular, are the coordinate
planes in $\cM$ where some moduli are set to zero. 
To be specific, let us consider a particular subspace of $\cM$,
with
three axes corresponding respectively to $\phi^{[(4)1]}_{p=1,2}$,
$\phi^{[(4)2]}_{q=1,2,3}$ and $\phi^{[(4)3]}_{r=1,2,3,4}$.
In the bulk of this space, $\rk(M_2) =
85$ generically. If we hit the plane $\phi^{[(4)1]}_{p=1,2} = 0$,
\eref{tabrk} tells us that $\rk(M_2) = 82$. If we hit the intersection
of the planes $\phi^{[(4)1]}_{p=1,2} = 0$ and $\phi^{[(4)2]}_{p=1,2,3}
= 0$, then the rank drops to 79. And so on.
We present this plot in
\fref{f:moduli}, indicating the various ranks as we restrict to
various intersections of the planes within
this region of moduli space.
We conclude that the rank of $M_2$ is highly sensitive
to where one evaluates it within the moduli space $\cM$.

There is one final issue that must  be addressed.
Recall that
the spectral cover $\cC_V$ must be irreducible 
to ensure the stability of $V$.
Now, $\cC_V$ is a section of
\beq
H^0(X, \cO_X(\cC_V)) = H^0(X, \cO_X(5 \sigma + \pi^* \eta)) \ .
\eeq
We can decompose this vector space into
cohomology groups on $\IP^1$ exactly as was done in the explicit
construction of the matrix $M_1$. 
This was carried out in Appendix D and presented in
\eref{basis3}.
For convenience, we remind the reader that
\beq
H^0(X, \cO_X(\cC_V))
= \bigoplus\limits_{R=1,2,3,4,6}
\bigoplus\limits_{i=R-3}^{3R-3} \hat{m}_{(R)i} (\hat{s}_{(R)i}-3 
	\pi^*\cE) A_{(R)},
\eeq
with
\bea
&&\hat{m}_{(R)i} = \pi^*m_{(R)i}, \quad
\hat{s}_{(R)i} = \pi^*s_{(R)i}, \quad
A_{(R)i} = 5\sigma+\pi^*((6-R)c_1(T\IF_1)) \nn \\
&&m_{(R)i}=\beta^*M_{(R)i}, \quad
s_{(R)i}=R c_1(T\IF_1)- i \cE \nn \\
&& M_{(R)i}= H^0(\IP^1, \cO_{\IP^1}(i)) \ .
\eea
It turns out that
a sufficient criterion for the irreducibility of $\cC_V$
lies only in the first step in the projection $\pi_*$ to the base $B =
\IF_1$. 
The conditions for the irreducibility of $\cC_V$ are found to be
\bea\label{mod-con}
(1) && m_{(1)i} \ne 0, \nn \\
(2) && m_{(6)i} \mbox{~generic},
\eea
where $m_{(R)i} \in M_{(R)i}$ are the sub-matrices defined in
\eref{mat-m}.
Therefore, in our matrices $M_1$ and $M_2$, whose constituent blocks
are the $m_{(R)i}$ sub-matrices, we must make sure that the two
conditions \eref{mod-con} are satisfied when making our choices of the
moduli. Now, $M_2$ does not depend on $m_{(1)i}$ and we have
been careful to set $m_{(6)i}$ to generic non-zero values in the
above discussions. 
Furthermore, $M_1$ does not depend on $m_{(6)i}$ and in our
choice of the maximal rank of 28, we have always made sure that 
$m_{(1)0}$, the only $m_{(1)i}$ sub-block occurring in $M_1$,
is non-zero. Therefore, \eref{mod-con} is indeed satisfied
in our choices and the spectral cover $\cC_V$ remains irreducible
throughout.

\bibliographystyle{JHEP}

\begin{thebibliography}{99}
\bibitem{mori}M.~Andreatta, ``An introduction to Mori theory: 
the case of surface,''\\
alpha.science.unitn.it/~andreatt/scuoladott1.pdf
\end{thebibliography}

\end{document}